\documentclass[twocolumn,showpacs,aps,prd,floatfix]{revtex4-1}

\usepackage{graphicx}
\usepackage{dcolumn}
\usepackage{latexsym}
\usepackage{amsmath}
\usepackage{amsfonts}
\usepackage{amssymb}
\usepackage{epsfig}
\usepackage{longtable}
\usepackage{rotating}
\usepackage{multirow}
\usepackage{enumerate}

\graphicspath{{figures/}}

\newcommand{\BABARPubYear}{15}
\newcommand{\BABARPubNumber}{006}
\newcommand{\SLACPubNumber}{16443}

\input{babarsym}

\newcommand{\deltat} {\ensuremath{\Delta t}\xspace}
\newcommand{\deltamd} {\ensuremath{\Delta m_d}\xspace}

\newcommand{\swave}{(K \pi)^{*0}_0}
\newcommand{\swavep}{(K \pi)^{*+}_0}
\newcommand{\swavem}{(K \pi)^{*-}_0}

\def\cerenkov{\v{C}erenkov}
\newcommand{\Belle}		{Belle}

\def\Btag{\ensuremath{B_{\rm tag}}\xspace}
\def\Brec{\ensuremath{B_{\rm rec}}\xspace}

\def\Fbar {\kern 0.18em\overline{\kern -0.18em \mathcal{F}}{}\xspace}
\def\Mbar {\kern 0.18em\overline{\kern -0.18em \mathcal{M}}{}\xspace}
\def\Ampbar {\kern 0.18em\overline{\kern -0.18em A}{}\xspace}
\newcommand{\splot}    {\mbox{$_s{\cal P}lot$}\xspace}
\newcommand{\sweights} {\mbox{$_s{\cal W}eights$}\xspace}

\newcommand{\laura}               {\mbox{\tt Laura++}\xspace}

\newcommand{\fisher}        {\mbox{$\cal F$}}

\def\Kpi   {\ensuremath{K^+\pi^-}\xspace}

\def\pipi   {\ensuremath{\pip\pim}\xspace}

\newcommand{\Rhoz}             {\mbox{$\rho(770)^{0}$} \xspace}

\newcommand{\Kl}            {\mbox{$K_{1}(1270)$} \xspace}
\newcommand{\Klp}            {\mbox{$K_{1}(1270)^{+}$} \xspace}

\newcommand{\Kll}            {\mbox{$K_{1}(1400)$} \xspace}

\newcommand{\KstarX}            {\mbox{$\Kstar(1410)$} \xspace}

\newcommand{\KstarIIz}            {\mbox{$\Kstar_{0}(1430)^{0}$} \xspace}

\newcommand{\Kstarlllmy}           {\mbox{$\Kstar_{2}(1430)$} \xspace}

\newcommand{\KstarlVmy}            {\mbox{$\Kstar(1680)$} \xspace}

\newcommand{\rhoz}               {\mbox{$\rho^0$} \xspace}

\newcommand{\D}                  {\mbox{$D$} \xspace}

\def \TDBbkgAa {\ensuremath{\Bz \to \Kstarz (\to K^{\pm} \pi^{\mp}) \g }\xspace}
\def \TDBbkgAb {\ensuremath{\Bz \to\Xsd(\to K^{\pm} \pi^{\mp}) \g}\xspace}
\def \TDBbkgAc {\ensuremath{\Bz \to \Kstarz (\to \KS \piz) \g }\xspace}
\def \TDBbkgAd {\ensuremath{\Bz \to\Xsd(\to  \KS \piz) \g}\xspace}
\def \TDBbkgB {\ensuremath{\Bp \to \Kstarp (\to \KS \pip) \g\, + \,\Bp \to \Xsu(\to \KS \pip) \g}\xspace}
\def \TDBbkgBa {\ensuremath{\Bp \to \Kstarp (\to \KS \pip) \g}\xspace}
\def \TDBbkgBb {\ensuremath{\Bp \to \Xsu(\to \KS \pip) \g}\xspace}

\def \TDBbkgCa {\ensuremath{\Bz \to \Xsd(\nrightarrow K \pi) \g}\xspace}
\def \TDBbkgCb {\ensuremath{\Bp \to \Xsu(\nrightarrow K \pi) \g}\xspace}

\def \BbkgA {\ensuremath{\Bz \to \Kstarz (\to K \pi) \g \,+ \,\Bz \to\Xsd(\to K \pi) \g}\xspace}

\def \BtoKspig {\ensuremath{\Bz \to \KS \piz \g}\xspace}
\def \MyChanel  {\ensuremath{\Bz \to \KS \rhoz \g}\xspace}
\def \MyChanelPiPi  {\ensuremath{\Bz \to \KS \pim \pip \g}\xspace}
\def \CPBdecays  {\ensuremath{\B \to f_{\CP} \g}\xspace}
\def \MyControlChanel  {\ensuremath{\Bp \to \Kp \pim \pip \g}\xspace}

\def \BztoKstarRhozg  {\ensuremath{\Bz \to K^{*\pm}(\KS \pi^{\pm})\pi^{\mp} \g}\xspace}

\def\mKpipi  {\ensuremath{m_{K  \pi  \pi}}\xspace}
\def\mKpi  {\ensuremath{m_{K \pi}}\xspace}
\def\mpipi  {\ensuremath{m_{\pi  \pi}}\xspace}
\def\Kpipi  {\ensuremath{\Kp \pim \pip}\xspace}
\def\Kspipi  {\ensuremath{\KS \pim \pip}\xspace}

\def\Kpipig  {\ensuremath{\Kp \pip \pim \g}\xspace}

\def\conti  {\ensuremath{udsc}\xspace}

\def \splot {$_s\mathcal{P}lot$\xspace}

\def \sweights {$_s$Weights\xspace}

\def \Xsu {\ensuremath{X_{su}}\xspace}
\def \Xsd {\ensuremath{X_{sd}}\xspace}

\def\Kres {{\ensuremath{K_{\rm res}}}\xspace}

\def\Seff {{\ensuremath{\mathcal{S}_{\KS\pip\pim\g}}}\xspace}
\def\Ceff {{\ensuremath{\mathcal{C}_{\KS\pip\pim\g}}}\xspace}
\def\S {{\ensuremath{\mathcal{S}}}\xspace}
\def\C {{\ensuremath{\mathcal{C}}}\xspace}
\def\Srho {{\ensuremath{\mathcal{S}_{\KS\rho\g}}}\xspace}
\def\Crho {{\ensuremath{\mathcal{C}_{\KS\rho\g}}}\xspace}
\def\D {{\ensuremath{\mathcal{D}_{\KS\rho\g}}}\xspace}

\newcommand{\twothirds}{\mbox{${2\over3}$}}

\long\def\inst#1{\par\nobreak\kern 4pt\nobreak
    {\it #1}\par\vskip 10pt plus 3pt minus 3pt}

\begin{document}

\begin{flushleft}
\babar-PUB-\BABARPubYear/\BABARPubNumber \\
SLAC-PUB-\SLACPubNumber
\end{flushleft}

\vspace*{1cm}
\title{
\large \bf
\boldmath
Time-dependent analysis of $B^0 \to \Kspipi\gamma$ decays and studies of the $\Kp\pim\pip$ system in \MyControlChanel decays
} 

%
\author{P.~del~Amo~Sanchez}
\author{J.~P.~Lees}
\author{V.~Poireau}
\author{V.~Tisserand}
\affiliation{Laboratoire d'Annecy-le-Vieux de Physique des Particules (LAPP), Universit\'e de Savoie, CNRS/IN2P3,  F-74941 Annecy-Le-Vieux, France}
\author{E.~Grauges}
\affiliation{Universitat de Barcelona, Facultat de Fisica, Departament ECM, E-08028 Barcelona, Spain }
\author{A.~Palano$^{ab}$ }
\affiliation{INFN Sezione di Bari$^{a}$; Dipartimento di Fisica, Universit\`a di Bari$^{b}$, I-70126 Bari, Italy }
\author{G.~Eigen}
\author{B.~Stugu}
\affiliation{University of Bergen, Institute of Physics, N-5007 Bergen, Norway }
\author{D.~N.~Brown}
\author{L.~T.~Kerth}
\author{Yu.~G.~Kolomensky}
\author{M.~J.~Lee}
\author{G.~Lynch}
\affiliation{Lawrence Berkeley National Laboratory and University of California, Berkeley, California 94720, USA }
% ...
\collaboration{The \babar\ Collaboration}
\noaffiliation

\begin{abstract}
   \noindent
   We measure the time-dependent $C\!P$ asymmetry in the radiative-penguin decay $B^0 \to {\ensuremath{K^0_{\scriptscriptstyle S}}}  \pi^- \pi^+ \gamma$, using a sample of $471$ million $\Upsilon(4S) \to B{\kern 0.18em\overline{\kern -0.18em B}}$ events recorded with the $\mbox{\slshape B\kern-0.1em{\smaller A}\kern-0.1em
    B\kern-0.1em{\smaller A\kern-0.2em R}}$ detector at the PEP-II $e^+e^-$ storage ring at SLAC. 
Using events with $m_{K\pi\pi}<1.8{\mathrm{\,Ge\kern -0.1em V\!/}c^2}$, we measure the branching fractions of $B^+ \to K^+ \pi^- \pi^+ \gamma$ and $B^0 \to K^0 \pi^- \pi^+ \gamma$, the branching fractions of the kaonic resonances decaying to $K^+ \pi^- \pi^+$, as well as the overall branching fractions of the $B^+ \to \rho^0 K^+ \gamma$, $B^+ \to K^{*0} \pi^+ \gamma$ and S-wave $B^+ \to (K \pi)^{*0}_0 \pi^+ \gamma$ components.
For events from the $\rho$ mass band, we measure the $C\!P$-violating parameters ${\mathcal{S}_{{\ensuremath{K^0_{\scriptscriptstyle S}}}\pi^+\pi^-\gamma}} = 0.14 \pm 0.25 \pm 0.03$ and ${\mathcal{C}_{{\ensuremath{K^0_{\scriptscriptstyle S}}}\pi^+\pi^-\gamma}} = -0.39 \pm 0.20^{+0.03}_{-0.02}$, where the first uncertainties are statistical and the second are systematic. 
We extract from this measurement the time-dependent $C\!P$ asymmetry related to the $C\!P$ eigenstate $\rho^0 {\ensuremath{K^0_{\scriptscriptstyle S}}}$ and obtain ${\mathcal{S}_{{\ensuremath{K^0_{\scriptscriptstyle S}}}\rho\gamma}} = -0.18 \pm 0.32^{+ 0.06}_{-0.05}$, which provides information on the photon polarization in the underlying $b \to s \gamma$ transition.
\end{abstract}


\maketitle

\section{INTRODUCTION}
\label{sec:introduction}

The $V\!-\!A$ structure of the Standard Model (SM) weak interaction implies that the circular polarization of the photon emitted in $b \to s\g$ transitions is predominantly left-handed, with contamination by oppositely-polarized photons suppressed by a factor $m_s/m_b$~\cite{Atwood:1997zr,Atwood:2004jj}.
Thus, \Bz mesons decay mostly to right-handed photons while decays of \Bzb mesons produce mainly left-handed photons.
Therefore, the mixing-induced \CP asymmetry in \CPBdecays decays, where $f_{\CP}$ is a \CP eigenstate, is expected to be small. 
This prediction may be altered by new-physics (NP) processes in which opposite helicity photons are involved. 
Especially, in some NP models~\cite{Fujikawa:1993zu, Babu:1993hx, Cho:1993zb}, the right-handed component may be comparable in magnitude to the left-handed component, without affecting the SM prediction for the inclusive radiative decay rate.
The present branching fraction measurement of ($\BR(\B \to X_{s}\g)_{\rm exp.} = (3.43 \pm 0.21 \pm 0.07) \times 10^{-4}$~\cite{Amhis:2014hma}) agrees with the SM prediction of ($\BR(\B \to X_{s}\g)_{\rm th.} = (3.15 \pm 0.23) \times 10^{-4}$~\cite{Misiak:2006zs}) calculated at next-to-next-to-leading order. 
Further information on right-handed photon could be obtained by measuring 
\CP asymmetries in different exclusive radiative decay modes.
Furthermore, \B meson decays to $K\pi\pi\g$ can display an interesting hadronic structure, since several resonances decay to three-body $K \pi \pi$ final state (referred to as ``kaonic resonances'' throughout the article). 
The decays of these resonances themselves exhibit a resonant structure, with contributions from $\Kstar\pi$, $K\rho$, and a $\swave \pi$ S-wave component.

In the present analysis, we extract information about the $K\pi\pi$ resonant structure by means of an amplitude analysis of the \mKpipi and \mKpi spectra in \MyControlChanel decays. 
Assuming isospin symmetry, we use these results to extract the mixing-induced \CP parameters of the process \MyChanel from the time-dependent analysis of \MyChanelPiPi decays without an explicit amplitude analysis of this mode.
Charge conjugation is implicit throughout the document.

The Belle Collaboration has previously reported a time-dependent \CP asymmetry measurement of \MyChanel decays~\cite{Li:2008qma}. 
Similar measurements with \BtoKspig decays have been reported by \babar~\cite{Aubert:2008gy} and \Belle~\cite{Ushiroda:2006fi}.
No evidence for NP was found in these measurements. 
The observed \CP asymmetry parameters are compatible with the SM predictions.
LHCb has recently reported a non-zero value of the photon polarization in $\Bp\to\Kp\pim\pip\g$ decays via the distribution of the angle of the photon with respect to the plane defined by the final state hadrons~\cite{Aaij:2014wgo}.
Studies of the processes $\Bp \to  \Kp \pim\pip\g$ and $\Bz \to  \KS\pim\pip\g$ including measurements of the branching fractions have been performed by both \babar~\cite{Aubert:2005xk} and Belle~\cite{Yang:2004as} using samples of 232 and 152 million \BB pairs, respectively. 
The latter analysis also determined the branching fraction of the resonant decay $\Bp \to \Klp\g$.

The article is organized as follows. In Sec.~\ref{sec:babar} we briefly describe the \babar\ detector and the data set. 
In Sec.~\ref{sec:CC_analysis_strategy} we describe the analysis strategy.
The amplitude analysis of \MyControlChanel decays and the time-dependent analysis of \MyChanelPiPi decays are described in Sec.~\ref{sec:CC_analysis} and Sec.~\ref{sec:TD_analysis}, respectively. Finally, we summarize the results in Sec.~\ref{sec:Summary}.

\section{THE \babar\ DETECTOR AND DATASET}
\label{sec:babar}

The data used in this analysis were collected with the \babar\ detector at the \pep2\ asymmetric-energy $e^+e^-$ storage ring at SLAC.
The sample consists of an integrated luminosity of $426.0\;\mathrm{fb}^{-1}$\cite{lumiPaper}, corresponding to $(470.9\pm 2.8)\times10^{6}$ $B\Bbar$ pairs collected at the \FourS resonance (``on-resonance''), and $44.5$~\invfb collected about $40$~\mev below the~\FourS (``off-resonance'').
A detailed description of the \babar\ detector is presented in Refs.~\cite{Aubert:2001tu,TheBABAR:2013jta}. 
The tracking system used for track and vertex reconstruction has two  components: a silicon vertex tracker (SVT) and a drift chamber (DCH), both operating within a 1.5~T magnetic field generated by a superconducting solenoidal magnet. 
A detector of internally reflected \cerenkov\ light (DIRC) is used for charged particle identification.
The energies of photons and electrons are determined from the measured light produced in electromagnetic showers inside a CsI(Tl) crystal electromagnetic calorimeter (EMC). 
Muon candidates are identified with the use of the instrumented flux return of the solenoid.
\section{ANALYSIS STRATEGY}
\label{sec:CC_analysis_strategy}

The main goal of the present study is to perform a time-dependent analysis of \MyChanelPiPi decays to 
extract the decay and mixing-induced
\CP asymmetry parameters, \Crho and \Srho, in the \MyChanel mode.
However, due to the large natural width of the $\Rhoz$, a non negligible number of \BztoKstarRhozg events, which do not contribute to \Srho, are expected to lie under the $\Rhoz$ resonance and modify \Srho. 
Using the formalism developed in Ref.~\cite{emi_kou}, which assumes the SM, the ``so called'' dilution factor \D can be expressed as
\begin{widetext}
\begin{eqnarray}
\label{eq:dilutionExpression}
\hspace*{-0.7cm}\D \equiv \frac{\Seff}{\Srho}
	&=&	\frac{
				\mathop{\mathlarger{\int}} \left[ 
							\left| A_{\rho\KS}\right|^2 - 
							\Big| A_{\Kstarp\pim}\Big|^2 -
							\left| A_{\swavep\pim}\right|^2 +	
							2\Re\!\left( A^{\ast}_{\rho\KS} A_{\Kstarp\pim}\right) +	
							2\Re\!\left( A^{\ast}_{\rho\KS} A_{\swavep\pim}\right) 
					\right]dm^2
			}
			{
				\mathop{\mathlarger{\int}} \left[ 
							\left| A_{\rho\KS}\right|^2 + 
							\Big| A_{\Kstarp\pim}\Big|^2 +
							\left| A_{\swavep\pim}\right|^2 +	
							2\Re\!\left( A^{\ast}_{\rho\KS} A_{\Kstarp\pim}\right) +	
							2\Re\!\left( A^{\ast}_{\rho\KS} A_{\swavep\pim}\right) 
					\right]dm^2
			}
\end{eqnarray}
\end{widetext}
where \Seff is the effective value of the mixing-induced \CP asymmetry measured for the whole \MyChanelPiPi dataset and $A_{RP}$ is the (complex) amplitude of the mode $RP$, where $R$ represents a hadronic resonance and $P$ a pseudoscalar particle. 
Here, $\Re(A)$ denotes the real part of the complex number $A$. 
We assume the final state $\KS \pip \pim$ to originate from a few resonant decay modes where $R$ corresponds to $\rhoz$, $\Kstarp$, $\Kstarm$, $\swavep$ or $\swavem$ S-wave. 
Since a small number of events is expected in this sample, the extraction of the $A_{RP}$ amplitudes from the \MyChanelPiPi sample is not feasible.
Instead, the amplitudes of the resonant modes are extracted from a fit to the \mKpi spectrum in the decay channel \MyControlChanel, which has more signal events and is related to \MyChanelPiPi by isospin symmetry. 
Assuming that the resonant amplitudes are the same in both modes, the dilution factor is calculated from those of \MyControlChanel.
While the entire phase-space region is used to extract the amplitudes in the charged decay channel, the integration region over the plane of the \Kpi and \pipi invariant masses in the calculation of \D is optimized in order to maximize the sensitivity on \Srho. 
Note that the expression of \D used in the present analysis slightly differs from the one used in the previous analysis performed by the Belle Collaboration~\cite{Li:2008qma}.

Moreover, the decay to the $\Kpipi\gamma$ final state proceeds in general through resonances with a three-body \Kpipi final state. 
Although the contributions of some of these states to the \MyControlChanel decay, such as \Kl or \Kstarlllmy have been measured, not all the contributions have been identified~\cite{Agashe:2014kda}. 
Since each of these resonances has different $\Kstar\pi$ and $K \rho$ mass spectra (see Sec.~\ref{sec:mKpi_model}), it is necessary to first determine the three-body resonance content of the \mKpipi spectrum by fitting the charged \MyControlChanel sample.

Two types of Monte Carlo (MC) samples are used to characterize signal and background, and to optimize the selection in both analyses of \MyControlChanel and \MyChanelPiPi. 
Generic \BB MC and MC samples for specific exclusive final states are used to study backgrounds from \B-meson decays, whereas only MC samples for specific exclusive final states are used to study signal events.
The size of the generic \BB MC sample approximately corresponds to three times that of the data sample.

\section{AMPLITUDE ANALYSIS OF \boldmath{\MyControlChanel} DECAYS}
\label{sec:CC_analysis}

In Sec.~\ref{sec:CCEventSelection}, we describe the selection requirements used to obtain the signal candidates and to suppress backgrounds.
In Sec.~\ref{sec:CC_likeDiscrim}, we describe the unbinned extended maximum-likelihood fit method used to extract the yield of \MyControlChanel correctly-reconstructed (CR) signal events from the data.
Using information from this fit, the \Kpipi, \Kpi, and \pipi invariant-mass spectra (\mKpipi, \mKpi, and \mpipi) for CR signal events are extracted by means of the \splot technique~\cite{Pivk:2004ty}. 
In a second step, we perform a binned maximum-likelihood fit to the CR signal \splot of \mKpipi to determine from data the branching fractions of the various kaonic resonances decaying to \Kpipi. 
We finally perform a binned maximum-likelihood fit to the CR signal \splot of \mKpi to extract from data the amplitudes and the branching fractions of the two-body resonances decaying to \Kpi and \pipi. 
The \mKpipi and the \mKpi fit models are described in Sec.~\ref{sec:mKpipi_model} and Sec.~\ref{sec:mKpi_model}, respectively.

In Sec.~\ref{sec:CC_results}, we present the results of the three fits described above, and finally, we discuss systematic uncertainties on the results in Sec.~\ref{sec:CC_systematics}.

\subsection{Event selection and backgrounds}
\label{sec:CCEventSelection}

We reconstruct \MyControlChanel candidates from a high-energy photon, a pair of oppositely-charged tracks consistent with pion hypotheses and one charged track consistent with a kaon hypothesis, based on information from the tracking system, from the EMC and from the DIRC.
The center-of-mass energy of the photon is required to be between  $1.5$ and $3.5 \gev$, as expected in a \B radiative decays. 
The system formed by the final state particles is required to have a good-quality vertex.

A \B-meson candidate is characterized kinematically by the energy-substituted mass $\mes \equiv \sqrt{(s/2+\mathrm{\bold{p}}_i\cdot\mathrm{\bold{p}}_{B})^2/E_{i}^2-\mathrm{\bold{p}}_{B}^{2}} $ and energy difference $\DeltaE = E_{B}^{*}-\sqrt{s}/2$, 
where $\left(E_B,\mathrm{\bold{p}}_B\right)$ and $\left(E_i,\mathrm{\bold{p}}_i\right)$ are the four-vectors of the \B candidate and of the initial electron-positron system, respectively, in the laboratory frame. 
The asterisk denotes the center-of-mass frame, and $s$ is the square of the invariant mass of the electron-positron system. 
We require $5.200 < \mes < 5.292 \gevcc$ and $ \left| \DeltaE \right|< 0.200 \gev$.

Since the \FourS is only just above the threshold for \BB production, the decay products from such events are approximately spherical in the center-of-mass frame, whereas $\epem\to q\bar q$ ($q=u,d,s,c$) continuum background events have a di-jet-like structure.
To enhance discrimination between signal and the continuum background we use a Fisher discriminant~\cite{Fisher:1936et} to combine six discriminating variables: 
the angle between the momentum of the \B candidate and the beam ($z$) axis in the center-of-mass frame, 
the angles between the \B thrust axis~\cite{Bagan:1991sg,Ball:1994uh} and the $z$ axis and between the \B thrust axis and that of the rest of the event, 
 the zeroth-order momentum-weighted Legendre polynomial $L_{0}$ and the second-to-zeroth-order Legendre polynomials ratio $L_{2}/L_{0}$ of the energy flow about the \B thrust axis, 
 and the second-to-zeroth-order Fox-Wolfram moments~\cite{Blasi:1993fi} ratio. 
The momentum-weighted Legendre polynomials are defined by $L_0 = \sum_i |\textbf{p}_i|$ and $L_2  =  \sum_i |\textbf{p}_i| \frac{1}{2} \left( 3\cos^2\theta_i -1 \right)$, 
where $\theta_i$ is the angle with respect to the \B thrust axis of track or neutral cluster $i$ and $\textbf{p}_i$ is its momentum. 
The sums exclude the \B candidate and all quantities are calculated in the \Y4S frame. 
The Fisher discriminant is trained using off-resonance data for the continuum and a mixture of simulated exclusive decays for the signal. 
The final sample of candidates is selected with a requirement on the Fisher discriminant output value ($\fisher$) that retains $90\%$ of the signal and rejects $73\%$ of the continuum background.

We use simulated events to study the background from \B decays other than our signal (\B background). 
In preliminary studies, a large number of channels were considered, of which only those with at least one event expected after selection are considered here. 
The main \B backgrounds originate from \btosgam processes. 
\B background decays are grouped into classes of modes with similar kinematic and topological properties.

In order to reduce backgrounds from photons coming from \piz and $\eta$ mesons, we construct \piz and $\eta$ likelihood ratios, $\mathcal{L}_{\mathcal{R}}$, 
for which the photon candidate, $\gamma_1$ is associated with all other photons in the event, $\gamma_2$, such that
\begin{equation}
\label{eq:LR}
\mathcal{L}_{\mathcal{R},h^0} = \frac{p(m_{\g_1\g_2},E_{\g_2}|h^0)}{p(m_{\g_1\g_2},E_{\g_2}|\Kpipig) + p(m_{\g_1\g_2},E_{\g_2}|h^0)},
\end{equation}
where $h^0$ is either \piz or $\eta$, and $p$ is a probability density function in terms of $m_{\g_1\g_2}$ and the energy of $\g_2$ in the laboratory frame, $E_{\g_2}$. 
The value of $\mathcal{L}_{\mathcal{R},(\piz/\eta)}$ corresponds to the probability for a photon candidate to originate from a $\piz/\eta$ decay.
We require $\mathcal{L}_{\mathcal{R},\piz} < 0.860 $($\mathcal{L}_{\mathcal{R},\eta} < 0.957$), resulting, if applied before any other selection cut, in a signal efficiency of $\sim 93\%$($95\%$) and in background rejection factors of $\sim 83\%$($87\%$) for continuum events and $\sim 63\%$($10\%$) for \B-background events.

The optimization of the selection criteria was done using the \texttt{Choudalakis:2011qn} algorithm~\cite{Choudalakis:2011qn}. 
We optimized the $S/\sqrt{S + B}$ figure of merit using several selection variables from which the kaon and pion particle identification levels, the $\piz$ and $\eta$ likelihood ratios and the vertex $\chi^2$ of the system formed by the final state particles. 
In the optimization, we used CR signal events from simulation, off-resonance data for combinatorial background and generic $\BB$ simulated events (filtered to remove signal) for $B$ backgrounds.

Table \ref{tab:CC_bbackground} summarizes the six mutually exclusive $B$-background classes that are considered in the present analysis.

\begin{table*}[htbp]
\begin{center}
\caption{ \label{tab:CC_bbackground}
Summary of \B-background classes included in the fit model to \MyControlChanel decays.
If the yield is a free parameter in the fit, the listed values correspond to the fit result. 
Otherwise the expected value is given, which is computed from the branching fraction and selection efficiency.
The terms ``$X_{su(sd)}(\nrightarrow K \pi)$'' designate all $X_{su(sd)}$ decays but the $K \pi$ final state.
The functions used to parametrize the \B-background probability density functions of \mes, \DeltaE and $\fisher$ are also given. 
The notations ``Exp.'', ``CB'' and ``$\tilde{\mathrm{G}}$'' correspond to the exponential function, the Crystal Ball function (given in Eq.~\eqref{eq:CB_function}) and the modified Gaussian function (given in Eq.~\eqref{eq:cruijff}), respectively.}
\setlength{\tabcolsep}{0.0pc}
\begin{tabular*}{\textwidth}{@{\extracolsep{\fill}}lccccc}
\hline\hline
\multirow{2}{*}{Class}                                 &   \multicolumn{3}{c}{PDFs}  & \multirow{2}{*}{Varied}                    & \multirow{2}{*}{Number of events} \\
& \mes &  \DeltaE & \fisher\ & \\
\hline\\[-9pt]
 $B^0 \to X_{sd}(\nrightarrow K \pi) \gamma $      &  $\tilde{\mathrm{G}}$ $+$  &  \multirow{2}{*}{Exp.} & \multirow{2}{*}{Gaussian} & \multirow{2}{*}{no}                              & \multirow{2}{*}{$ 2872 \pm 242  $} \\
$B^+ \to X_{su}(\nrightarrow K \pi) \gamma $& ARGUS & & & & \\
\hline\\[-9pt]
 $B^0 \rightarrow K^{*0}(\to K\pi) \gamma$          & \multicolumn{2}{c}{Two-dimensional} & \multirow{2}{*}{$\tilde{\mathrm{G}}$} & \multirow{2}{*}{yes}                       & \multirow{2}{*}{$ 1529	 \pm	116  $} \\
$B^0 \rightarrow X_{sd}(\to K\pi) \gamma$& \multicolumn{2}{c}{nonparametric} & & & \\
\hline\\[-9pt]
 $B^+ \rightarrow K^{*+}(\to K\pi) \gamma$          &   Linear $+$ & \multirow{2}{*}{Exp.}  & \multirow{2}{*}{$\tilde{\mathrm{G}}$} & \multirow{2}{*}{no}                  & \multirow{2}{*}{$ 442	 \pm	50  $} \\
$  B^+ \rightarrow X_{su}(\to K\pi) \gamma$&ARGUS  & & & & \\
\hline\\[-9pt]
\multirow{2}{*}{$B^0 \to K^{*0} \eta  $ }        &  $\tilde{\mathrm{G}}$ $+$  & Gaussian $+$   & \multirow{2}{*}{$\tilde{\mathrm{G}}$} & \multirow{2}{*}{no}                & \multirow{2}{*}{$ 56	 \pm	21  $} \\
&ARGUS & Constant & & & \\
\hline\\[-9pt]
 $B^+ \to a_1^{+}(\to\rhoz \pip )\piz \gamma $  		&   \multirow{2}{*}{CB} &  Asymmetric & Asymmetric & \multirow{2}{*}{no}                   & \multirow{2}{*}{$ 17  \pm 9  $} \\
$ B^+ \to K^{*0}(\to K \pi) \pip \piz \gamma $& &Gaussian &Gaussian & & \\
\hline\\[-9pt]
\multirow{2}{*}{$B \to \{\text{charged and neutral generic decays}\}$} 	& \multirow{2}{*}{ARGUS}  &  \multirow{2}{*}{Exp.} & \multirow{2}{*}{Gaussian} & \multirow{2}{*}{yes}  & \multirow{2}{*}{$3270 \pm 	385$}  \\
& & & & & \\
\hline\hline
\end{tabular*}
\end{center}
\end{table*}

\subsection{The maximum-likelihood fit and extraction of the physical observables}
\label{sec:CC_likelihoodFit}

\subsubsection{The \mes, \DeltaE, and $\fisher$ PDFs}
\label{sec:CC_likeDiscrim}

We perform an unbinned extended maximum-likelihood fit to extract the \MyControlChanel event yield. 
We further obtain the signal \mKpipi, \mKpi and \mpipi spectra, where the background is statistically subtracted using the \splot technique.
Note that this technique may produce bins with negative entries. 
The fit is performed using the \laura package~\cite{lauraCite}.
The fit uses the variables \mes, \DeltaE, and the Fisher-discriminant output $\fisher$, to discriminate CR signal events from other event categories.
The likelihood function ${\cal L}_i$ for the event $i$ is the sum
\begin{equation}
\label{eq:PropDenSingle}
 {\cal L}_i=\sum_j{N_j{\cal P}^i_j(\mes,\Delta E, \fisher)},
\end{equation}
where $j$ stands for the event species (signal, continuum and the various \B backgrounds) and $N_j$ is the corresponding yield. 
The CR yield is a free parameter in the fit to the data, while the mis-reconstructed signal yield is fixed, defined as the product of the mis-reconstructed signal ratio obtained from simulation and the signal branching fraction taken from Ref.~\cite{Agashe:2014kda}.
If no correlation is seen among the fitting variables, the probability density function (PDF) ${\cal P}^i_j$ is the product of three individual PDFs:
\begin{equation}
\label{eq:PropProd}
 {\cal P}^i_j={\cal P}^i_j(\mes) \; {\cal P}^i_j(\Delta E) \; {\cal P}^i_j(\fisher).
\end{equation}
Otherwise, the correlations are taken into account through multi-dimensional PDFs that depend on the correlated variables.
The total likelihood is given by
\begin{equation}
\label{eq:dp_Likelihood}
 {\cal L}=\exp(-\sum_j N_j)\prod_i {\cal L}_i.
\end{equation}

The \mes\ distribution of CR signal events is parametrized by a Crystal Ball function (CB)~\cite{Skwarnicki:1986xj,Oreglia:1980cs,Gaiser:1982yw} defined as 
\begin{eqnarray}
\label{eq:CB_function}
 \lefteqn{\textrm{CB}(x;\mu,\sigma,\alpha,n) = } \\
&&   \left\{ \begin{array}{ll}
  \left(\frac{n}{\alpha}\right)^n \frac{\textrm{exp}(-\alpha^2/2)}{((\mu - x) / \sigma + n/\alpha -\alpha)^n} & x \leq \mu - \alpha\sigma,\\
 & \\
 \, \textrm{exp} \left[ - \frac{1}{2} \left(\frac{x-\mu}{\sigma}\right)^2 \right] & x > \mu - \alpha\sigma.
       \end{array} \right. \nonumber
\end{eqnarray}
where the parameters $\mu$ and $\sigma$ designate the mean and width of a Gaussian distribution that is joined at $\mu - \alpha\sigma$ to a power law tail.
The \DeltaE\ distribution of CR signal events is parametrized by a modified Gaussian ($\tilde{\mathrm{G}}$) defined as
\begin{eqnarray}
\label{eq:cruijff}
& &  \tilde{\mathrm{G}}(x;\mu,\sigma_l,\sigma_r,\alpha_l,\alpha_r)=\\
\nonumber & &  \exp\left(-\frac{(x-\mu)^2}{2\sigma_k^2+\alpha_k(x-\mu)^2}\right)
\left\{
\begin{array}{ll}
x-\mu<0 :    & k=l,\\
x-\mu\geq0 : & k=r.
\end{array}
\right.
\end{eqnarray} 
The $\mu$ and $\sigma_l$ parameters are free in the fit to the data, while the other parameters are fixed to values determined from simulations.
Correlations between \mes and \DeltaE in CR signal are taken into account through a two-dimensional  PDF. 
It is constructed as the product of a conditional PDF (CB for \mes) by a marginal PDF ($\tilde{\mathrm{G}}$ for \DeltaE). 
The dependences on \DeltaE of the CB parameters $\mu$ and $\sigma$ are parametrized by two second-order polynomials, while those of the parameters $\alpha$ and $n$ are parametrized by two first-order polynomials. 
The three parameters of both second-order polynomials are determined by the fit, while the parameters of the first-order polynomials are fixed in the fit to the values determined from simulations.
The $\fisher$ PDF of CR signal events is parametrized by a Gaussian, for which the mean and variance are left free in the fit to the data.
No significant correlations were found between $\fisher$ and either \mes or \DeltaE.

The shape parameters of the PDFs of mis-reconstructed signal events are fixed to values determined from simulations.
The \mes PDF is parametrized by the sum of an asymmetric Gaussian and of an ARGUS shape function~\cite{Albrecht:1990cs}, while the \DeltaE and $\fisher$ PDFs are parametrized by a first-order polynomial and a Gaussian, respectively. 

The \mes, \DeltaE and $\fisher$ PDFs for continuum events are parametrized by an ARGUS shape function, a second-order Chebychev polynomial and an exponential function, respectively, with parameters determined by the fit, except for the exponential shape parameter, which is fixed to the value determined from a fit to off-resonance data.

The \mes, \DeltaE and $\fisher$ PDFs for all the classes of $B$-background events are described by parametric functions, given in Table~\ref{tab:CC_bbackground}, except for the $B^0 \to K\pi \gamma$ background \mes and \DeltaE PDFs, for which significant correlations are present. 
These are taken into account through a nonparametric two-dimensional PDF, defined as a histogram constructed from a mixture of simulated events. 
No significant correlations were found among the fit variables for the other species in the fit.
The distributions of the combined \BzBzb and \BpBm generic \B backgrounds were studied using generic \BB MC from which all other \B background class contributions were filtered out.
The shape parameters of the $B$-background PDFs are fixed to values determined from simulated events.
If the yield of a class is allowed to vary in the fit, the number of events listed in Table~\ref{tab:CC_bbackground} corresponds to the fit results.
For the other classes, the expected numbers of events are computed by multiplying the selection efficiencies estimated from simulations by the world average branching fractions~\cite{Agashe:2014kda,Amhis:2014hma}, scaled to the data set luminosity. 
The yield of the $B^0 \rightarrow K\pi \gamma$ class, which has a clear signature in \mes, and that of the generic $B$-background class are left free in the fit to the data. The remaining background yields are fixed.

\subsubsection{The \mKpipi spectrum}
\label{sec:mKpipi_model}

We model the \mKpipi distribution as a coherent sum of five resonances described by relativistic Breit--Wigner ($R_k$) lineshapes~\cite{Agashe:2014kda}, with widths that are taken to be constant.
The total decay amplitude is then defined as:
\begin{equation}
\left|A(m;c_k)\right|^2 = \sum_J \left| \sum_k c_k\, R^J_k(m)\right|^2, 
\label{equ:mKpipi_amplitude}
\end{equation}
with
\begin{equation}
R^J_k(m) = \frac{1}{(m^0_k)^2 - m^2 - im^0_k\Gamma^0_k},
\label{equ:RBW_formula}
\end{equation}
and where $c_k = \alpha_k \,e^{i\phi_k}$ and $m=\mKpipi$.
In Eq.~\eqref{equ:mKpipi_amplitude}, the index $J$ runs over the different spin-parities ($J^P$) and the index $k$ runs over the \Kpipi resonances of same $J^P$. The coefficients $\alpha_k$ and $\phi_k$ are the magnitude and the phase of the complex coefficients, $c_k$, corresponding to a given resonance. 
Due to the fact that helicity angles are not explicitly taken into account in the fit model, it only has to account for interference between resonances with same spin-parity $J^P$.
Table~\ref{tab:mKpipi_model} details the resonances in the \mKpipi fit model.
The $\Kl$ magnitude is fixed to 1, and the $\Kl$,  $\KstarlVmy$, and $\Kstarlllmy$ phases are fixed to 0. It has been checked that the choice of reference does not affect the results. 
The remaining parameters of the complex coefficients are left free in the fit: namely the $\Kll$, $\KstarX$, $\KstarlVmy$, and $\Kstarlllmy$ magnitudes as well as the two relative phases; that between the two $J^P=1^+$ resonances and that between the two $J^P=1^-$ resonances.

In addition to the complex coefficients, the widths of the two resonances, $\Kl$ and $\KstarlVmy$, are left free in the fit.
In the case of the $\Kl$, this is motivated by the fact that the width quoted in Ref.~\cite{Agashe:2014kda} might be underestimated according to the measurements reported in Ref.~\cite{Daum:1981hb}. 
In the case of the $\KstarlVmy$, the uncertainty on the width quoted in Ref.~\cite{Agashe:2014kda} is large. 
In total, eight parameters are kept free in the fit.

Note that we do not take phase-space effects into account here. 
However, distortions of lineshapes of the \Kpipi resonances may occur from two sources: the available energy in the production process (i.e. $\B \to K_{\mathrm{res.}} \g$), and a mass of intermediate-state particles close to threshold, as for instance in the case of $\Kl \to K \Rhoz$.
For each \Kpipi resonance, the first source of distortion is studied by comparing the invariant-mass distribution generated by {\tt EvtGen}~\cite{Lange:2001uf} to the $R_k$ mass used as an input to the generator.
We see no significant distortion. 
For each resonance, the second source of distortion is estimated from the known properties of all decaying processes. 
Ideally, one should perform an iterative procedure in which the input values of the decaying processes are compared to the results of the fit to the data repeating the procedure until the fit results converge to values compatible with the inputs.
However, due to the limited size of the data sample, we use the effective model described in Eqs.~\eqref{equ:mKpipi_amplitude} and~\eqref{equ:RBW_formula} where no correction is applied to the lineshapes. As described in Sec.~\ref{sec:CC_results_mes_deltae_fisher}, this approach describes the data well.
 
The fit fractions FF($k$) extracted for each resonance, as well as the interference fit fractions FF$(k,l)$ between same $J^P$ resonances, are calculated as:
\begin{eqnarray}
\label{equ:fit_fractions_res}
\mathrm{FF}(k) &=& \frac{\left| c_{k} \right|^2 \left<R_k R_k^* \right>}{\sum_{\mu\nu}(c_{\mu}c_{\nu}^*)\left<R_{\mu}R_{\nu}^* \right>}, \\
\label{equ:fit_fractions_inter}
\mathrm{FF}(k,l) &=& \frac{2\,\Re\{(c_{k}c_l^*) \left<R_k R_l^* \right>\}}{\sum_{\mu\nu}(c_{\mu}c_{\nu}^*)\left<R_{\mu}R_{\nu}^* \right>}, 
\end{eqnarray}
where the terms $\left<R_{\mu}R_{\nu}^* \right>$ are:
\begin{equation}
\left<R_{\mu}R_{\nu}^* \right>  = \int R_{\mu}R_{\nu}^* dm.
\label{equ:fit_fractions_norm}
\end{equation}
The sum of fit fractions is defined as the algebraic sum of all fit fractions. 
This quantity is not necessarily unity due to the possible presence of net constructive or destructive interference.

The branching fraction to the $\Kp \pip \pim \g$ final state is determined from the fitted yield of the CR signal event category, $N^{\rm CR}_{\rm sig}$, the weighted CR signal efficiency, $\langle{\epsilon}^+\rangle$, and the number of charged \B events, $N_{\Bpm}$ 
\begin{equation}
\BR(\Bp \to \Kp \pip \pim \g) = \frac{N^{\rm CR}_{\rm sig}}{\langle{\epsilon}^+\rangle \times N_{\Bpm}},
\label{equ:CC_BR_tot_formula}
\end{equation}
with 
\begin{equation}
\langle{\epsilon}^+\rangle = \sum_{k} \epsilon^+_{k} \frac{{\rm FF}(k)}{\sum_l {\rm FF}(l)}.
\label{equ:weighted_effi_formula}
\end{equation}
Here, $k$ and $l$ run over the kaonic resonances, $\epsilon^+_{k}$ represents the efficiency without requirement on \mKpipi for resonance $k$ listed in Table~\ref{tab:CC_Cut_Efficiency_By_Mode} and ${\rm FF}$ are the fit fractions extracted from a binned maximum-likelihood fit to the CR signal \splot of \mKpipi plotted in 80 bins.
The term $N_{\Bpm}$ is obtained from the total number of \BB pairs in the full \babar\ dataset, $N_{\BB}$, and the corresponding $\Y4S$ branching fraction taken from Ref.~\cite{Agashe:2014kda}
\begin{eqnarray}
\label{eq:n_BBbar}
N_{\Bpm} & = & 2\times N_{\BB} \times \BR(\Y4S \to \BpBm) \\
& =& (483.2 \pm 6.4) \times 10^6. \nonumber
\end{eqnarray}
The branching fraction of each kaonic resonance
\begin{equation}
\label{eq:BR_Kres}
\!\! \BR(\Bp \to \Kres (\to \Kp \pip \pim) \g) = {\rm FF}(k) \frac{N^{\rm CR}_{\rm sig} }{\epsilon^{+\prime}_k \times N_{\Bpm}},
\end{equation}
is computed using the corresponding fit fraction ${\rm FF}(k)$ and efficiency accounting for the requirement on \mKpipi, $\epsilon^{+\prime}_k$, listed in Table~\ref{tab:CC_Cut_Efficiency_By_Mode}.

\begin{table}[htbp]
\begin{center}
\caption{\label{tab:mKpipi_model} The five kaonic resonances decaying to \Kpipi included in the model used to fit the \mKpipi spectrum. The pole mass $m^0$ and the width $\Gamma^0$ are taken from Ref.~\cite{Agashe:2014kda}.}
\vspace{5pt}
\begin{tabular*}{\columnwidth}{@{\extracolsep{\fill}}l | c c c}
\hline
\hline
\multirow{2}{*}{$J^P$} 	             	& \multirow{2}{*}{$K_{\textrm{\footnotesize{res}}}$} & Mass $m^0$	& Width $\Gamma^0$ \\
	             	&     	& 	(\mevccnosp) &  (\mevccnosp)\\
\hline
\multirow{2}{*}{$1^+$} 			&\Kl   	&	$1272\pm7\phantom{0}$	&	$90\pm20$	\\
	 		&	\Kll	&	$1403\pm7\phantom{0}$	&	$174\pm13\phantom{0}$	\\
\hline
 \multirow{2}{*}{$1^-$}	&\KstarX		&	$1414\pm15$	&	$232\pm21\phantom{0}$	\\
	&	\KstarlVmy 	&	$1717\pm27$	&	$322\pm110$	\\
\hline
 $2^+$	&	\Kstarlllmy	&	$1425.6\pm1.5\phantom{0}$	& 	$98.5\pm2.7\phantom{0}$	\\
\hline	
\hline	
\end{tabular*}
\end{center}
\end{table}
\begin{table}[htbp]
\caption{\label{tab:CC_Cut_Efficiency_By_Mode} 
Efficiencies $\epsilon^+_k$($\epsilon^{+\prime}_k$) for correctly-reconstructed signal candidates for each kaonic resonance from simulations without(with) the applied requirement $\mKpipi< 1.8\gevcc$. 
}
\vspace{10pt}
\begin{tabular*}{\columnwidth}{@{\extracolsep{\fill}}l c c}
\hline
\hline
\multirow{2}{*}{\Kres}  &	\multirow{2}{*}{$\epsilon^+_k$} &	\multirow{2}{*}{$\epsilon^{+\prime}_k$} \\
&\\
\hline
$\Kl^+$			& $0.2190 \pm 0.0006$ &	$0.2130 \pm 0.0006$ \\
$\Kll^+$		& $0.2250 \pm 0.0013$ &	$0.2110 \pm 0.0013$ \\
$\KstarX^+$	 	& $0.2056 \pm 0.0012$ &	$0.1926 \pm 0.0013$ \\
$\Kstarlllmy^+$	& $0.2130 \pm 0.0015$ &	$0.2092 \pm 0.0016$ \\
$\KstarlVmy^+$ 	& $0.1878 \pm 0.0022$ &	$0.1276 \pm 0.0020$	\\
\hline
\hline
\end{tabular*}
\end{table}
\subsubsection{The \mKpi spectrum}
\label{sec:mKpi_model}

In a third step, we perform a binned maximum-likelihood fit to the efficiency-corrected CR signal \mKpi \splot with 90 bins to extract amplitudes and branching fractions of the intermediate resonances decaying to \Kpi and \pipi.
The branching fractions of the intermediate state resonances are obtained using the averaged efficiency, $\langle{\epsilon}^+\rangle$, such that
\begin{equation}
\label{eq:BR_hRes}
\BR(\Bp \to R h \g) = {\rm FF}(R) \frac{N^{\rm CR}_{\rm sig} }{\langle{\epsilon}^+\rangle N_{\Bpm}},
\end{equation}
where $R$ denotes an intermediate state resonance and $h$ is either a kaon or a pion, and ${\rm FF}(R)$ is the corresponding fit fraction. 
\begin{figure}[b]
	\includegraphics[height= 6.2cm] {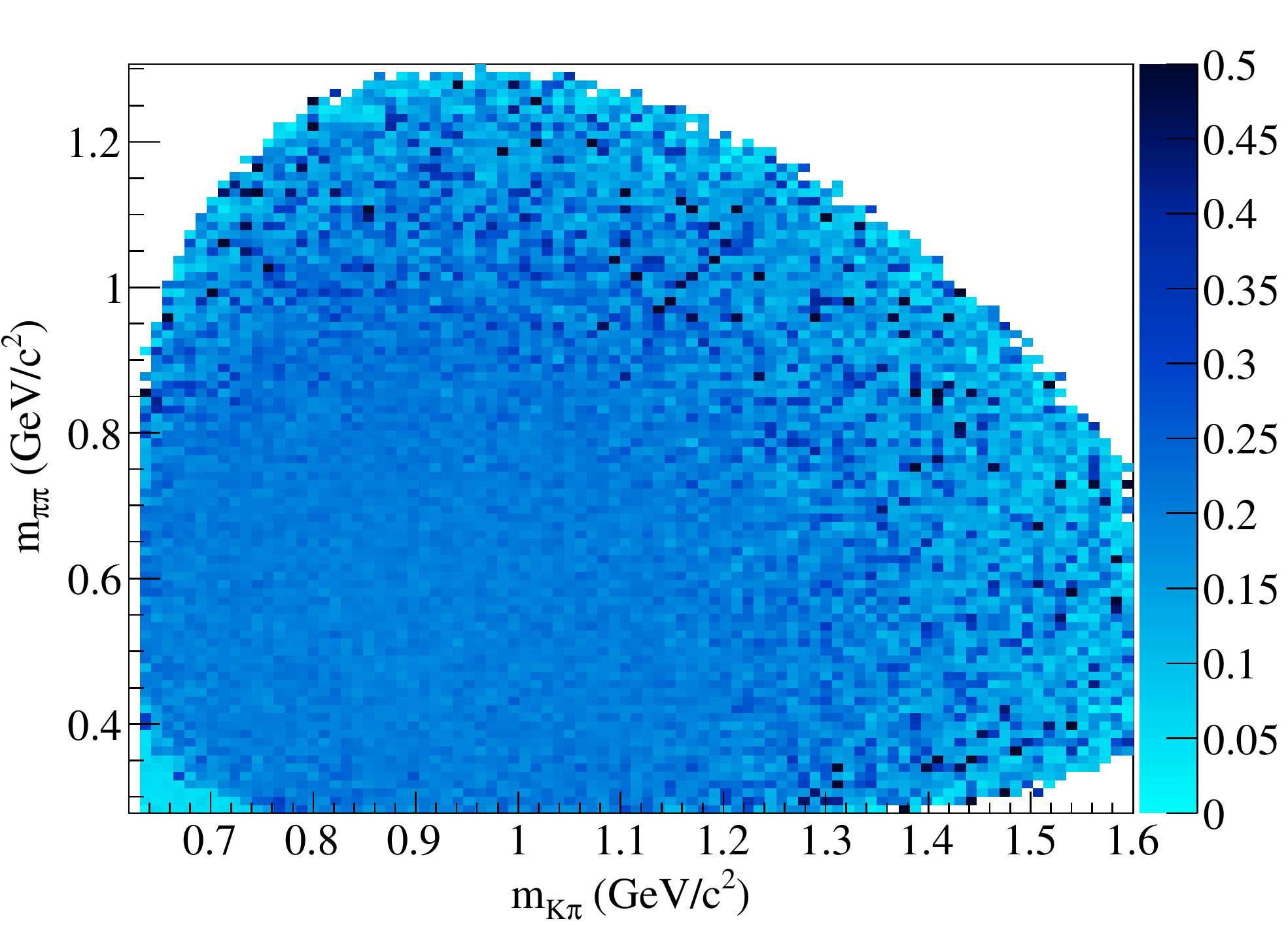}
\vspace{-10pt}
\caption{Combination of the efficiency maps for each kaonic resonance. The relative weights used for the combination are extracted from a fit to the \mKpipi spectrum (Sec.~\ref{sec:CC_results}). Large fluctuations at high \mKpi or \mpipi are due to small number of events. 
\label{fig:mKpi_efficiencies}}
\end{figure}
The resonance $R$ is decaying either to $K^+\pim$ when $h=\pip$ or to $\pip\pim$ when $h=\Kp$.
To correct for efficiency effects, we construct efficiency maps in the \mKpi-\mpipi plane for each kaonic resonance in the fit model. 
For each exclusive decay, the efficiency map is determined from the phase space decay of that resonance ($\Bp \to K_{res}(\Kpipi) \gamma$). 
The efficiency map of the combined sample shown in Fig.~\ref{fig:mKpi_efficiencies} is obtained by applying weights to the individual maps, which were extracted from the fit to the \mKpipi spectrum.
The \mKpi spectrum is corrected for efficiency effects by dividing the ($\mKpi,\,\mpipi$) \splot distribution by the combined efficiency map and integrating over the \mpipi dimension.
The approach of projecting the \mpipi-\mKpi phase space of \MyControlChanel onto the \mKpi axis was chosen since the sample size was too small for a two-dimensional fit.
This is further complicated by the four-body nature of the decay: since the value of \mKpipi can vary from event to event, the kinematic boundaries for the \mpipi-\mKpi plane vary as well.
We model the \mKpi spectrum as the projection of two $1^-$ P-wave and one $0^+$ S-wave components.
The two P-wave components, namely the $K^{*}(892)^{0}$ and the $\Rhoz$ resonances, are described by relativistic Breit--Wigner (RBW) and Gounaris-Sakurai (GS)~\cite{Gounaris:1968mw} lineshapes, respectively. 
The $0^+$ (S-wave) component of the $K\pi$ spectrum, designated by $\swave$, is modeled by the LASS parametrization~\cite{Aston:1987ir}, which consists of the $\KstarIIz$ resonance together with an effective range nonresonant (NR) component. 

Due to the relatively low mass of the \Kpipi resonances, the lineshapes of the two-body resonances are distorted; the phase space is noticeably different for events below and above the resonance pole mass. 
To account for this effect, we model the invariant-mass-dependent magnitude of each resonance $R_j$ by: $\sqrt{\textrm{H}_{\scriptsize{R}_j}(m_{K\pi},m_{\pi\pi})}$, where H is a two-dimensional histogram. 
The $K^{*}(892)^{0}$ and $\Rhoz$ histograms are directly generated from the Monte Carlo event generator~\cite{Lange:2001uf}, while the LASS parametrized S-wave histogram is obtained by applying weights to the sample of phase-space generated events, as described below. 
To take into account the interference between the components, invariant-mass-dependent phases $\Phi_{\scriptsize{R}_j}(m)$ are required. 
We make the hypothesis that the phases can be directly taken from the analytical expression of the corresponding lineshape
\begin{eqnarray}
&\Phi_{\scriptsize{R}_j}(m) = \arccos\left(\frac{\Re[R_j(m)]}{\left|R_j(m)\right|}\right) \label{eq:phi_res}\\
& \nonumber \\
& \left\{ \begin{array}{llll}
& & R_j(\mKpi)\textrm{  is taken as } & \nonumber\\
m = \mKpi & \Rightarrow & \textrm{RBW for }K^{*}(892)^{0} \textrm{ and}& \nonumber\\
 & & \textrm{as LASS for S-wave,} & \nonumber\\
& &  & \\
m = \mpipi &\Rightarrow & R_j(\mpipi)\textrm{  is taken as a GS} & \nonumber\\
& & \textrm{lineshape for } \Rhoz,&\nonumber 
       \end{array} \right.
\label{equ:phase_formula}
\end{eqnarray}
where the line-shapes are taken from the following expressions.

The RBW parametrization used to determine the corresponding invariant-mass-dependent phase, $\Phi_{\scriptsize{\textrm{\Kstar}}}(\mKpi)$, is defined as
\begin{equation}
R_j(m) = \frac{1}{(m^2_0 - m^2) - i m_0 \Gamma(m)},
\label{eqn:BreitWigner}
\end{equation}
where $m_0$ is the nominal mass of the resonance and $\Gamma(m)$ is the
mass-dependent width.
In the general case of a spin-$J$ resonance, the latter can be expressed as
\begin{equation}
\Gamma(m) = \Gamma_0 \left( \frac{|\textbf{q}|}{|\textbf{q}|_0} \right)^{2J+1} \left( \frac{m_0}{m}\right) \frac{X^2_J(|\textbf{q}|r)}{X^2_J(|\textbf{q}|_0 r)}.
\label{equ:RBW_Width_formula}
\end{equation}
The symbol $\Gamma_0$ denotes the nominal width of the resonance. The values of $m_0$ and $\Gamma_0$ are listed in Table~\ref{tab:mKpi_model}. The symbol $\textbf{q}$ is the momentum of one of the resonance daughters, evaluated in the resonance rest frame. The modulus of $\textbf{q}$ is a function of $m$ and the resonance daughter masses, $m_a$ and $m_b$, given by
\begin{equation}
|\textbf{q}| = \frac{m}{2}\left( 1-\frac{(m_a+m_b)^2}{m^2}\right)^{1/2} \left( 1-\frac{(m_a-m_b)^2}{m^2}\right)^{1/2}.
\label{equ:q_formula}
\end{equation}
The symbol $|\textbf{q}|_0$ denotes the value of $|\textbf{q}|$ when $m = m_0$. The $X_J(|\textbf{q}|r)$ function describes the Blatt-Weisskopf barrier factor~\cite{blatt_weisskopf} with a barrier radius of $r$. Defining the quantity $z = |\textbf{q}|r$, the Blatt-Weisskopf barrier function for a spin-1 resonance is given by
\begin{eqnarray}
X_{J = 1}(z) & = & \sqrt{\frac{1+z_0^2}{1+z^2}},
\end{eqnarray}
where $z_0$ represents the value of $z$ when $m = m_0$. 

\begin{table}[t]
\begin{center}
\caption{\label{tab:mKpi_model} 
The three resonances included in the model used in the fit to the \mKpi spectrum and their line-shape parameters.
The nominal mass and width of the resonance, $m_0$ and $\Gamma_0$, which are expressed in $\mevcc$, are taken from the references given in the table. The parameter $r$ for $\Rhoz$ and $K^{*}(892)^{0}$ is the Blatt-Weisskopf barrier radius, expressed in $(\gevcnosp)^{-1}$. The parameters $a$ and $r$ of the $\swave$ are the scattering length and the effective range, respectively, both expressed in $(\gevcnosp)^{-1}$.}
\vspace{5pt}
\setlength{\tabcolsep}{0.47pc}
\begin{tabular*}{\columnwidth}{@{\extracolsep{\fill}}l|lccc}
\hline
\hline
   \multirow{2}{*}{$J^P$}  &  \multirow{2}{*}{Resonance} 	      	&  \multirow{2}{*}{Parameters} & Analytical	& \multirow{2}{*}{ Ref.} \\
	           &  	&     	& 	Expression &  \\
\hline
\multirow{6}{*}{$1^-$} 	&	\multirow{3}{*}{$K^{*}(892)^{0}$} 		&$m_0 = 895.94\pm0.22 $   	&	\multirow{3}{*}{RBW} & \multirow{3}{*}{\cite{Agashe:2014kda}} \\
 		&	&$\Gamma_0 = 50.8\pm0.9 $  	& & \\
 		&	&$r = 3.6\pm0.6$    	& & \\
\cline{2-5}
 & \multirow{3}{*}{\Rhoz} 		&$m_0 = 775.49\pm0.34 $ & \multirow{3}{*}{GS} & \multirow{3}{*}{\cite{Agashe:2014kda}} \\
 			& &$\Gamma_0 = 149.1\pm0.8 $ & & \\
 			& &$r = 5.3^{+0.9}_{-0.7}$  	& & \\
\hline
\multirow{4}{*}{$0^+$}& \multirow{4}{*}{$\swave$}& $m_0 = 1425\pm50 $   	&	\multirow{4}{*}{LASS} & \multirow{2}{*}{\cite{Agashe:2014kda}}	 \\
 			& &$\Gamma_0 = 270\pm80 $ 	& & \\
 			& &$a= 2.07\pm0.10$   	& &\multirow{2}{*}{\cite{Aston:1987ir}} \\
 			& &$r = 3.32\pm0.34$ 	& & \\
\hline	
\hline	
\end{tabular*}
\end{center}
\end{table}

For the \Rhoz we use the GS parametrization, which describes the $P$-wave
scattering amplitude for a broad resonance decaying to two pions
\begin{equation}
\label{eq:rhoGS}
R_{j}(m) = \frac{1+C\cdot\Gamma_0/m_0}
               	{(m_0^2 - m^2) + f(m) - i\, m_0 \Gamma(m)},
\end{equation}
where
\begin{eqnarray}
f(m) =
\Gamma_0 \frac{m_0^2}{q_0^3}
       \bigg[&&
             q^2 \left(\,h(m)-h(m_0)\,\right) + \\ \nonumber
&&           \left(\,m_0^2-m^2\,\right)\,q^2_0\,
             \frac{dh\ }{dm^2}\bigg|_{m=m_0}\,
       \bigg],
\end{eqnarray}
and $C$ is a constant that depends on the pion mass $m_\pi$ and the $\rho$ mass $m_0$ such that
\begin{equation}
C = \frac{3}{\pi}\frac{m_\pi^2}{q_0^2}\,
    \ln\left(\frac{m_0+2q_0}{2m_\pi}\right) 
    + \frac{m_0}{2\pi\,q_0} 
    - \frac{m_\pi^2 m_0}{\pi\,q_0^3}.
\end{equation}
The function $h(m)$ is defined for $m > 2 m_\pi$ such that
\begin{equation}
h(m) = \frac{2}{\pi}\,\frac{q}{m}\,
       \ln\left(\frac{m+2q}{2m_\pi}\right),
\end{equation}
with 
\begin{equation}
\frac{dh\ }{dm^2}\bigg|_{m=m_0} =
h(m_0)\left(\frac{1}{8q_0^2}-\frac{1}{2m_0^2}\right) \,+\, \frac{1}{2\pi m_0^2}. 
\end{equation}

The $0^+$ component of the $K\pi$ spectrum is described by the LASS parametrization
\begin{eqnarray}
\label{eq:LASSEqn}
R_j(m)  &=& \frac{m_{K\pi}}{q \cot{\delta_B} - iq} \\ &+& e^{2i \delta_B} 
\frac{m_0 \Gamma_0 \frac{m_0}{q_0}}
     {(m_0^2 - m_{K\pi}^2) - i m_0 \Gamma_0 \frac{q}{m_{K\pi}} \frac{m_0}{q_0}},
\nonumber
\end{eqnarray}
where $\cot{\delta_B} = \frac{1}{aq} + \frac{1}{2} r q$.

Table~\ref{tab:mKpi_model} gives the parameters of the lineshapes used to derive the invariant-mass-dependent phase of the components entering the fit model.
The total amplitude for describing the \mKpi distribution can be written as
\begin{eqnarray}
\label{equ:total_mKpi_Amplitude}
&& \quad |A(m_{K\pi};c_j)|^2 =\\
\nonumber &&  \int_{m_{\pi\pi}^{min}}^{m_{\pi\pi}^{max}} \left| \left( \sum_j c_j \sqrt{\textrm{H}_{\scriptsize{\textrm{R}}_j}(m_{K\pi},m_{\pi\pi})}\,e^{i\Phi_{\tiny{\textrm{R}}_j}(m)} \right) \right|^2 dm_{\pi\pi} \\
\nonumber && =  \left|c_{K^{*}}\right|^{2}\mathcal{H}_{K^{*}} + \left|c_{\rho^{0}}\right|^{2}\mathcal{H}_{\rho^{0}} + \left|c_{\swave}\right|^{2}\mathcal{H}_{\swave} + I,
\nonumber
\end{eqnarray}
with
\begin{equation}
c_j = \alpha_j \,e^{i\phi_j},
\end{equation}
and 
\begin{equation}
\mathcal{H}_{\scriptsize{\textrm{R}}_j} =  \int_{m_{\pi\pi}^{min}}^{m_{\pi\pi}^{max}} \textrm{H}_{\scriptsize{\textrm{R}}_j}(m_{K\pi},m_{\pi\pi}) \,dm_{\pi\pi},
\label{equ:mKpi_hits_proj_def}
\end{equation}
(see below for the expression of $I$).

\begin{figure*}[htbp]
\hspace{-30pt}
\begin{tabular}{cc}
	\includegraphics[height= 5.5cm] {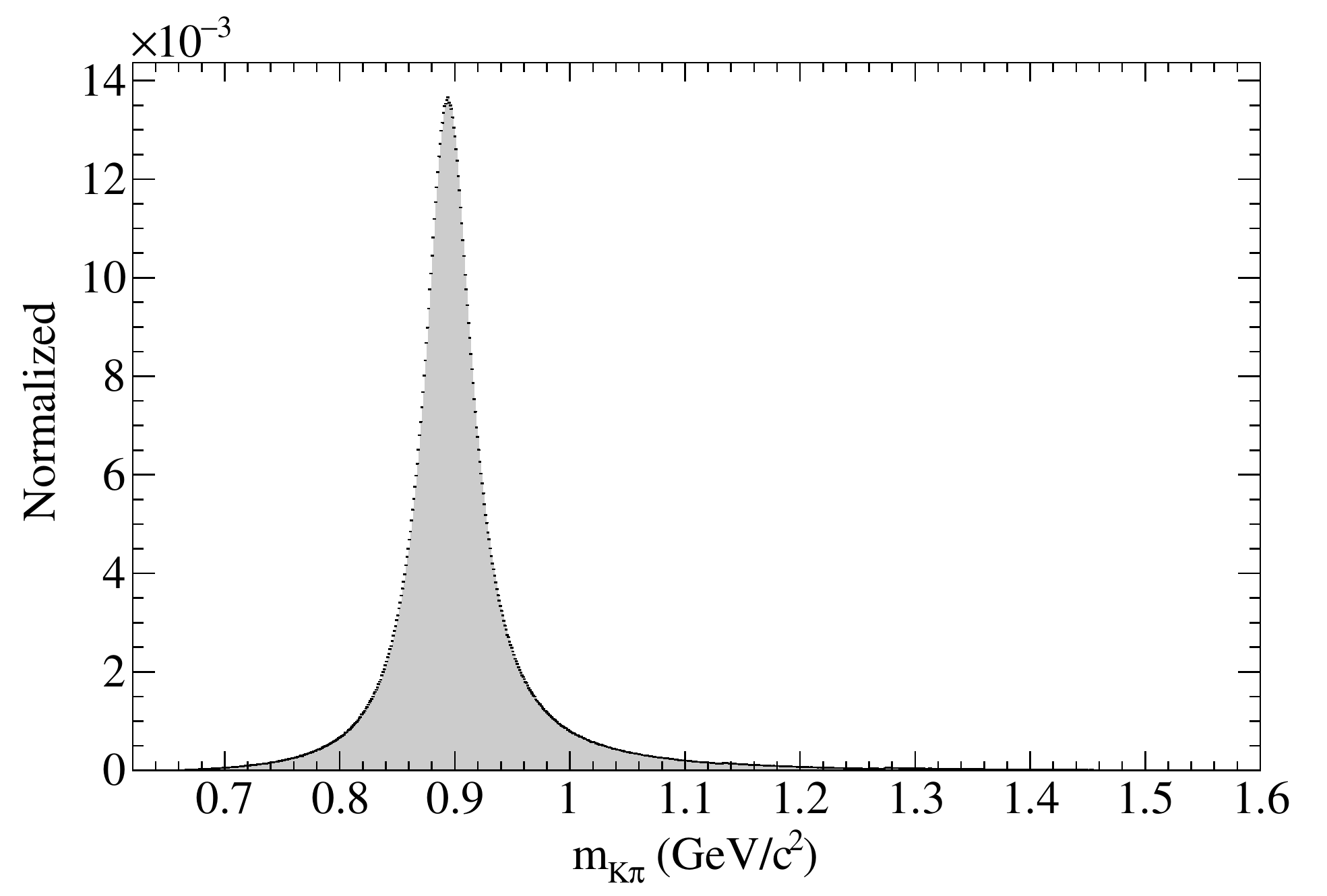} &\includegraphics[height= 5.5cm] {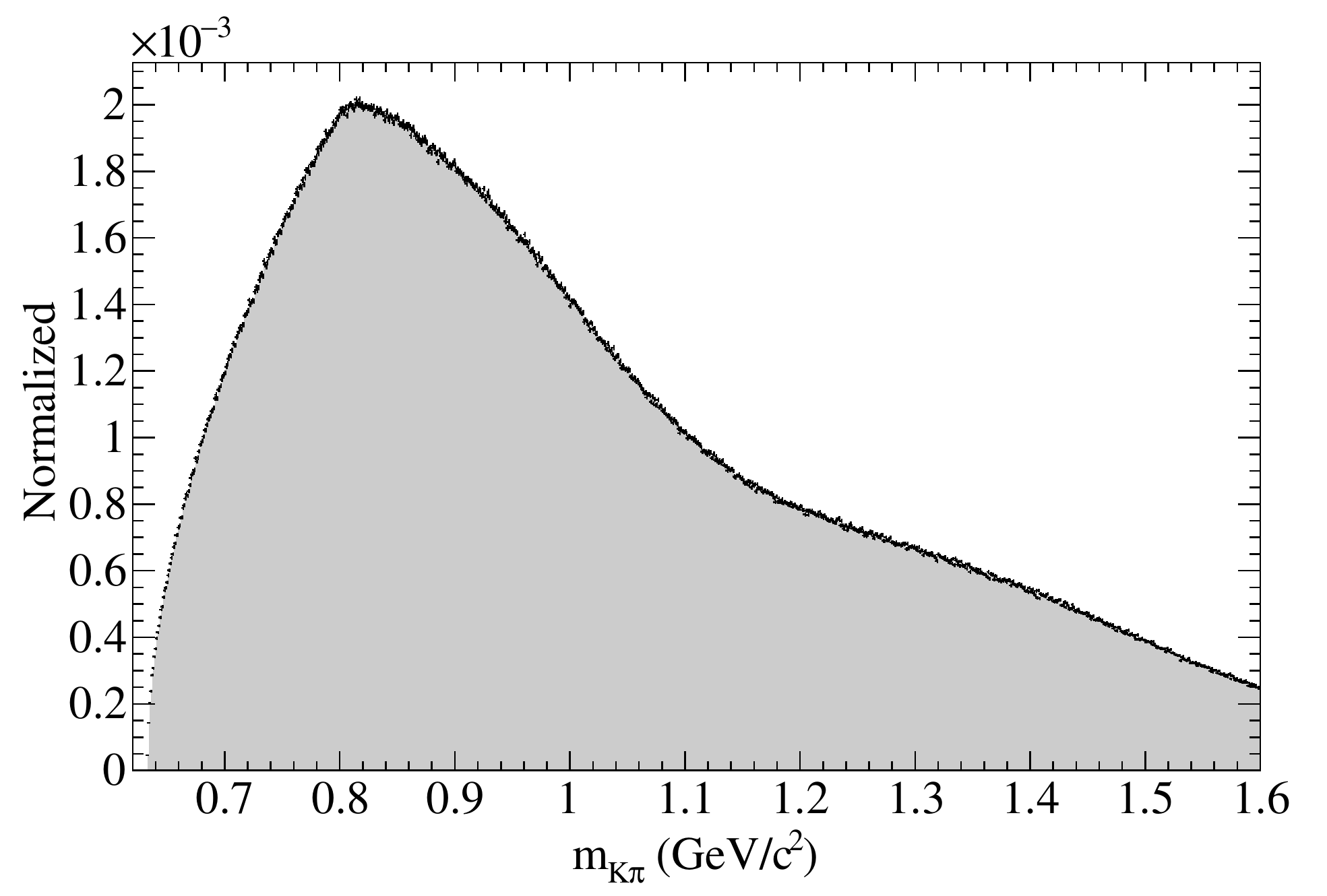} \\
\end{tabular}\\
\begin{center}
\begin{tabular}{c}
\includegraphics[height= 5.5cm] {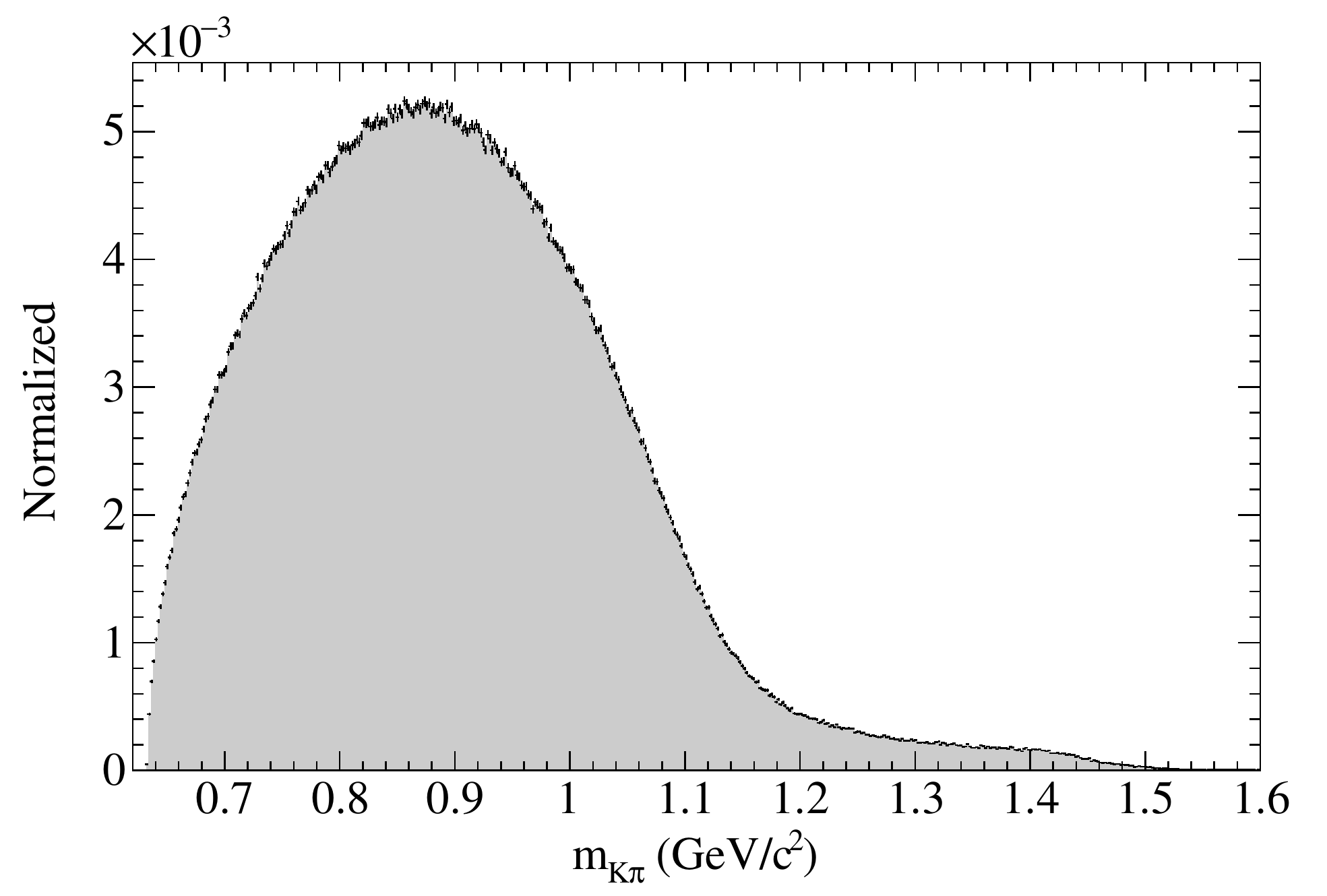}\\
\end{tabular}
\end{center}
\vspace{-10pt}
\caption{The \mKpi projections $\mathcal{H}_{\scriptsize{\textrm{R}}_j}$ of the two
dimensional histograms $\textrm{H}_{\scriptsize{\textrm{R}}_j}(m_{K\pi},m_{\pi\pi})$ describing the $K^{*}(892)^{0}$ (upper-left), the $\Rhoz$ (upper-right) and the $\swave$ (bottom) contributions in the \mKpi fit model.
The histograms, normalized to unit area, describe the expected \mKpi distributions of these components once reconstruction or resolution effects have been corrected. 
	\label{fig:CC_mKpi_hists}}
\end{figure*}

The histograms used to describe the \mKpi-dependent magnitudes of the $K^{*}(892)^{0}\pi$, $K\Rhoz$, and $\swave\pi$ decays are depicted in Fig.~\ref{fig:CC_mKpi_hists}. 
Those describing the $K^{*}(892)^{0}\pi$ and $K\Rhoz$ decay are both obtained from the projection onto the \mKpi axis of a two-dimensional histogram, constructed as the combination of the individual kaonic resonance contribution to the corresponding resonance (i.e. $K_{\scriptsize{\textrm{res}}} \to K^{*}(892)^{0}\pi$ or $K\Rhoz$). 
The combination is performed using the relative weights between each \Kpipi resonance extracted from the \mKpipi fit.
The unusual shape of the $\swave$ distribution, obtained from the phase space distribution of $\Bp \to \Klp(\to K^{+} \pim \pip) \g$ processes weighted by the LASS parametrization (see Eq.~\eqref{eq:LASSEqn}), is due to phase-space effects.
In the present analysis, the resonant part of the LASS is described by the $K^*_0(1430)$ scalar, which is very much suppressed. 
The dominant contribution comes from the nonresonant term that corresponds to the effective-range part.
For each histogram used to build the total PDF, the number of bins is 450 and 100 in the \mKpi and \mpipi dimensions, respectively.

\begin{widetext}
\begin{eqnarray}
\label{equ:mKpi_Interf_term}
\hspace*{-20pt}
I(m_{K\pi};c_{\rho^{0}},c_{\swave}) & =& 2\alpha_{\rho^{0}} \left[\cos(\phi_{\rho^{0}}-\Phi_{K^*})\int_{m_{\pi\pi}^{min}}^{m_{\pi\pi}^{max}} \sqrt{\mathrm{H}_{\rho^{0}}\mathrm{H}_{K^{*}}} \cos(\Phi_{\rho^0}) \,dm_{\pi\pi}\right.  \\
& & \quad\quad\; \left. - \sin(\phi_{\rho^{0}}-\Phi_{K^*}) \int_{m_{\pi\pi}^{min}}^{m_{\pi\pi}^{max}} \sqrt{\mathrm{H}_{\rho^{0}}\mathrm{H}_{K^{*}}} \sin(\Phi_{\rho^0}) \,dm_{\pi\pi} \right] \nonumber \\
& & + 2\alpha_{\rho^{0}}\alpha_{\swave}\left[\cos(\phi_{\rho^{0}} - \phi_{\swave} - \Phi_{\swave}) \int_{m_{\pi\pi}^{min}}^{m_{\pi\pi}^{max}} \sqrt{\mathrm{H}_{\rho^{0}}\mathrm{H}_{\swave}} \cos(\Phi_{\rho^0}) \,dm_{\pi\pi}\right. \nonumber \\
& & \quad\quad\quad\quad\quad \;\;\, \left. - \sin(\phi_{\rho^{0}} - \phi_{\swave} - \Phi_{\swave}) \int_{m_{\pi\pi}^{min}}^{m_{\pi\pi}^{max}} \sqrt{\mathrm{H}_{\rho^{0}}\mathrm{H}_{\swave}} \sin(\Phi_{\rho^0}) \,dm_{\pi\pi}\right]. \nonumber 
\end{eqnarray}
\end{widetext}

The term $I$ in Eq.~\eqref{equ:total_mKpi_Amplitude} describes the interference among the components in the model. 
In the two-dimensional \mKpi-\mpipi plane, the interference between the $\swave$ and the $(K\pi)$ P-wave components are proportional to a term containing the cosine of the helicity angle. 
Therefore, when integrating over the \mpipi dimension, this interference term vanishes. 
Since the fit is to be performed to an efficiency-corrected \mKpi distribution, we do not allow for $(K\pi)$ S-wave and P-wave interference in the model. 
The remaining source of interference comes from the $(K\pi)$ and $(\pi\pi)$ P-wave components, as well as from the $(K\pi)$ S-wave and the $(\pi\pi)$ P-wave components. 
The resulting expression for the interference term of the total PDF is given by Eq.~\eqref{equ:mKpi_Interf_term}, where $\Phi_{\scriptsize{R}}$ are the invariant-mass-dependent phases defined in Eq.~\eqref{eq:phi_res}.

We use the $K^{*}(892)^{0}$ coefficients as a reference, setting $\alpha_{K^{*}}=1$ and $\phi_{K^{*}}=0$. 
We checked that other choices do not affect the results. 
This leads to four free parameters in the fit: $\alpha_{\rho^{0}}$, $\phi_{\rho^{0}}$, $\alpha_{\swave}$ and $\phi_{\swave}$. 
The fit fractions FF($j$) extracted for each component in the model are defined in the same way as in the \mKpipi fit model (see Eqs.~\eqref{equ:fit_fractions_res} and~\eqref{equ:fit_fractions_inter}).
\subsection{Results}
\label{sec:CC_results}

\subsubsection{Event yield in \MyControlChanel and \mKpipi spectrum.}
\label{sec:CC_results_mes_deltae_fisher}

\begin{figure*}[htbp]
\begin{tabular}{cc}
	\includegraphics[height= 5.6cm] {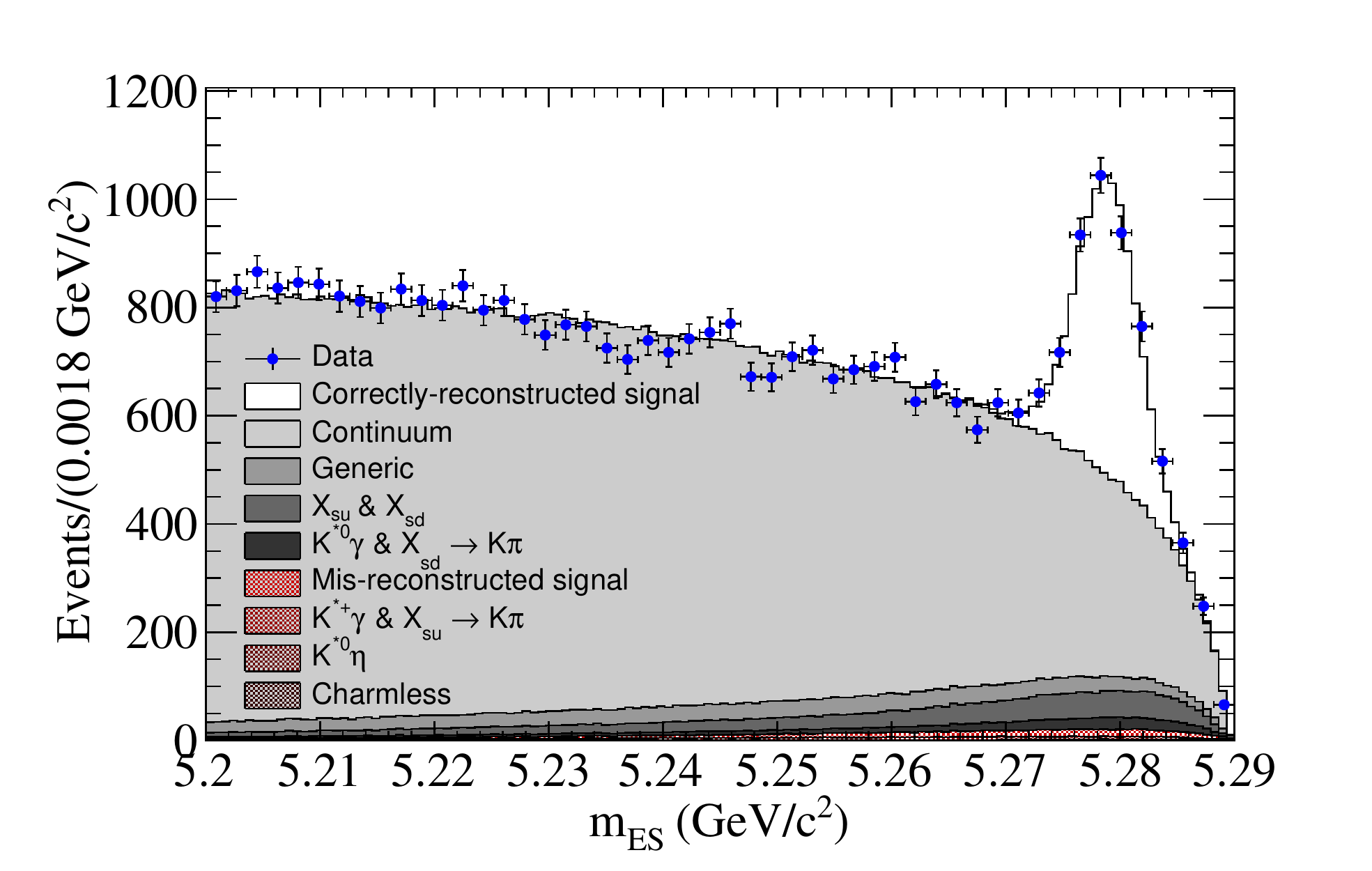}
	&
	\includegraphics[height= 5.6cm] {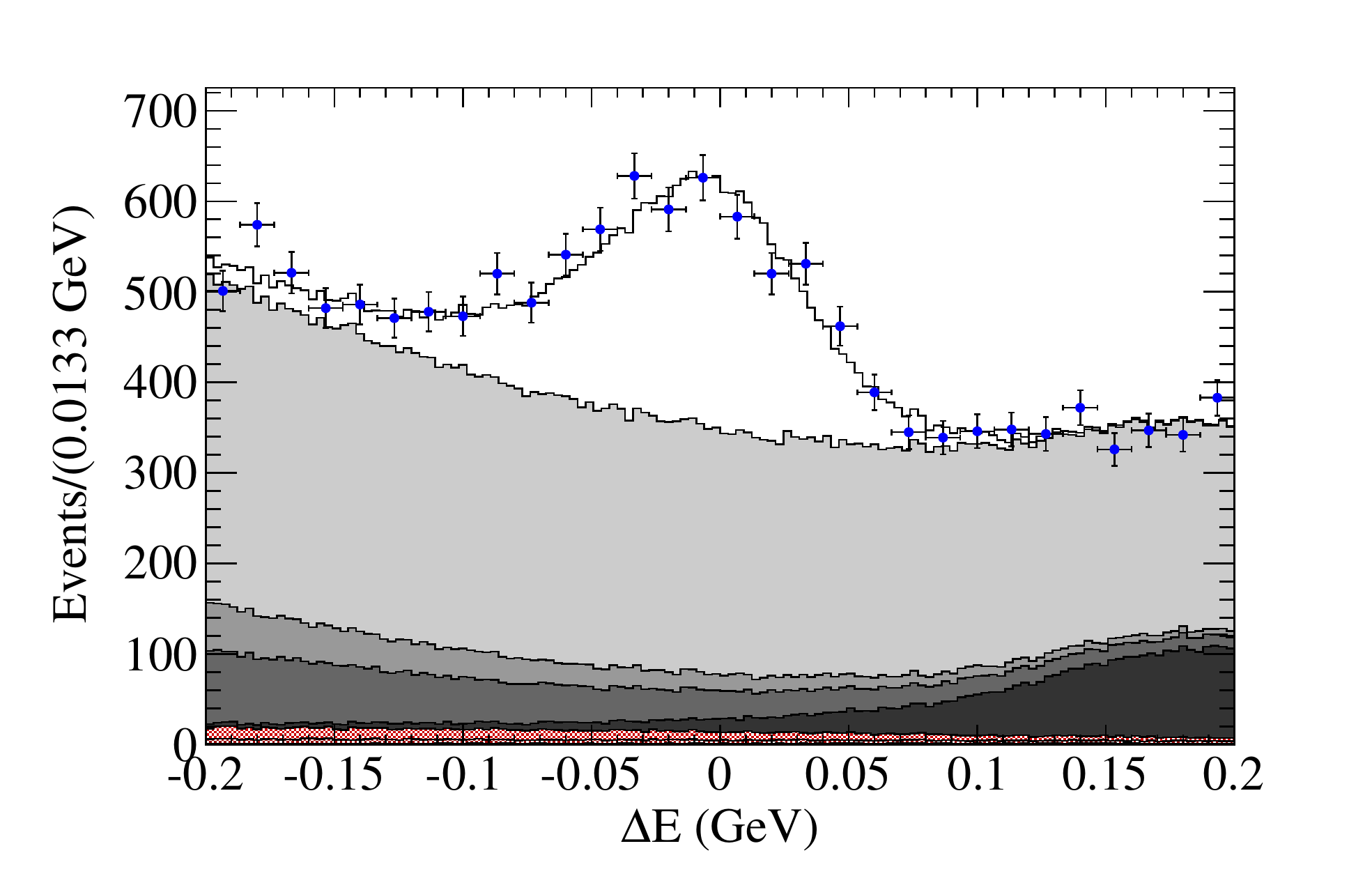}\\
\end{tabular}\\
\vspace{-18pt}
\begin{center}
\includegraphics[height= 5.6cm] {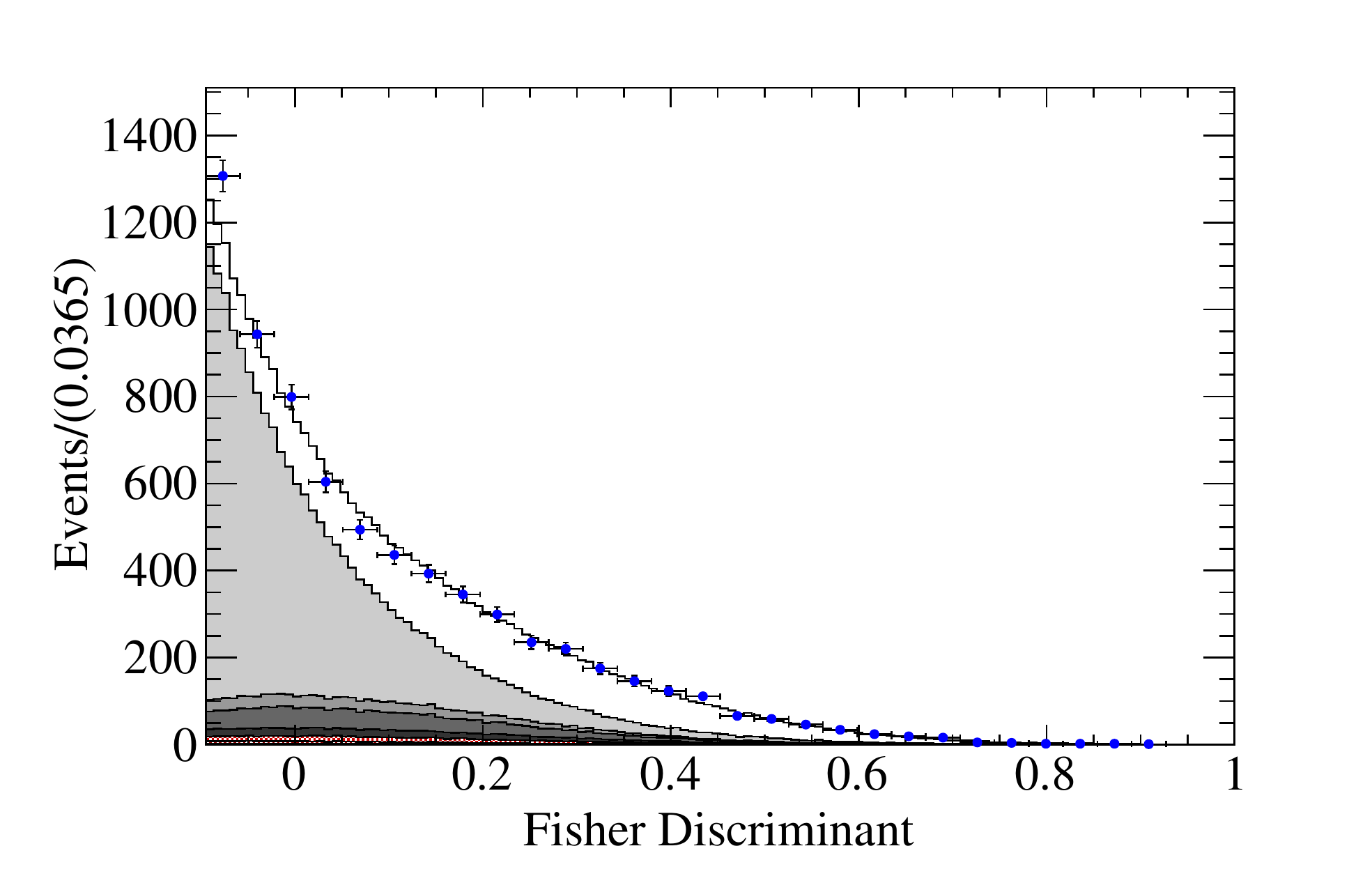}\\
\end{center}
\vspace{-10pt}
\caption{Distributions of \mes (top left), \DeltaE (top right) and the Fisher Discriminant (bottom) showing the fit results on the \MyControlChanel data sample. 
The distributions have their signal/background ratio enhanced by means of the following requirements: $ -0.10 \leq \DeltaE \leq 0.075 \gev$ (\mes); $\mes > 5.27 \gevcc$ (\DeltaE); $\mes > 5.27 \gevcc \, , \, -0.10 \leq \DeltaE \leq 0.075 \gev $ (Fisher). 
Points with error bars show data. The projection of the fit result is represented by stacked histograms, where the shaded areas represent the background contributions, as described in the legend. 
Some of the contributions are hardly visible due to their small fractions.
Note that the same order is used for the various contributions in both the stacked histograms and the corresponding legend, in which the ``Generic'' and ``Charmless'' entries correspond to the generic \B background and the sum of $B^+ \to a_1^{+}(\to\rhoz \pip )\piz \gamma $ and $ B^+ \to K^{*0}(\to K \pi) \pip \piz \gamma $ event categories, respectively, as defined in Table~\ref{tab:CC_bbackground}.}
\label{fig:FitProjCharged}
\end{figure*}

In the charged \B-meson decay mode for \break $m_{K\pi\pi}<1.8\gevcc$, the unbinned maximum-likelihood fit of \mes, \DeltaE, and \fisher, as described in Sec.~\ref{sec:CC_likeDiscrim}, yields $2441 \pm 91^{+41}_{-54}$ correctly reconstructed signal \break \MyControlChanel events in data.
This translates into a branching fraction of
\begin{equation}
\BR(\MyControlChanel) = (24.5	\pm	0.9	\pm 1.2)\times 10^{-6}.
\end{equation}
In both cases, the first uncertainty is statistical and the second is systematic. The latter is discussed in Sec.~\ref{sec:CC_BRSyst}.
This result is in good agreement with the previous world average~\cite{Agashe:2014kda} and supersedes that of Ref.~\cite{Aubert:2005xk}.
Figure~\ref{fig:FitProjCharged} shows signal-enhanced distributions of the three discriminating variables in the fit: \mes, \DeltaE, and $\fisher$. 
Using 331 generated pseudo-experiments with embedded signal events drawn from fully simulated MC samples, we checked that the parameters of interest exhibit no significant biases.

\begin{figure*}[htbp]
\begin{center}
	\includegraphics[height= 7.5cm] {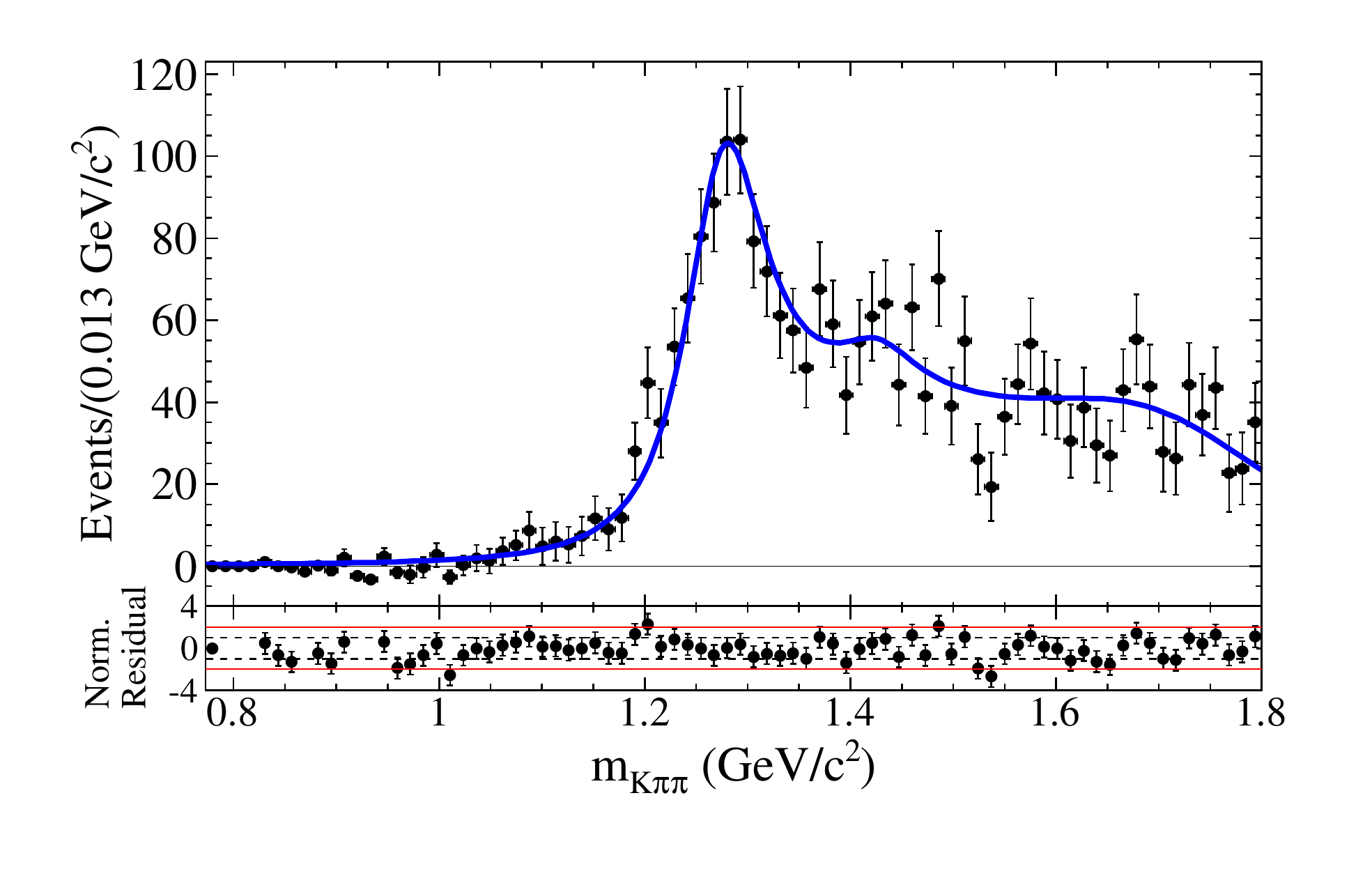}\\
\vspace{-10pt}
\caption{Distribution of \mKpipi for correctly-reconstructed \MyControlChanel signal events (\splot), extracted from the maximum likelihood fit to \mes, \DeltaE, and $\fisher$. 
Points with error bars give the sum of \sweights. 
The blue solid curve is the result of the fit performed directly to this \mKpipi distribution to extract the contributions from kaonic resonances decaying to \Kpipi. 
Below each bin are shown the residuals, normalized in error units. The parallel dotted and full lines mark the one and two standard deviation levels, respectively. 
\label{fig:FitProjmKpipi}}
\end{center}
\end{figure*}

Figure~\ref{fig:FitProjmKpipi} shows the extracted \mKpipi \splot distribution. 
The magnitudes and phases of the signal model components, as well as the widths of the $\Kl$ and $\KstarlVmy$ resonances, are extracted directly from a binned maximum likelihood fit to the \splot distribution of \mKpipi. 
Using Eqs.~\eqref{equ:fit_fractions_res} and~\eqref{equ:fit_fractions_inter}, we further compute the FF corresponding to the different resonances and the interference among those with the same $J^P$. 
The fitted parameters and FFs are listed in Table~\ref{tab:AMPmKpipi}. 
The statistical uncertainties on the magnitudes and phases, as well as on the widths of the $\Kl$ and $\KstarlVmy$ resonances, come directly from the fit. 
The central values of these widths are in good agreement with the corresponding world average values~\cite{Agashe:2014kda}.

\begin{figure*}[htbp]
\hspace{-20pt}
\begin{tabular}{cc}
	\includegraphics[height= 6.0cm] {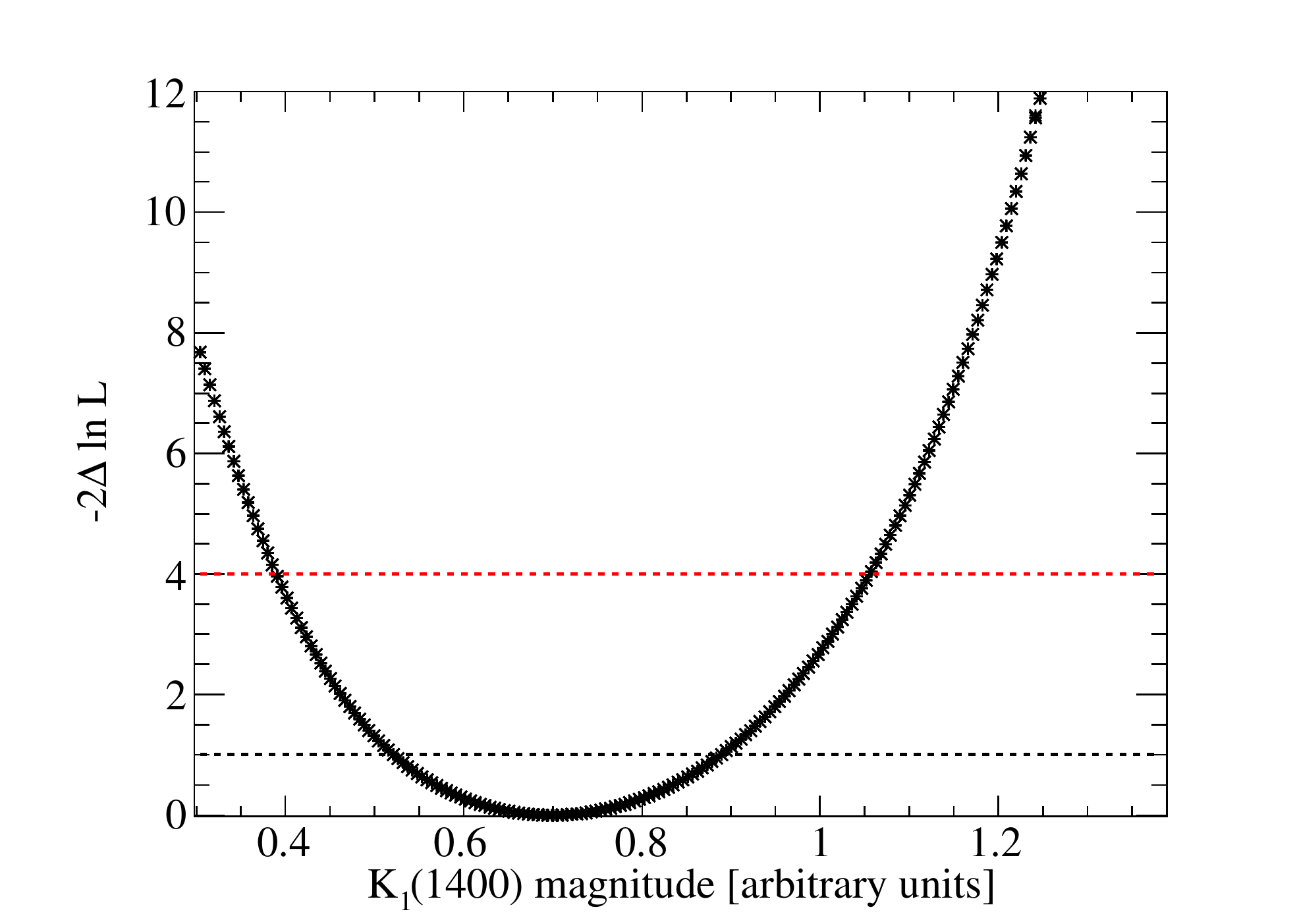}
&
	\includegraphics[height= 6.0cm] {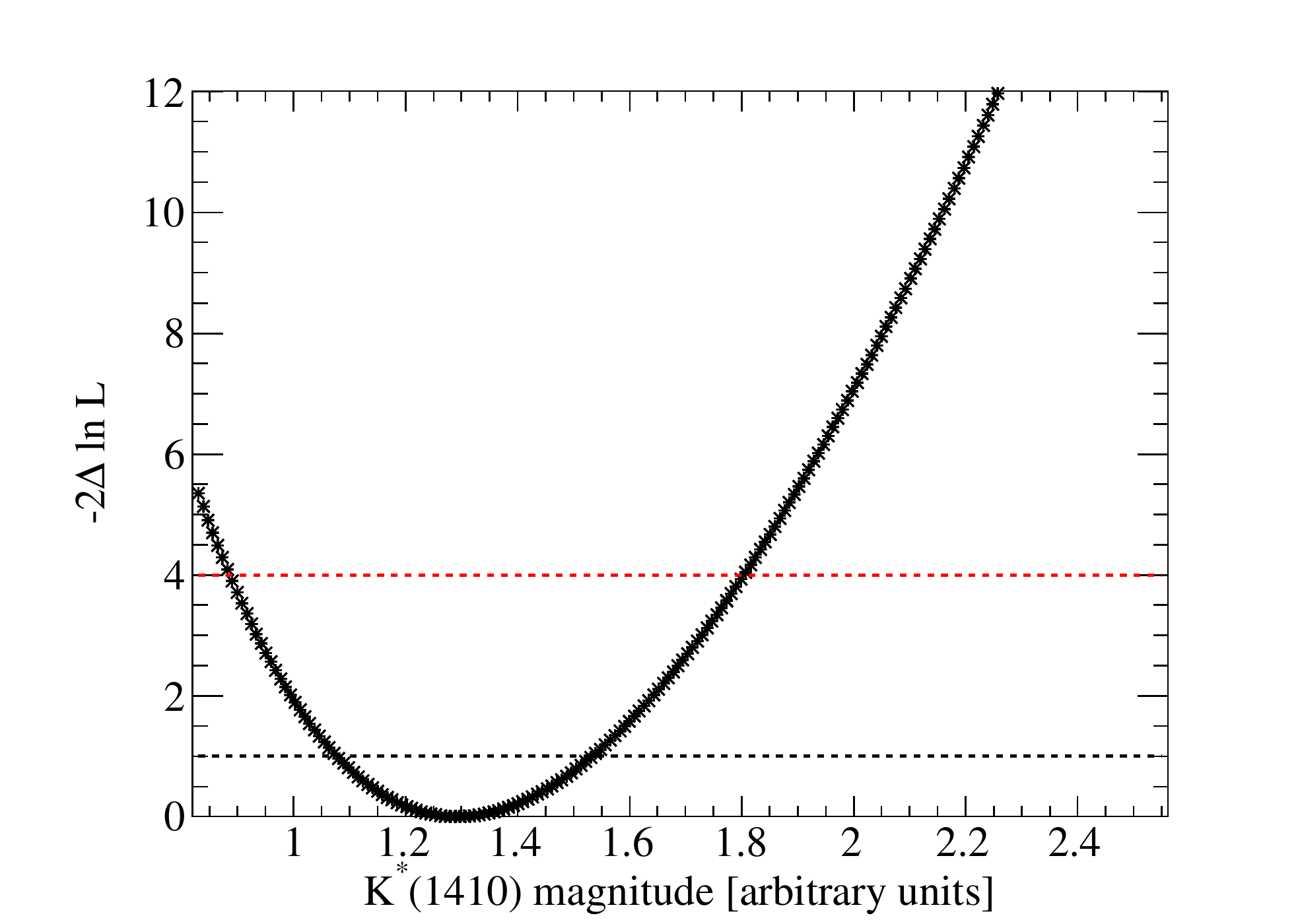}\\
	\includegraphics[height= 6.0cm] {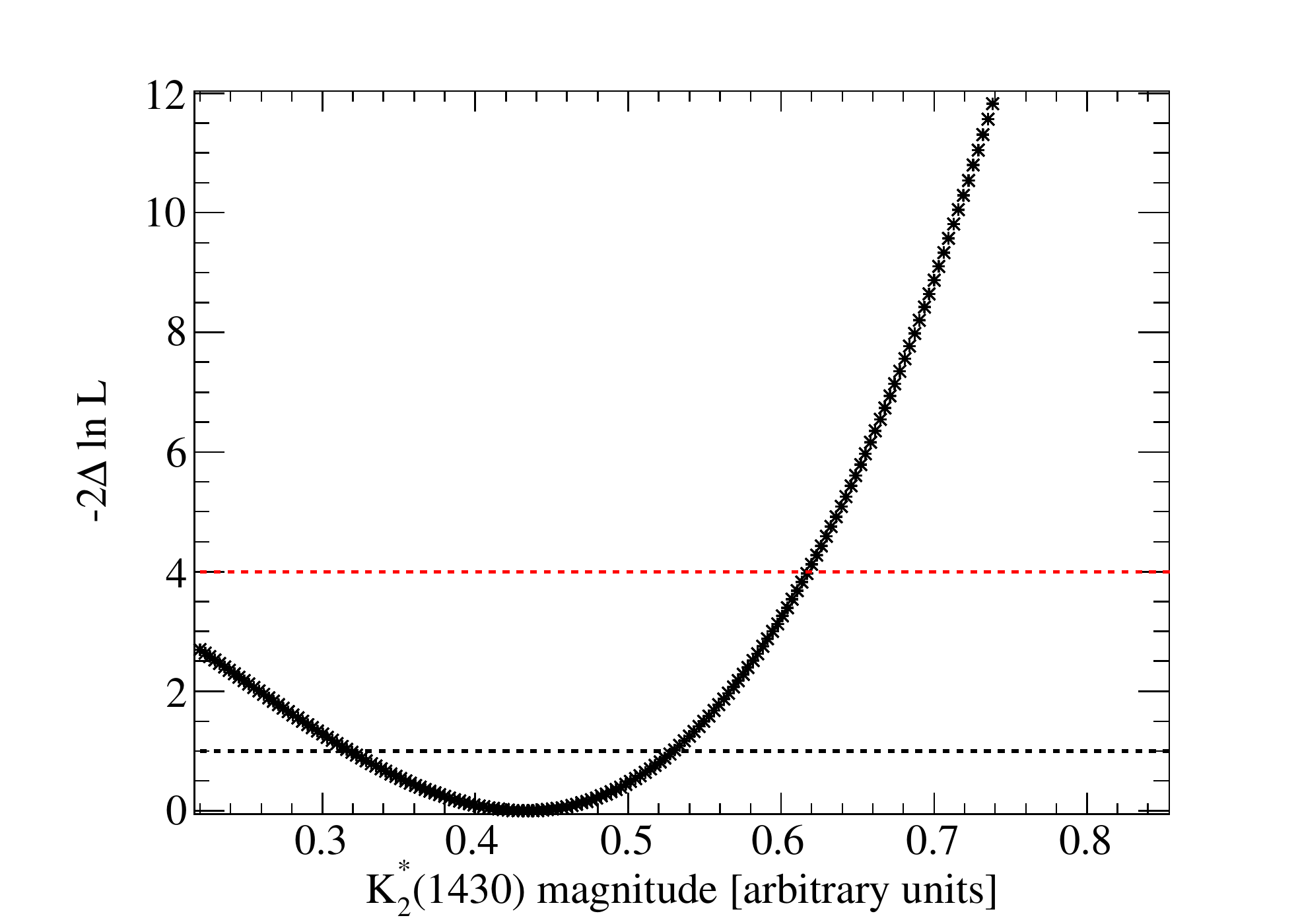}
&
	\includegraphics[height= 6.0cm] {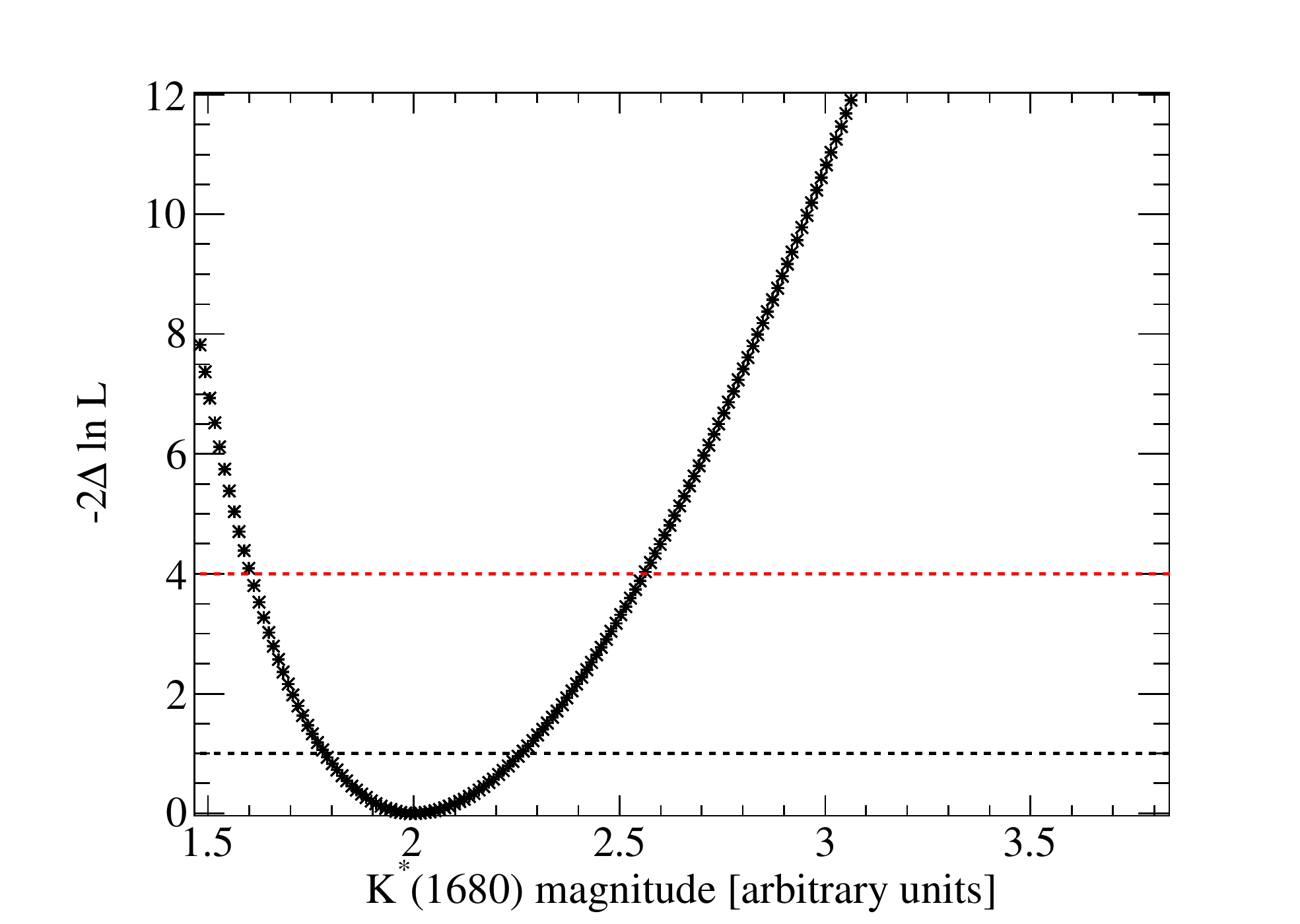}\\
	\includegraphics[height= 6.0cm] {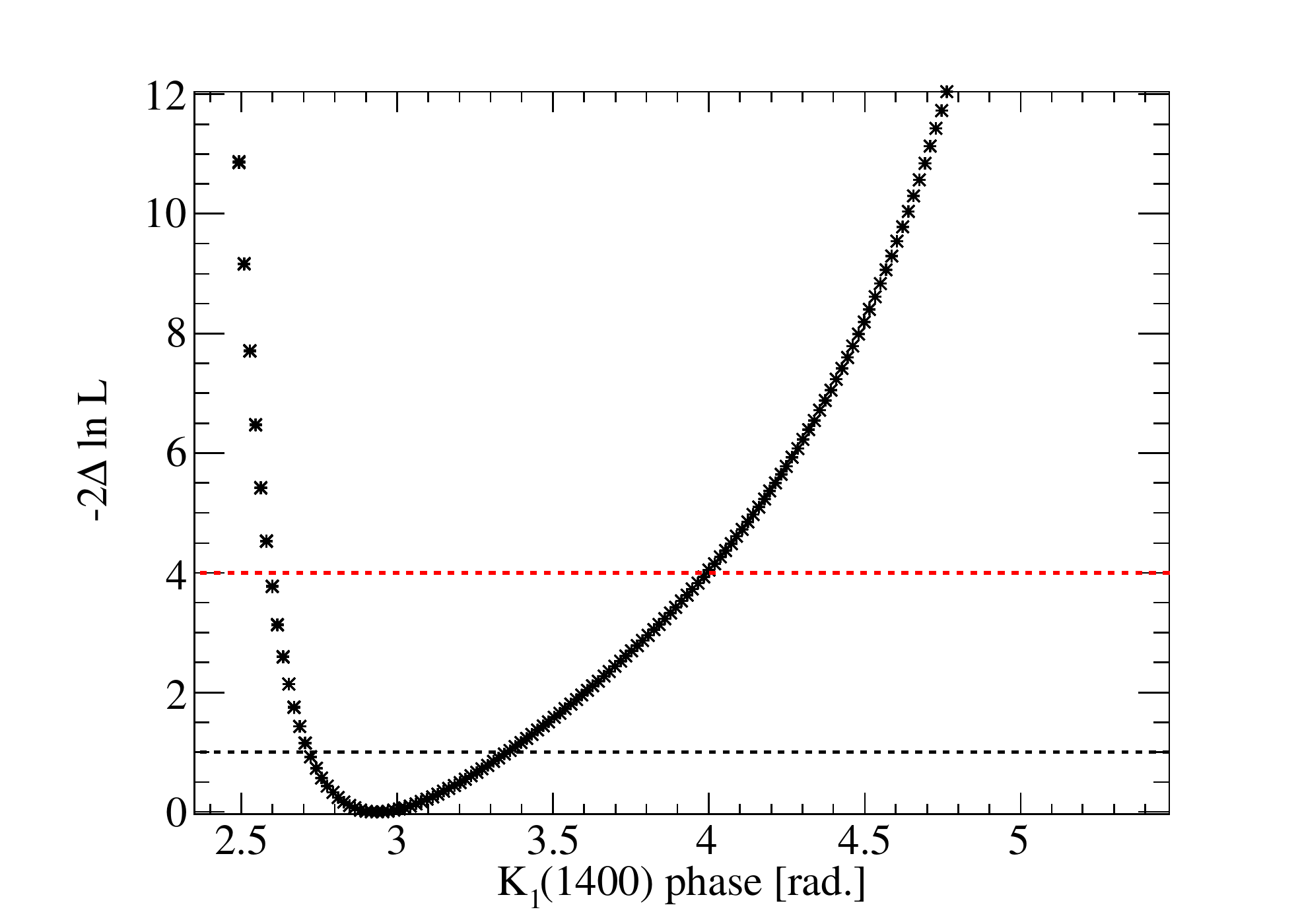}
&
	\includegraphics[height= 6.0cm] {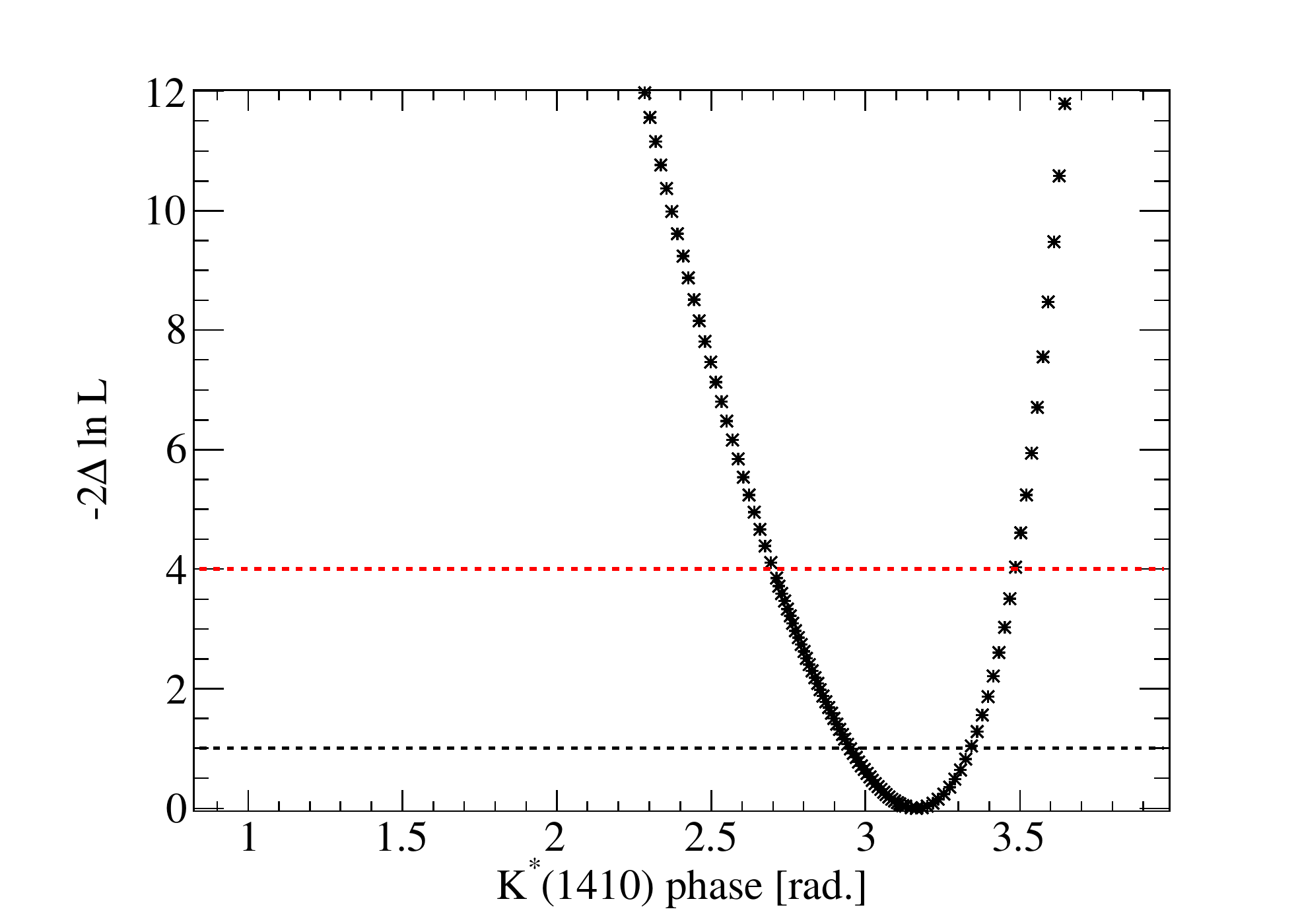}\\
\end{tabular}
\caption[One-dimensional scans of $-2\Delta \textrm{ln}\mathcal{L}$ as a function of magnitudes and phases]{One-dimensional scans of $-2\Delta \textrm{ln}\mathcal{L}$ as a function of magnitudes (top and middle) and phases (bottom). The horizontal dashed lines mark the one- and two-standard deviation levels.\label{fig:CC_mKpipi_Scans}}
\end{figure*}

As the fit fractions are functions of the complex amplitudes $c_k$, the statistical uncertainties on the FF are estimated in a different way. 
From the full fit result information (including correlations between fitted parameters) obtained using the nominal model, $10^5$ sets of values of the resonance amplitudes $c_k$ are randomly generated. 
We then compute the corresponding fit fractions for each set and obtain the FF($k$) distributions. 
The $\pm1 \sigma$ statistical uncertainties are taken as the values at $\pm 34.1\%$ of the FF distribution integral around the FF value extracted from the nominal fit results.
We also performed likelihood scans of the fitted parameters, as shown in Fig.~\ref{fig:CC_mKpipi_Scans}, in order to check for the presence of multiple solutions. 
It appears that the fitted solution is unique. 
Each of these scans is obtained by fixing the corresponding parameter at several consecutive values and refitting the
rest of the parameters. 
Each of the fits is repeated 30 times with random initial values of the varying parameters and always converge to the same likelihood solution.

Inserting the FF values listed in Table~\ref{tab:AMPmKpipi} into Eqs.~\eqref{equ:weighted_effi_formula} and~\eqref{eq:BR_Kres}, we obtain the weighted efficiency, 
\break $\langle{\epsilon}^+\rangle = 0.2068^{+ 0.0010}_{-0.0017}$ 
and the branching fractions listed in Table~\ref{tab:BFmKpipi}.
In the calculation of the branching fractions, we use both the fitted signal yield and the corresponding fit fraction. 
Since these two quantities come from measurements on the same data sample, we assume that the corresponding statistical uncertainties are 100\% correlated when calculating the statistical uncertainty on each branching fraction. 
This is a conservative approach of determining the total statistical uncertainty.

\begin{table*}[htbp]
\caption[]{\label{tab:AMPmKpipi} Results of the fit to the correctly-reconstructed signal \splot of \mKpipi. 
The first uncertainty is statistical and the second is systematic (see Sec.~\ref{sec:CC_mKpipiSyst}). The uncertainties on the $\Kl$ and $\KstarlVmy$ widths are statistical only. Interferences for both $J^P=1^+$ and $1^-$ resonances are destructive.}
\setlength{\tabcolsep}{2.25pc}
\begin{tabular*}{\textwidth}{@{\extracolsep{\fill}} c|c|c c|c}
\hline
\hline
\multirow{2}{*}{$J^P$} & \multirow{2}{*}{$K_{\textrm{\footnotesize{res}}}$} & \multirow{2}{*}{Magnitude $\alpha$} & \multirow{2}{*}{Phase $\phi$ (rad.)} & \multirow{2}{*}{Fit fraction} \\ 
 & & & & \\ 
\hline 
\multirow{4}{*}{$1^+$} & \multirow{2}{*}{$\Kl$} & \multirow{2}{*}{1.0 (fixed)} & \multirow{2}{*}{0.0 (fixed)}& \multirow{2}{*}{$\phantom{-}0.61^{+0.08}_{-0.05}{}^{+ 0.05}_{-0.05}\phantom{-} $} \\ 
  & & & & \\ 
\cline{2-5}
 & \multirow{2}{*}{$\Kll$} & \multirow{2}{*}{$\phantom{-}0.72 \pm 0.10{}^{+ 0.12}_{-0.08}{}$} & \multirow{2}{*}{$2.97 \pm 0.17 {}^{+ 0.11}_{-0.12}{}$}& \multirow{2}{*}{$\phantom{-}0.17^{+ 0.08}_{-0.05}{}^{+ 0.05}_{-0.04} \phantom{-} $} \\ 
  & & & & \\ 
\hline 
\multirow{4}{*}{$1^-$} & \multirow{2}{*}{$\KstarX$} & \multirow{2}{*}{$\phantom{-}1.36 \pm 0.16{}^{+ 0.20}_{-0.16}{}$} & \multirow{2}{*}{$3.14 \pm 0.12{}^{+ 0.02}_{-0.04}{}$}& \multirow{2}{*}{$\phantom{-}0.42^{+ 0.08}_{-0.07}{}^{+ 0.08}_{-0.04} \phantom{-}$} \\ 
  & & & & \\ 
\cline{2-5}
 & \multirow{2}{*}{$\KstarlVmy$} & \multirow{2}{*}{$\phantom{-}2.10 \pm 0.28{}^{+ 0.27}_{-0.26}{}$} & \multirow{2}{*}{0.0 (fixed)}& \multirow{2}{*}{$\phantom{-}0.40^{+ 0.05}_{-0.04}{}^{+ 0.08}_{-0.06} \phantom{-}$} \\ 
  & & & & \\ 
\hline
\multirow{2}{*}{$2^+$} & \multirow{2}{*}{$\Kstarlllmy$} & \multirow{2}{*}{$\phantom{-}0.29 \pm 0.09{}^{+ 0.09}_{-0.11}{}$} & \multirow{2}{*}{0.0 (fixed)}& \multirow{2}{*}{$\phantom{-}0.05^{+ 0.04}_{-0.03}{}^{+ 0.05}_{-0.06} \phantom{-}$} \\ 
  & & & & \\ 
\hline
\hline
\multicolumn{4}{l|}{\multirow{2}{*}{Sum of fit fractions}} & \multirow{2}{*}{$\phantom{-}1.65 ^{+ 0.18}_{-0.14}{}^{+ 0.12}_{-0.08}\phantom{-} $} \\ 
\multicolumn{4}{c|}{\multirow{2}{*}{} } & \\ 
\hline
\multicolumn{2}{l|}{\multirow{4}{*}{interference}} & \multicolumn{2}{c|}{\multirow{2}{*}{$J^P = 1^+ : \{\Kl \,\textrm{--}\, \Kll \}$} } & \multirow{2}{*}{$-0.35^{+0.10}_{-0.16}{}^{+0.05}_{-0.05}\phantom{-} $} \\ 
\multicolumn{2}{c|}{\multirow{2}{*}{ } } & \multicolumn{2}{c|}{\multirow{2}{*}{} } & \\ 
\cline{3-5}
\multicolumn{2}{c|}{\multirow{2}{*}{ } } & \multicolumn{2}{c|}{\multirow{2}{*}{$J^P = 1^- : \{\KstarX \,\textrm{--}\, \KstarlVmy \}$} } & \multirow{2}{*}{$-0.30 ^{+ 0.08}_{-0.11}{}^{+0.09}_{-0.06}\phantom{-} $} \\ 
\multicolumn{2}{c|}{\multirow{2}{*}{ } } & \multicolumn{2}{c|}{\multirow{2}{*}{} } & \\ 
\hline
\hline
\multicolumn{5}{c}{\multirow{2}{*}{Line-shape parameters}} \\ 
\multicolumn{5}{c}{\multirow{2}{*}{}} \\ 
\hline
\hline
\multicolumn{2}{c|}{\multirow{2}{*}{$K_{\textrm{\footnotesize{res}}}$}} & \multirow{2}{*}{Mean (\gevccnosp)} & \multirow{2}{*}{Width (\gevccnosp) $\phantom{-}$} & \\ 
\multicolumn{2}{c|}{} & & & \\ 
\cline{1-4}
\multicolumn{2}{c|}{\multirow{2}{*}{$\Kl$}} & \multirow{2}{*}{1.272 (fixed)} & \multirow{2}{*}{$0.098 \pm 0.006\phantom{--}$} & \\ 
\multicolumn{2}{c|}{} & & & \\ 
\cline{1-4}
\multicolumn{2}{c|}{\multirow{2}{*}{$\KstarlVmy$}} & \multirow{2}{*}{1.717 (fixed)} & \multirow{2}{*}{$0.377 \pm 0.050\phantom{--}$} & \\ 
\multicolumn{2}{c|}{} & & & \\ 
\hline
\hline
\end{tabular*}
\end{table*}
\begin{table*}[htbp] 
\begin{center}
\caption{\label{tab:BFmKpipi} Branching fractions of the different \Kpipi resonances extracted from the fit to the \mKpipi spectrum. The listed numbers are averaged over charge-conjugate states. They are obtained using the fit fraction of each component and the corresponding efficiency. To correct for the secondary branching fractions, we use the values from Ref.~\cite{Agashe:2014kda}. The first uncertainty is statistical, the second is systematic (see Sec.~\ref{sec:CC_BRSyst}), and the third, when present, is due to the uncertainties on the secondary branching fractions. When the symbol ``n/a'' is quoted, it indicates that the corresponding branching fraction was not previously reported.}
\setlength{\tabcolsep}{0.0pc}
\begin{tabular*}{\textwidth}{@{\extracolsep{\fill}}l c c c}
\hline
\hline
\multirow{2}{*}{Mode}	&			$\BR(\Bp \to {\rm Mode}) \times $								&\multirow{2}{*}{			$\;\;\BR(\Bp \to {\rm Mode}) \times 10^{-6}\;\;$	}							&	$\;\;$ Previous world$\;\;$	\\
	&			$\BR(\Kres \to \Kp \pip \pim) \times 10^{-6}$							&													&	average~\cite{Agashe:2014kda} $(\times 10^{-6})$	\\
\hline																										
\multirow{2}{*}{ $\Bp \to \Kp \pip \pim \g$  }	& \multirow{2}{*}{$			\cdot\cdot\cdot								$}&\multirow{2}{*}{$		24.5	\pm	0.9	\pm	1.2 $}&	\multirow{2}{*}{ $27.6 \pm 2.2$ }	\\
	&											&													&		\\
\hline																											
\multirow{2}{*}{$ K_1(1270)^+ \g $  }	& \multirow{2}{*}{$	14.5	^{+	2.1	}_{-	1.4	}$$^{+	1.2	}_{-	1.2	}	$}&\multirow{2}{*}{$		44.1	^{+	6.3	}_{-	4.4	}$$^{+	3.6	}_{-	3.6	}\pm	4.6	$}&	\multirow{2}{*}{ $43 \pm 13 $ }	\\
	&											&													&		\\
\multirow{2}{*}{$ K_1(1400)^+ \g $  }	& \multirow{2}{*}{$	4.1	^{+	1.9	}_{-	1.2	}$$^{+	1.2	}_{-	1.0	}	$}&\multirow{2}{*}{$		9.7	^{+	4.6	}_{-	2.9	}$$^{+	2.8	}_{-	2.3	}\pm	0.6	$}&	\multirow{2}{*}{ $<15 $ at $90\%$ CL }	\\
	&											&													&		\\
\multirow{2}{*}{$ K^*(1410)^+ \g $  }	& \multirow{2}{*}{$	11.0	^{+	2.2	}_{-	2.0	}$$^{+	2.1	}_{-	1.1	}	$}&\multirow{2}{*}{$		27.1	^{+	5.4	}_{-	4.8	}$$^{+	5.2	}_{-	2.6	}\pm	2.7	$}&	\multirow{2}{*}{ n/a }	\\
	&											&													&		\\
\multirow{2}{*}{$ K_2^*(1430)^+ \g $  }	& \multirow{2}{*}{$	1.2	^{+	1.0	}_{-	0.7	}$$^{+	1.2	}_{-	1.5	}	$}&\multirow{2}{*}{$		8.7	^{+	7.0	}_{-	5.3	}$$^{+	8.7	}_{-	10.4	}\pm	0.4	$}&	\multirow{2}{*}{ $14 \pm 4 $ }	\\
	&											&													&		\\
\multirow{2}{*}{$ K^*(1680)^+ \g $  }	& \multirow{2}{*}{$	15.9	^{+	2.2	}_{-	1.9	}$$^{+	3.2	}_{-	2.4	}	$}&\multirow{2}{*}{$		66.7	^{+	9.3	}_{-	7.8	}$$^{+	13.3	}_{-	10.0	}\pm	5.4	$}&	\multirow{2}{*}{ $<1900$ at $90\%$ CL }	\\
	&											&													&		\\
\hline
\hline
\end{tabular*} 
\end{center}
\end{table*}

%
%
\begin{figure*}[htbp]
\begin{center}
	\includegraphics[height= 8.5cm] {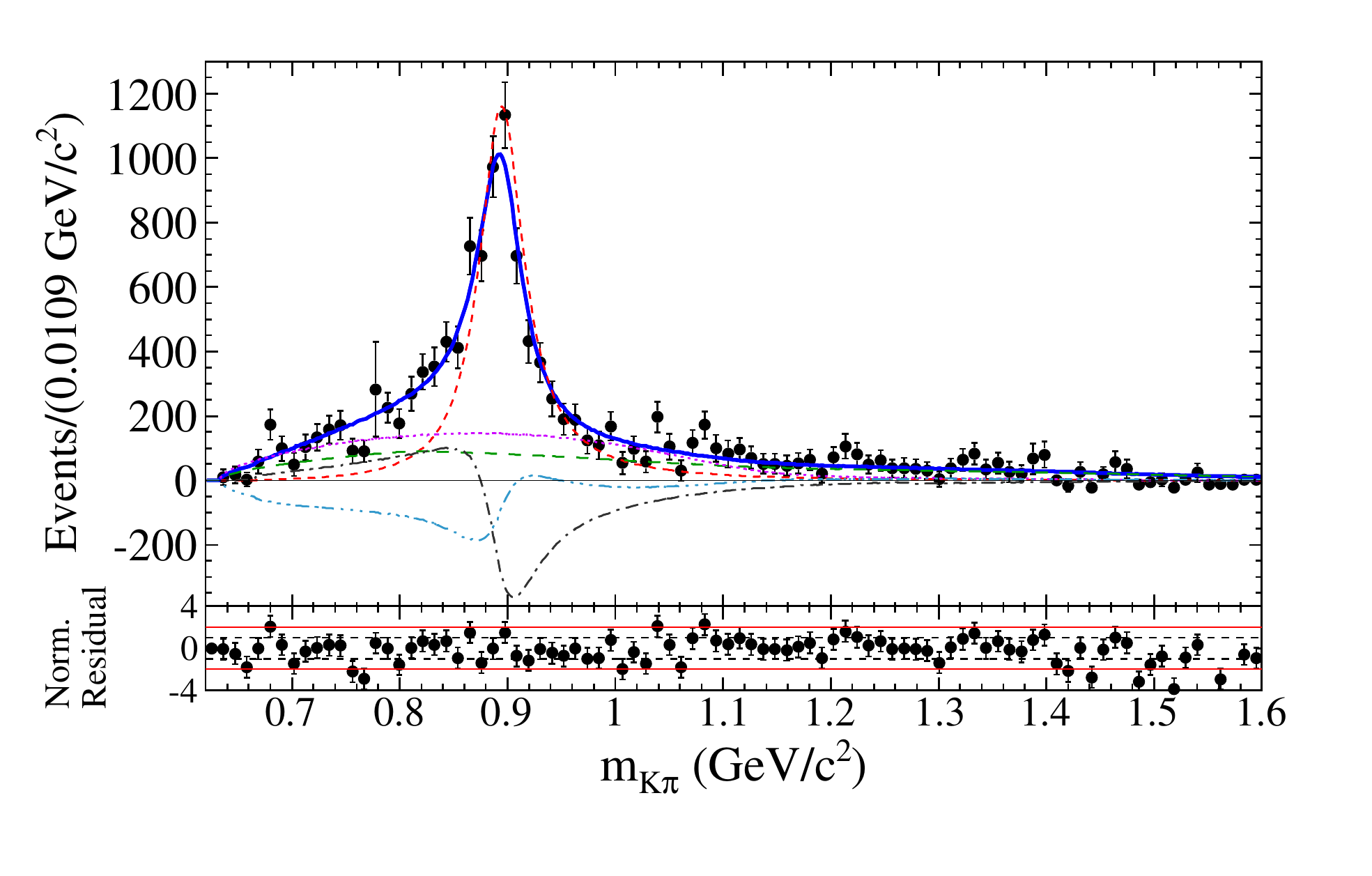}\\
\vspace{-10pt}
\caption{Distribution of \mKpi for correctly-reconstructed \MyControlChanel signal events (\splot), extracted from the maximum likelihood fit to \mes, \DeltaE, and $\fisher$. Points with error bars give the sum of \sweights. The blue solid curve corresponds to the total PDF fit projection.
The small-dashed red, medium-dashed green and dotted magenta curves correspond to the $K^{*}(892)^{0}$, $\Rhoz$ and $\swave$ contributions, respectively. 
The dashed-dotted gray curve corresponds to the interference between the two P-wave components, i.e. the $K^{*}(892)^{0}$ and the $\Rhoz$, and the dashed-triple-dotted light blue curve corresponds to the interference between the $\swave$ and the $\Rhoz$.
Below the \mKpi spectrum, we also show the residuals normalized in units of
standard deviations, where the parallel dotted and full lines mark the one and two standard deviation levels, respectively.
\label{fig:FitProjmKpi}}
\end{center}
\end{figure*}
\begin{table*}[htbp]
\centering
\caption[Results of the fit to the correctly-reconstructed signal \splot of \mKpi]{\label{tab:mKpi_fit_results} Results of the fit to the correctly-reconstructed signal \splot of \mKpi. The first uncertainty is statistical and the second is systematic (see Sec.~\ref{sec:CC_mKpiSyst}).}
\vspace{5pt}
\setlength{\tabcolsep}{2.7pc}
\begin{tabular*}{\textwidth}{@{\extracolsep{\fill}} l|c c|c}
\hline
\hline
 & \multirow{2}{*}{Module $lpha$} & \multirow{2}{*}{Phase $\phi$ (rad.)} & \multirow{2}{*}{ Fit Fraction} \\ 
 & & & \\ 
\hline 
\multirow{2}{*}{$K^{*}(892)^{0}$} & \multirow{2}{*}{1.0 (fixed)} & \multirow{2}{*}{0.0 (fixed)} & \multirow{2}{*}{$\phantom{-}0.637^{+ 0.011}_{-0.009}{}^{+ 0.017}_{-0.013}\phantom{-} $} \\ 
 & & & \\ 
\hline
\multirow{2}{*}{$\Rhoz$} & \multirow{2}{*}{$ \phantom{-}0.717\pm 0.015^{+ 0.017}_{-0.022}$} & \multirow{2}{*}{$\phantom{-}3.102^{+ 0.036}_{-0.035}{}^{+ 0.055}_{- 0.066}\phantom{-}$} & \multirow{2}{*}{$\phantom{-}0.331^{+ 0.015}_{-0.013}{}^{+ 0.031}_{-0.028}\phantom{-} $} \\ 
& & & \\ 
\hline
\multirow{2}{*}{$\swave$} & \multirow{2}{*}{$\phantom{-}0.813^{+ 0.044}_{-0.050}{}^{+ 0.048}_{-0.060}{}$} & \multirow{2}{*}{$\phantom{-}3.182^{+ 0.132}_{-0.125}{}^{+ 0.117}_{-0.108}\phantom{-} $} & \multirow{2}{*}{$\phantom{-}0.423^{+ 0.039}_{-0.041}{}^{+ 0.055}_{-0.076}\phantom{-} $} \\ 
 & & & \\ 
\hline
\hline
\multicolumn{3}{l|}{\multirow{2}{*}{Sum of fit fractions} } & \multirow{2}{*}{$\phantom{-}1.391^{+ 0.048}_{-0.042}{}^{+ 0.094}_{-0.057}\phantom{-} $}\\ 
\multicolumn{3}{c|}{\multirow{2}{*}{} } & \\ 
\hline
\multirow{4}{*}{Interference} & \multicolumn{2}{c|}{\multirow{2}{*}{\{$K^{*}(892)^{0}$---$\Rhoz$\}}} & \multirow{2}{*}{$-0.176^{+ 0.004}_{-0.006}{}^{+ 0.010}_{-0.008}\phantom{-} $} \\ 

& \multicolumn{2}{c|}{\multirow{2}{*}{$$}} & \\ 
\cline{2-4}
& \multicolumn{2}{c|}{\multirow{2}{*}{\{$\swave$---$\Rhoz$\}} } & \multirow{2}{*}{$-0.215^{+ 0.029}_{-0.044}{}^{+ 0.047}_{-0.033}\phantom{-} $} \\ 
&\multicolumn{2}{c|}{\multirow{2}{*}{} } & \\ 
\hline
\hline
\end{tabular*}
\end{table*}
\begin{table*}[htbp]
\begin{center}
\caption{\label{tab:BFmKpi}
Branching fractions of the resonances decaying to $K\pi$ and $\pi\pi$ extracted from the fit to the \mKpi spectrum. The listed results are averaged over charge-conjugate states. They are obtained using the ``fit fraction'' of each component and the corresponding efficiency.
$R$ denotes an intermediate resonant state and $h$ stands for a final state hadron: a charged pion or kaon. 
To correct for the secondary branching fractions, we used the values from Ref.~\cite{Agashe:2014kda} and $\BR(K^{*}(892)^{0} \to \Kp \pim) = \twothirds$. 
The first uncertainty is statistical, the second is systematic (see Sec.~\ref{sec:CC_BRSyst}), and the third (when applicable) is due to the uncertainties on the secondary branching fractions. 
The last two rows of the table are obtained by separating the contributions from the resonant and the nonresonant part of the LASS parametrization. 
Integrating separately the resonant part, the nonresonant part, and the coherent sum we find that the nonresonant part accounts for $95.6\%$, the resonant contribution $7.92\%$, and the destructive interference $-3.52\%$. 
When the symbol ``n/a'' is quoted, it indicates that the corresponding branching fraction was not previously reported.}
\vspace{5pt}
\setlength{\tabcolsep}{0.0pc}
\begin{tabular*}{\textwidth}{@{\extracolsep{\fill}}l c c c}
\hline
\hline
\multirow{2}{*}{Mode}	&			$\BR(\Bp \to {\rm Mode}) \times $								&\multirow{2}{*}{			$\;\;\BR(\Bp \to {\rm Mode}) \times 10^{-6}\;\;$	}							&	$\;\;$ Previous world$\;\;$	\\
	&			$\BR(R \to h\pi ) \times 10^{-6}$							&													&	average~\cite{Agashe:2014kda} $(\times 10^{-6})$	\\
\hline
\multirow{2}{*}{$ K^{*}(892)^{0} \pip \g $  }	& \multirow{2}{*}{$	15.6 \pm 0.6 \pm 0.5	$}&\multirow{2}{*}{$		23.4 \pm 0.9^{+	0.8	}_{-	0.7	}		$}&	\multirow{2}{*}{ $20 ^{+7}_{-6} $ }	\\
	&											&													&		\\
\multirow{2}{*}{$ \Kp \rho(770)^0 \g $  }	& \multirow{2}{*}{$	8.1	\pm 0.4 ^{+	0.8	}_{-	0.7	}	$}&\multirow{2}{*}{$		8.2	\pm 0.4 \pm 0.8 \pm	0.02	$}&	\multirow{2}{*}{ $<20 $ at $90\%$ CL }	\\
	&											&													&		\\
\multirow{2}{*}{$ (K \pi)^{*0}_0 \pip \g $  }	& \multirow{2}{*}{$	10.3	^{+	0.7	}_{-	0.8	}$$^{+	1.5	}_{-	2.0	}	$}&\multirow{2}{*}{$		\cdot\cdot\cdot											$}&	\multirow{2}{*}{ n/a }	\\
	&											&													&		\\
\hline
\multirow{2}{*}{$ (K \pi)^{0}_0 \pip \g$ (NR)  }	& \multirow{2}{*}{$	\cdot\cdot\cdot										$}&\multirow{2}{*}{$		9.9	\pm 0.7^{+	1.5	}_{-	1.9	}		$}&	\multirow{2}{*}{ $<9.2 $ at $90\%$ CL }	\\
	&											&													&		\\
\multirow{2}{*}{$ K^*_0(1430)^0 \pip \g $  }	& \multirow{2}{*}{$	0.82	\pm 0.06 ^{+	0.12	}_{-	0.16	}	$}&\multirow{2}{*}{$		1.32	^{+	0.09	}_{-	0.10	}$$^{+	0.20	}_{-	0.26	}\pm	0.14	$}&	\multirow{2}{*}{ n/a }	\\
	&											&													&		\\
\hline
\hline
\end{tabular*}	
\end{center}
\end{table*}
\begin{figure*}[htbp]
\hspace{-25pt}
\begin{tabular}{cc}
	\includegraphics[height= 5.5cm] {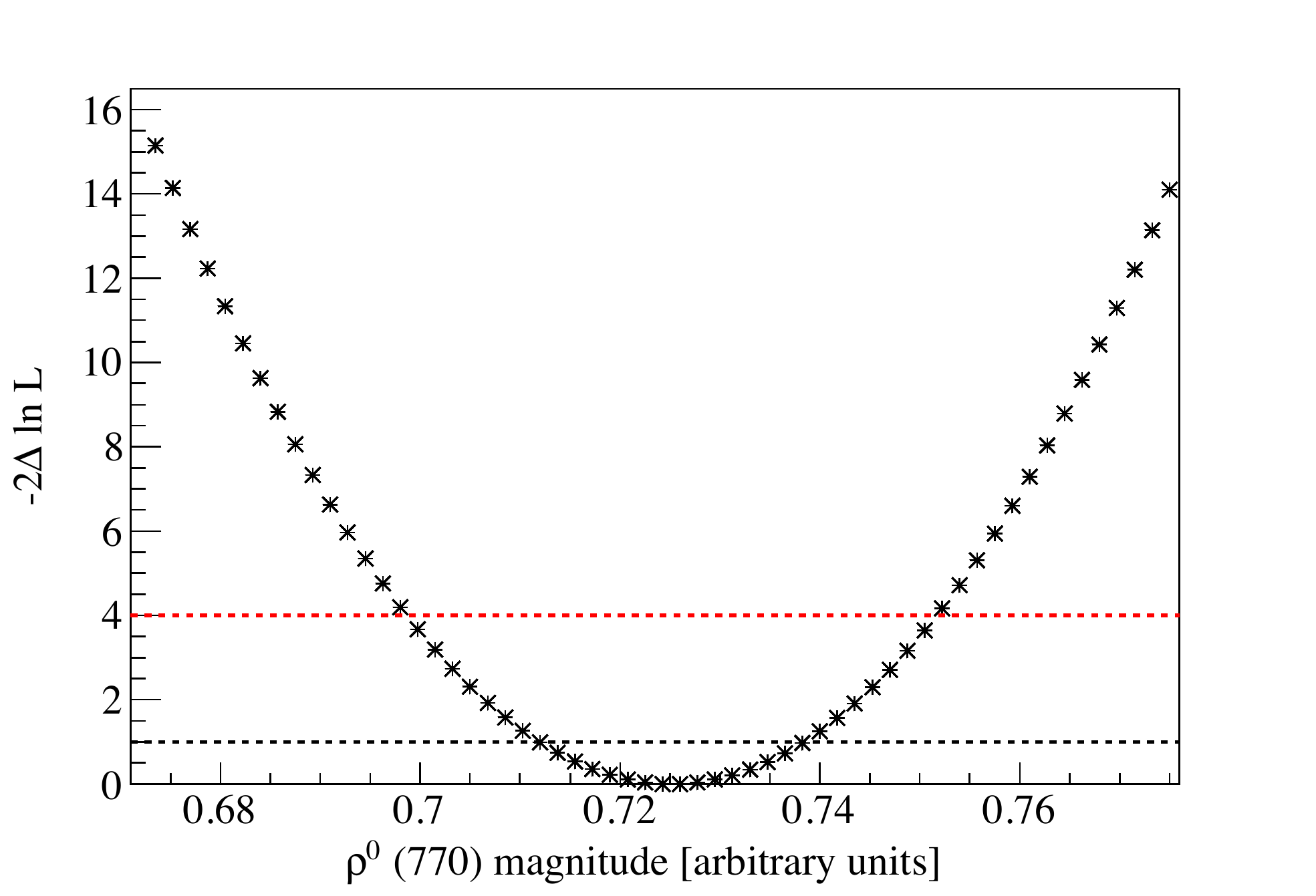} 
	& 
	\includegraphics[height= 5.5cm] {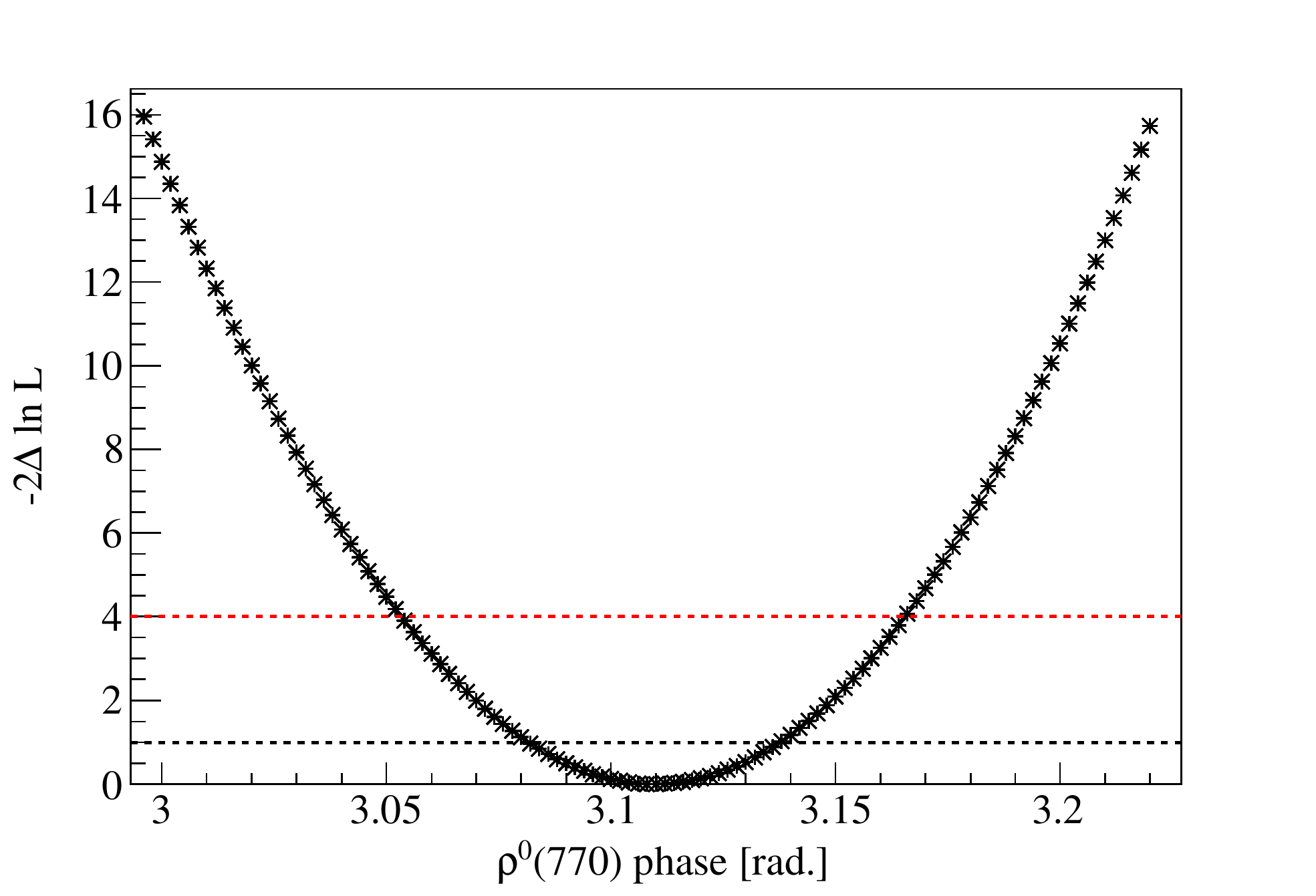}\\
\includegraphics[height= 5.5cm] {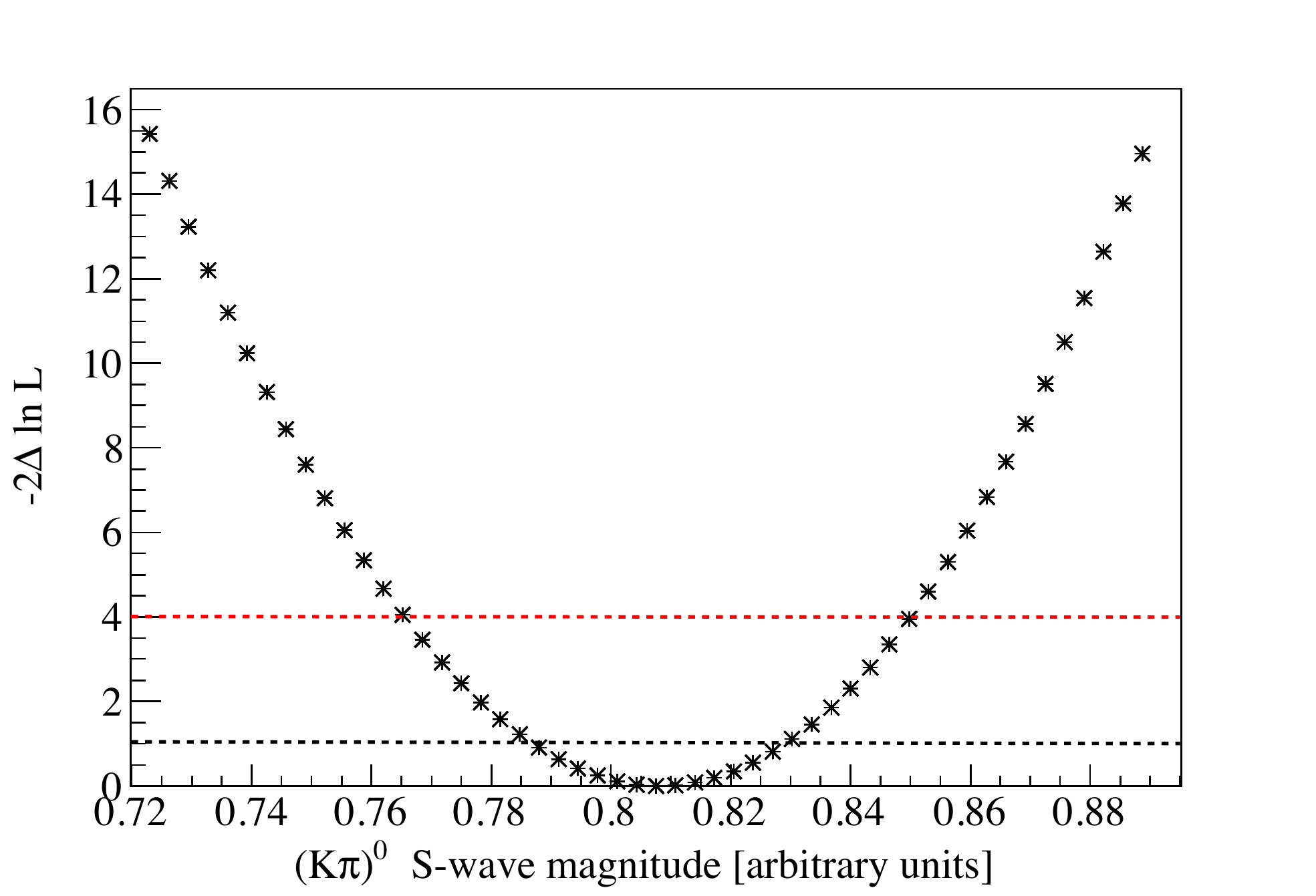}
& 
\includegraphics[height= 5.5cm] {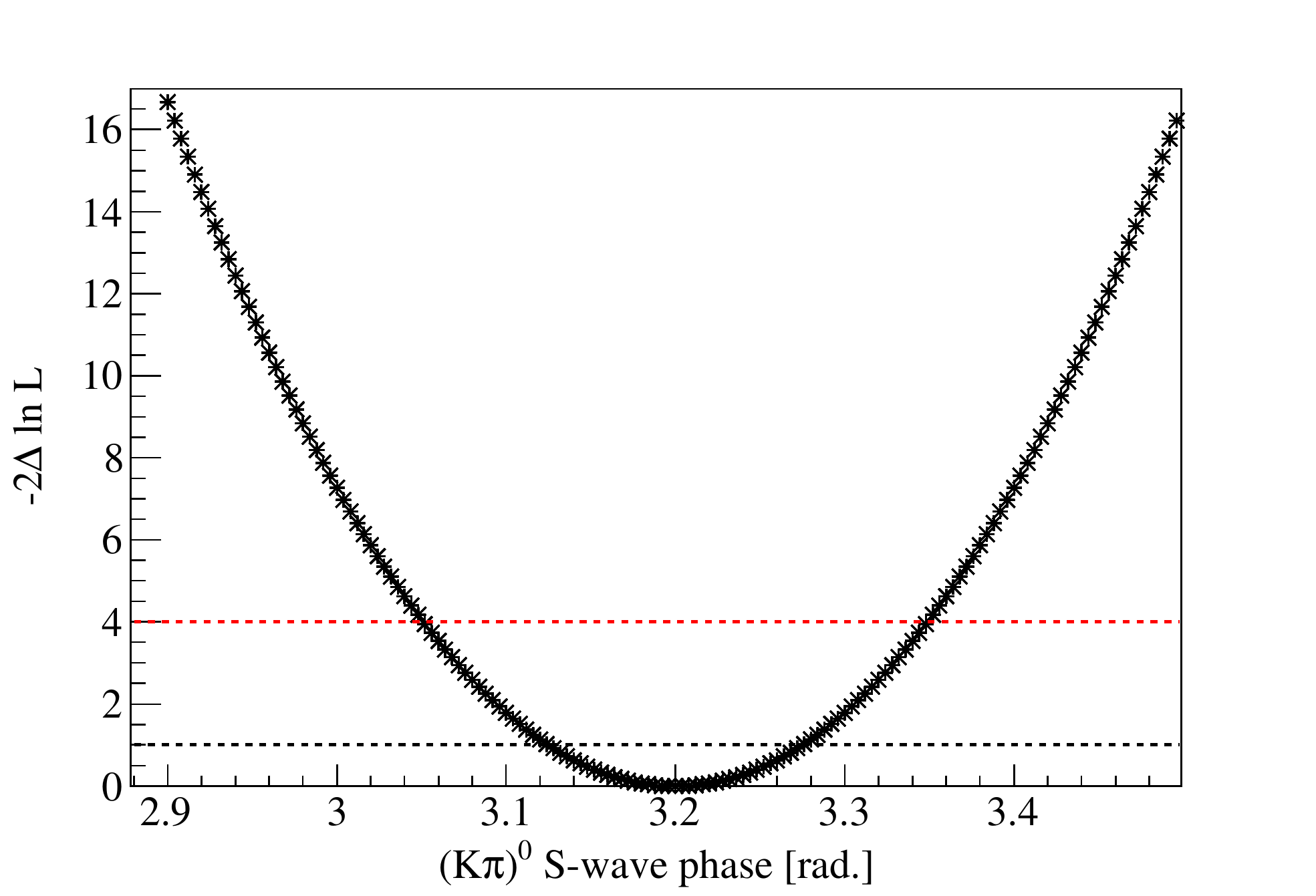}\\
\end{tabular}
\caption[One-dimensional scans of $-2\Delta \textrm{ln}\mathcal{L}$ as a function of $\Rhoz$ and $\swave$ magnitudes and $\Rhoz$ phase]{One-dimensional scans of $-2\Delta \textrm{ln}\mathcal{L}$ as a function of the magnitudes (left) and phases (right) of the $\Rhoz$ and $\swave$ components. The horizontal dashed lines mark the one- and two-standard deviation levels.
\label{fig:CC_mKpi_likelihood_Scans}}
\end{figure*}

\subsubsection{The \mKpi spectrum}
\label{sec:CC_results_mKpi}

Figure~\ref{fig:FitProjmKpi} shows the efficiency-corrected \mKpi \splot distribution that is also extracted from the unbinned maximum-likelihood fit to \mes, \DeltaE, and $\fisher$ and is corrected for efficiency effects (see Sec.~\ref{sec:mKpi_model}).
The figure shows the contributions of the different two-body resonances, as extracted from the fit to the \mKpi spectrum itself. 
Table~\ref{tab:mKpi_fit_results} summarizes the relative magnitudes and phases of the different components of the signal model, measured directly from the fit to the \mKpi spectrum, as well as the corresponding fit fractions computed using Eqs.~\eqref{equ:fit_fractions_res} and~\eqref{equ:fit_fractions_inter}. 
The statistical uncertainties on the magnitudes and phases come directly from the fit while the statistical uncertainties on the fit fractions are estimated in the same way as those obtained in the fit to the \mKpipi spectrum. 
As in the fit to the \mKpipi spectrum, we perform likelihood scans of the fitted parameters, shown in Fig.~\ref{fig:CC_mKpi_likelihood_Scans}, in order to check for multiple solutions. 
The fitted solution appears to be unique.

Table~\ref{tab:BFmKpi} summarizes the branching fractions via intermediate $\Kp \rho(770)^0$, $K^{*}(892)^{0} \pip$ and $(K \pi)^{*0}_0 \pip$ decays that are obtained after inserting the two-body resonance fit fractions into Eq.~\eqref{eq:BR_hRes}. 
Since the $(K \pi)^{*0}_0$ component is modeled by the LASS parametrization, which consists of a NR effective range term plus a relativistic Breit--Wigner term for the $K^*_0(1430)^0$ resonance, we report a separate branching fraction for the $K^*_0(1430)^0$ of $\BR(\Bp \to K^*_0(1430)^0 \pip \g) = (1.44 \pm 0.19 	^{+	0.26 }_{- 0.34} \pm 0.14) \times 10^{-6}$ after correction for the $\BR(K^*_0(1430) \to K \pi)$~\cite{Agashe:2014kda} and the isospin factor of $2/3$.
The first uncertainty is statistical, the second is systematic, and the third is due to the uncertainty on the secondary branching fraction. 
Since in this analysis the $K^*_0(1430)^0$ contribution is modeled exclusively in the decay process $\Bp \to K_1(1270)^+ (\to K^*_0(1430)^0 \pip) \g$, we extract a branching fraction of $\BR( K_1(1270)^+ \to K^*_0(1430)^0 \pip ) = (3.34 ^{+	0.62 + 0.64	}_{- 0.54 - 0.82 }) \times 10^{-2}$, where the first uncertainty is statistical and the second is systematic. 
This result is in good agreement with the measurement performed by the Belle collaboration in the analysis of $B \to \jpsi (\psi')K\pi\pi$ decays~\cite{Guler:2010if}, while it is significantly smaller than the value given in Ref.~\cite{Agashe:2014kda}.
In the present analysis, the relative fraction between the resonant and NR part of the LASS is fixed while the overall $(K \pi)^{*0}_0$ contribution is a free parameter in the fit. 
The NR contribution, described by the effective range part of the LASS parametrization, is found to be $ (11.0	^{+	1.4	}_{-	1.5	}$$^{+	2.0	}_{-	2.5	}) \times 10^{-6}$.
As in the case of the three-body resonance branching fraction measurement, we assume a 100\% correlation between the fitted signal yield and the fit fraction when calculating the statistical uncertainty on each branching fraction.

We compute the dilution factor defined in Eq.~\eqref{eq:dilutionExpression} by inserting the FFs extracted from the fit to the \mKpi spectrum into the expressions listed in Appendix~A, which show the relations between amplitudes and the FFs. 
To optimize the sensitivity to \Srho, we impose in the dilution factor calculation the mass requirements $600 \leq \mpipi \leq 900\mevcc$ and $\mKpi^{\rm min} \leq \mKpi \leq 845\mevcc$ or $945\mevcc \leq \mKpi \leq \mKpi^{\rm max}$, where $\mKpi^{\rm min}$ and $\mKpi^{\rm max}$ denote the allowed phase-space boundaries in the \mKpi dimension.
The \mpipi mass requirement accounts for the distortion of the $\Rhoz$ line shape towards the low invariant mass region due to phase-space effects.
Using the integration region defined above in the \mpipi and \mKpi dimensions, we obtain
\begin{eqnarray}
&& 	\int \left| A_{\rho\KS}\right|^2 d\mpipi d\mKpi = 0.269 \pm 0.028, \nonumber \\
&&  \int \Big| A_{\Kstarp\pim}\Big|^2 d\mpipi d\mKpi = 0.078 \pm 0.002, \nonumber \\
&&  \int \left| A_{\swavep\pim}\right|^2 d\mpipi d\mKpi = 0.141^{+0.029}_{-0.027}, \nonumber \\
&& 	\int 2\Re\!\left( A^{\ast}_{\rho\KS} A_{\Kstarp\pim}\right) d\mpipi d\mKpi = -0.090 \pm 0.006,\nonumber \\
&& 	\int 2\Re\!\left( A^{\ast}_{\rho\KS} A_{\swavep\pim}\right)  d\mpipi d\mKpi = -0.149^{+0.052}_{-0.040}, \nonumber
\end{eqnarray}
where the uncertainties account for both statistical and systematic uncertainties, which are summed in quadrature.
Inserting the above results into Eq.~\eqref{eq:dilutionExpression}, yields 
\begin{equation}
\D = -0.78^{+ 0.19}_{-0.17},
\label{eq:dilutionFactor}
\end{equation}
where the uncertainties are statistical and systematic uncertainties added in
quadrature. 
The systematic uncertainties contribution are discussed in Sec.~\ref{sec:CC_systematics}.

\subsection{Systematic uncertainties}
\label{sec:CC_systematics}

\begin{table*}[t!]
\begin{center}
\caption[Systematic uncertainties of the parameters of thefit to the \mKpipi]{\label{tab:CC_mKpipiAmp_syst} Systematic uncertainties of the parameters of thekaonic resonance amplitudes extracted from a fit to the \mKpipi spectrum. 
The symbol $\emptyset$ denotes a systematic uncertainty of zero, while $0.0$ indicates that the corresponding systematic uncertainty is less than $0.05\%$.
}
\setlength{\tabcolsep}{1.25pc}
\begin{tabular*}{\textwidth}{@{\extracolsep{\fill}} l | c c c c | c c}
\hline
\hline

\multirow{6}{*}{Source}	&	\multicolumn{6}{c}{\multirow{2}{*}{$+/-$ signed deviation (\%)}}		\\
 & \multicolumn{6}{c}{} \\

\cline{2-7}
					
            &  \multicolumn{4}{c|}{\multirow{2}{*}{Magnitude}} & \multicolumn{2}{c}{\multirow{2}{*}{Phase}}   \\
 & \multicolumn{4}{c|}{\multirow{2}{*}{}} & \multicolumn{2}{c}{\multirow{2}{*}{}}  \\

\cline{2-7}

            &  \multirow{2}{*}{$ \Kll$} & \multirow{2}{*}{$ \KstarX$} & \multirow{2}{*}{$ \Kstarlllmy$} & \multirow{2}{*}{$ \KstarlVmy$} &  \multirow{2}{*}{$ \Kll$} &  \multirow{2}{*}{$ \KstarX$}  \\
& & & & & &  \\
            
\hline

Fixed parameters in  &\multirow{3}{*}{$	2.7	/	2.3	$}&\multirow{3}{*}{$	3.7	/	2.1	$}&\multirow{3}{*}{$	5.8	/	6.4	$}&\multirow{3}{*}{$	4.2	/	2.2	$}&\multirow{3}{*}{$	0.6	/	0.5	$}&\multirow{3}{*}{$	0.3	/	0.2	$}\\
the fit performed to & & & & & & \\
 \mes, \DeltaE and Fisher & & & & & & \\

\hline

Fixed line-shape  &\multirow{3}{*}{$	16	/	11	$}&\multirow{3}{*}{$	12	/	11	$}&\multirow{3}{*}{$	31	/	39	$}&\multirow{3}{*}{$	12	/	12	$}&\multirow{3}{*}{$	3.6	/	3.9	$}&\multirow{3}{*}{$	0.6	/	0.6	$}\\
 parameters of the & & & & & & \\
 kaonic resonances & & & & & & \\

\hline

	Number of bins  &\multirow{2}{*}{$	0.4	/	0.2	$}&\multirow{2}{*}{$	0.4	/	0.2	$}&\multirow{2}{*}{$	0.5	/	1.9	$}&\multirow{2}{*}{$	0.4	/	0.2	$}&\multirow{2}{*}{$	0.1	/	0.1	$}&\multirow{2}{*}{$	0.0	/	0.0	$}\\
in the fitted dataset & & & & & & \\

\hline

\multirow{2}{*}{	\splot procedure} &\multirow{2}{*}{$	0.4	/	\emptyset	$}&\multirow{2}{*}{$	\emptyset	/	1.3	$}&\multirow{2}{*}{$	\emptyset	/	2.0	$}&\multirow{2}{*}{$	\emptyset	/	2.5	$}&\multirow{2}{*}{$	0.1	/	\emptyset	$}&\multirow{2}{*}{$	0.0	/	\emptyset	$}\\
 & & & & & & \\

\hline

\mKpipi fit model  &\multirow{3}{*}{$	0.0	/	0.3	$}&\multirow{3}{*}{$	11.6	/	\emptyset	$}&\multirow{3}{*}{$	\emptyset	/	20.8	$}&\multirow{3}{*}{$	4.8	/	\emptyset	$}&\multirow{3}{*}{$	\emptyset	/	0.3	$}&\multirow{3}{*}{$	0.1	/	1.3	$}\\
(add and remove & & & & & & \\
kaonic resonances) & & & & & & \\

\hline					
\hline	
\end{tabular*}
\end{center}
\end{table*}
\begin{table*}[t!]
\begin{center}
\caption[Systematic uncertainties of the parameters of thefit to the \mKpipi]{\label{tab:CC_mKpipiFF_syst} Systematic uncertainties on the kaonic resonance fit fractions extracted from a fit to the \mKpipi spectrum. 
The symbol $\emptyset$ denotes a systematic uncertainty of zero, while $0.0$ indicates that the corresponding systematic uncertainty is less than $0.05\%$.
The term ``Sum'' represents the sum of all fit fractions without interference terms, which can deviate from unity.
}
\setlength{\tabcolsep}{0.67pc}
\begin{tabular*}{\textwidth}{@{\extracolsep{\fill}} l | c c c c c  c  c c}
\hline
\hline

\multirow{6}{*}{Source}	&	\multicolumn{8}{c}{\multirow{2}{*}{$+/-$ signed deviation (\%)}}		\\
 & \multicolumn{8}{c}{} \\

\cline{2-9}
					
            &  \multicolumn{8}{c}{\multirow{2}{*}{Fit Fraction}}  \\
 & \multicolumn{8}{c}{} \\

\cline{2-9}

            & \multirow{2}{*}{$ \Kl$} & \multirow{2}{*}{$ \Kll$} & \multirow{2}{*}{$ \KstarX$} & \multirow{2}{*}{$ \Kstarlllmy$} & \multirow{2}{*}{$ \KstarlVmy$} & \multirow{2}{*}{Sum} & \multicolumn{2}{c}{interference}    \\
& & & & & & & $J^P = 1^+$ &$J^P = 1^-$ \\
            
\hline

Fixed parameters in &\multirow{3}{*}{$	1.1	/	1.3	$}&\multirow{3}{*}{$	2.9	/	2.8	$}&\multirow{3}{*}{$	3.1	/	2.2	$}&\multirow{3}{*}{$	16	/	18	$}&\multirow{3}{*}{$	1.6	/	1.5	$}&\multirow{3}{*}{$	0.6	/	0.5	$}&\multirow{3}{*}{$	3.1	/	1.7	$}&\multirow{3}{*}{$	2.7	/	3.9	$}\\
the fit performed  to & & & & & & & & \\
 \mes, \DeltaE and Fisher & & & & & & & & \\

\hline

Fixed line-shape  &\multirow{3}{*}{$	8.0	/	8.2	$}&\multirow{3}{*}{$	28	/	20	$}&\multirow{3}{*}{$	10	/	7.6	$}&\multirow{3}{*}{$	79	/	87	$}&\multirow{3}{*}{$	18	/	11	$}&\multirow{3}{*}{$	7.0	/	4.8	$}&\multirow{3}{*}{$	15	/	15	$}&\multirow{3}{*}{$	17	/	29	$}\\
 parameters of the & & & & & & & & \\
 kaonic resonances & & & & & & & & \\

\hline

Number of bins &\multirow{2}{*}{$	0.1	/	1.4	$}&\multirow{2}{*}{$	4.0	/	0.6	$}&\multirow{2}{*}{$	1.3	/	1.4	$}&\multirow{2}{*}{$	5.0	/	3.1	$}&\multirow{2}{*}{$	1.4	/	0.1	$}&\multirow{2}{*}{$	0.1	/	0.1	$}&\multirow{2}{*}{$	0.6	/	0.4	$}&\multirow{2}{*}{$	0.3	/	0.3	$}\\
in the fitted dataset & & & & & & & & \\

\hline

\multirow{2}{*}{	\splot procedure} &\multirow{2}{*}{$	1.4	/	\emptyset	$}&\multirow{2}{*}{$	3.3	/	\emptyset	$}&\multirow{2}{*}{$	\emptyset	/	0.1	$}&\multirow{2}{*}{$	\emptyset	/	1.7	$}&\multirow{2}{*}{$	\emptyset	/	2.0	$}&\multirow{2}{*}{$	\emptyset	/	0.2	$}&\multirow{2}{*}{$	\emptyset	/	2.5	$}&\multirow{2}{*}{$	1.6	/	\emptyset	$}\\
 & & & & & & & & \\

\hline

\mKpipi fit model  &\multirow{3}{*}{$	0.0	/	2.1	$}&\multirow{3}{*}{$	0.1	/	4.2	$}&\multirow{3}{*}{$	20	/	\emptyset	$}&\multirow{3}{*}{$	\emptyset	/	41	$}&\multirow{3}{*}{$	0.2	/	12	$}&\multirow{3}{*}{$	1.0	/	\emptyset	$}&\multirow{3}{*}{$	3.2	/	0.1	$}&\multirow{3}{*}{$	\emptyset	/	9.3	$}\\
(add and remove & & & & & & & & \\
kaonic resonances) & & & & & & & & \\

\hline					
\hline	
\end{tabular*}
\end{center}
\end{table*}

Since the main purpose of the analysis of \MyControlChanel decays is to extract the dilution factor \D, we have studied the systematic effects that influence its value. 
The dilution factor uncertainties depend on uncertainties of the two-body amplitudes obtained from a fit to the \mKpi spectrum (see Sec.~\ref{sec:CC_mKpiSyst}), themselves depending on the uncertainties of the kaonic-resonance amplitudes obtained from a fit to the \mKpipi spectrum (see Sec.~\ref{sec:CC_mKpipiSyst}). 
Finally, in Sec.~\ref{sec:CC_BRSyst}, the systematic uncertainties corresponding to the branching fractions measurements are described. 
For the combination of asymmetric systematic uncertainties, the method described in Ref.~\cite{Barlow:2003sg} was used.

\subsubsection{Kaonic resonance amplitudes}
\label{sec:CC_mKpipiSyst}

Table~\ref{tab:CC_mKpipiAmp_syst} gives the systematic uncertainties on the kaonic resonance amplitude parameters and Table~\ref{tab:CC_mKpipiFF_syst} gives the systematic uncertainties on the corresponding fit fractions.
The dominant sources of systematic uncertainty are the fixed parameters of the resonance lineshapes in the \mKpipi fit model.
The large relative effect of fixed line-shape parameters on the magnitude and the fit fraction of the $\Kstarlllmy$ are due to its small contribution.

To assign systematic uncertainties due to the fixed parameters in the fit to \mes, \DeltaE and $\fisher$, we vary each of the fixed parameters within its uncertainty, based on a fit to the simulated event sample, and we repeat the fit. 
Since the \mes-\DeltaE distribution of \BbkgA background events is described by a two-dimensional histogram, the fit is performed fluctuating the bin contents according to a Gaussian distribution centered on the nominal bin content and with a width given by the corresponding statistical uncertainty. The procedure is repeated 50 times.
The root mean square (rms) of the resulting distribution of fitted parameter values is taken as the systematic uncertainty.
The fixed yields are varied according to the corresponding branching fraction uncertainties taken from Ref.~\cite{Agashe:2014kda}. 
For the categories describing a sum of modes, the fraction of each mode is varied according to the relative branching fraction uncertainties taken from Ref.~\cite{Agashe:2014kda}. 
The mis-reconstructed signal yield is varied according to the uncertainties due to the sample size of simulated events and the signal branching fraction uncertainty in Ref.~\cite{Agashe:2014kda}. 
The fixed yield of the generic \B-background category, describing a sum of several small contributions from various \B-background modes, is varied within the uncertainties due to the sample size of simulated events.
For each new fit performed this way, we derive the corresponding \mKpipi \splot distribution that we then fit using the nominal \mKpipi model. 
Assuming no correlations among the fixed parameters, we combine each of the negative (positive) difference between the new fit value and nominal fit value of each free parameter, and take the resulting values as negatively (positively) signed uncertainties.

To assign systematic uncertainties due to the choice of bin size in the fitted dataset, we perform new fits using either 60 or 100 bins, instead of 80 in the nominal fit model. 

To assign systematic uncertainties due to the fixed parameters of the line-shape resonances in the \mKpipi fit model, we vary each of the eight fixed parameters according to its uncertainties, taken from Ref.~\cite{Agashe:2014kda}, and redo the fit to the nominal CR signal \mKpipi \splot distribution. 

For the systematic uncertainties due to the fit model (i.e. the resonances describing the \mKpipi spectrum), we vary the nominal model by adding other kaonic resonances at high masses to the fit model. 
We considered three additional resonances, the $K_2(1770)$, the $K^*_3(1780)$, and the $K_2(1820)$, whose parameters are given in Table~\ref{tab:CC_mKpipi_model_Add_Res}.
We add each of these resonances in turn to the model and
re-perform the fit to the CR signal \mKpipi \splot distribution.
We observe no variations on the parameters of the fit to the \mKpipi spectrum when the $K_2(1820)$ is added to the resonance model. Using the method described in Ref.~\cite{Barlow:2003sg}, we combine each of the negative (positive) difference between the new fit value and nominal fit value due to the presence of either the $K_2(1770)$ or the $K^*_3(1780)$ in the resonance model. 

If the yields of one or more event categories are fixed in the fit to an \splot spectrum, a correction is necessary (see Ref.~\cite{Pivk:2004ty}) to extract the CR signal \splot.
This correction implies that the distributions of the variable of interest for the fixed categories are well known. 
The \mKpipi distributions of the event categories with fixed yields cannot be considered to completely fulfill this criterion since they are taken from simulation. 
A detailed description of the evaluation of the systematic uncertainties due to the \splot technique is given in Appendix~B.

\begin{table}[htbp]
\begin{center}
\caption[Additional resonances considered in the \mKpipi fit model]{\label{tab:CC_mKpipi_model_Add_Res} Additional resonances considered in the \mKpipi fit model. The pole mass $m^0_k$ and width $\Gamma^0_k$ are fixed to the values taken from Ref.~\cite{Agashe:2014kda}.}
\vspace{5pt}
\begin{tabular*}{\columnwidth}{@{\extracolsep{\fill}}l | c c c}
\hline
\hline
\multirow{2}{*}{$J^P$} & \multirow{2}{*}{\Kres} & Mass $m^0_k$	& Width $\Gamma^0_k$ \\
&& 	(\mevccnosp)&  (\mevccnosp)\\
\hline
\multirow{2}{*}{$2^-$}&$K_2(1770)$&	$1773 \pm 8$&$186 \pm 14$	\\
&	$K_2(1820)$&	$1816 \pm 13$	&	$276 \pm 35$	\\
\hline
 $3^-$&	$K^*_3(1780)$&	$1776 \pm 7$	& 	$159 \pm 21$	\\
\hline	
\hline	
\end{tabular*}
\end{center}
\end{table}
%

\subsubsection{Two-body resonances}
\label{sec:CC_mKpiSyst}

\begin{table*}[t!]
\begin{center}
\caption{\label{tab:CC_mKpi_syst} Systematic uncertainties of the parameters of theintermediate state resonance amplitudes and on the corresponding fit fractions extracted from a fit to the \mKpi spectrum. 
The symbol $\emptyset$ denotes a systematic uncertainty of zero, while $0.0$ indicates that the corresponding systematic uncertainty is less than $0.05\%$.
The term ``Sum'' represents the sum of all fit fractions without interference terms, which can deviate from unity.
The quoted systematic uncertainties due to the number of bins in the fitted PDF correspond to the combined systematic uncertainties from the bins in \mKpi and \mpipi, which were estimated separately as described in Sec.~\ref{sec:CC_mKpiSyst}.
}
\setlength{\tabcolsep}{0.25pc}
\begin{tabular*}{\textwidth}{@{\extracolsep{\fill}} l | c c | c c | c c c c c c }
\hline
\hline

\multirow{6}{*}{Source}	&	\multicolumn{10}{c}{\multirow{2}{*}{$+/-$ signed deviation (\%)}}		\\
 & \multicolumn{6}{c}{} \\

\cline{2-11}
					
            &  \multicolumn{2}{c|}{\multirow{2}{*}{Magnitude}} &  \multicolumn{2}{c|}{\multirow{2}{*}{Phase}} &  \multicolumn{6}{c}{\multirow{2}{*}{Fit Fraction}}  \\
 & \multicolumn{2}{c|}{} & \multicolumn{2}{c|}{} & \multicolumn{6}{c}{} \\

\cline{2-11}

            & \multirow{2}{*}{ $\rho(770)^0$} & \multirow{2}{*}{$(K \pi)^{*0}_0$} & \multirow{2}{*}{ $\rho(770)^0$} & \multirow{2}{*}{$(K \pi)^{*0}_0$} & \multirow{2}{*}{$K^{*}(892)^{0} $} & \multirow{2}{*}{ $\rho(770)^0$} & \multirow{2}{*}{$(K \pi)^{*0}_0$}  & \multirow{2}{*}{$\;\;\;$Sum$\;\;\;$} & \multicolumn{2}{c}{interference}    \\
& & & & & & & & &  $K^{*0}$--$\rho^0$ &$(K \pi)^{*0}_0$--$\rho^0$ \\
            
\hline

Fixed parameters in the fit  	&\multirow{3}{*}{$	1.5	/	2.2	$}&\multirow{3}{*}{$	4.0	/	3.5	$}&\multirow{3}{*}{$	0.6	/	0.5	$}&\multirow{3}{*}{$	1.8	/	1.1	$}&\multirow{3}{*}{$	0.8	/	0.7	$}&\multirow{3}{*}{$	3.1	/	4.2	$}&\multirow{3}{*}{$	7.9	/	6.7	$}&\multirow{3}{*}{$	0.3	/	0.4	$}&\multirow{3}{*}{$	2.5	/	1.9	$}&\multirow{3}{*}{$	5.3	/	4.5	$}\\				
performed  to  \mes, \DeltaE and & & & & & & & & & & \\
 $\fisher$  & & & & & & & & & & \\
\hline

Fixed line-shape parameters 	&\multirow{3}{*}{$	0.3	/	0.2	$}&\multirow{3}{*}{$	0.9	/	0.6	$}&\multirow{3}{*}{$	0.4	/	0.6	$}&\multirow{3}{*}{$	1.1	/	1.4	$}&\multirow{3}{*}{$	0.3	/	0.5	$}&\multirow{3}{*}{$	1.6	/	2.5	$}&\multirow{3}{*}{$	3.7	/	1.9	$}&\multirow{3}{*}{$	0.4	/	0.6	$}&\multirow{3}{*}{$	1.2	/	0.8	$}&\multirow{3}{*}{$	5.3	/	3.2	$}\\				
of the intermediate state & & & & & & & & & & \\
 resonances & & & & & & & & & & \\
\hline

Fixed line-shape parameters 	&\multirow{3}{*}{$	0.5	/	0.3	$}&\multirow{3}{*}{$	1.1	/	1.4	$}&\multirow{3}{*}{$	1.1	/	1.7	$}&\multirow{3}{*}{$	1.7	/	2.1	$}&\multirow{3}{*}{$	0.5	/	0.8	$}&\multirow{3}{*}{$	0.1	/	0.1	$}&\multirow{3}{*}{$	1.8	/	2.7	$}&\multirow{3}{*}{$	0.2	/	0.1	$}&\multirow{3}{*}{$	0.9	/	1.5	$}&\multirow{3}{*}{$	3.4	/	2.9	$}\\				
of the kaonic resonances & & & & & & & & & & \\
 (in \evtgen)  & & & & & & & & & & \\

\hline

\multirow{2}{*}{Number of bins in the PDF} 	&\multirow{2}{*}{$	0.0	/	0.6	$}&\multirow{2}{*}{$	2.4	/	0.0	$}&\multirow{2}{*}{$	0.4	/	0.0	$}&\multirow{2}{*}{$	0.4	/	0.0	$}&\multirow{2}{*}{$	0.0	/	1.0	$}&\multirow{2}{*}{$	0.0	/	0.8	$}&\multirow{2}{*}{$	3.6	/	0.0	$}&\multirow{2}{*}{$	0.0	/	1.6	$}&\multirow{2}{*}{$	0.6	/	0.0	$}&\multirow{2}{*}{$	3.5	/	0.0	$}\\				
 & & & & & & & & & & \\

\hline

Number of bins in the fitted 	&\multirow{2}{*}{$	0.8	/	0.0	$}&\multirow{2}{*}{$	0.0	/	4.3	$}&\multirow{2}{*}{$	0.0	/	0.3	$}&\multirow{2}{*}{$	0.0	/	0.5	$}&\multirow{2}{*}{$	1.8	/	0.0	$}&\multirow{2}{*}{$	4.2	/	0.0	$}&\multirow{2}{*}{$	0.0	/	7.1	$}&\multirow{2}{*}{$	3.8	/	0.0	$}&\multirow{2}{*}{$	0.0	/	3.3	$}&\multirow{2}{*}{$	0.0	/	9.4	$}\\				
dataset & & & & & & & & & & \\

\hline

\multirow{2}{*}{	\splot procedure} 	&\multirow{2}{*}{$	\emptyset	/	2.6	$}&\multirow{2}{*}{$	3.7	/	\emptyset	$}&\multirow{2}{*}{$	\emptyset	/	0.5	$}&\multirow{2}{*}{$	\emptyset	/	1.3	$}&\multirow{2}{*}{$	0.2	/	\emptyset	$}&\multirow{2}{*}{$	\emptyset	/	8.0	$}&\multirow{2}{*}{$	10	/	\emptyset	$}&\multirow{2}{*}{$	\emptyset	/	3.5	$}&\multirow{2}{*}{$	2.1	/	\emptyset	$}&\multirow{2}{*}{$	6.9	/	\emptyset	$}\\				
 & & & & & & & & & & \\

\hline

Kaonic resonance weights  	&\multirow{3}{*}{$	1.5	/	0.5	$}&\multirow{3}{*}{$	1.2	/	6.0	$}&\multirow{3}{*}{$	1.0	/	1.1	$}&\multirow{3}{*}{$	2.6	/	1.5	$}&\multirow{3}{*}{$	2.2	/	1.2	$}&\multirow{3}{*}{$	8.8	/	2.1	$}&\multirow{3}{*}{$	3.1	/	17	$}&\multirow{3}{*}{$	6.3	/	2.2	$}&\multirow{3}{*}{$	3.0	/	4.6	$}&\multirow{3}{*}{$	11	/	20	$}\\				
(taken from a fit to   & & & & & & & & & & \\
 the \mKpipi spectrum)  & & & & & & & & & & \\

\hline					
\hline	
\end{tabular*}
\end{center}
\end{table*}

Table~\ref{tab:CC_mKpi_syst} summarizes both the systematic uncertainties on the intermediate state resonance amplitude parameters, and those on the corresponding fit fractions.
The dominant sources of systematic uncertainty are the weights of the kaonic resonances extracted from the fit to the \mKpipi spectrum.
The relatively large systematic uncertainties on the $(K \pi)^{*0}_0$ parameters and fit fraction are due to the low sensitivity to this component.

We account for two sources of systematic uncertainties from the number of bins: the first in the fitted \splot (90 bins in the nominal fit model) and another in the two-dimensional histograms used to create the PDF ($450\times100$ bins in the nominal fit model for $\mKpi\times\mpipi$).
We estimate the effect of the bin size of the \splot from fits performed with 75 and 105 bins, while the bin size of the PDF is fixed to its nominal value. 
We associate one systematic uncertainty to the bin size in \mKpi and another to that in \mpipi. 
We estimate the effect of the bin sizes of the PDF, in the $\mKpi(\mpipi)$ dimension, from fits performed with alternative PDFs with 270(50) and 630(150) bins in $\mKpi(\mpipi)$, and the nominal number of bins in the other dimension.
For each of these sources we take the lower and upper deviations from the nominal value of each FF as the corresponding uncertainty. 
We add the uncertainties coming from the bin size in $\mKpi(\mpipi)$ in quadrature assuming no correlations between them. 

To assign systematic uncertainties due to the fixed parameters in the fit to \mes, \DeltaE and $\fisher$, we use the procedure described in Sec.~\ref{sec:CC_mKpipiSyst}. We derive a set of new \mKpi \splot distributions that we fit using the nominal model. 

To account for systematic effects due to the fixed parameters of the resonances in the \mKpi fit model, we vary each of them according to the uncertainties given in Table~\ref{tab:mKpi_model}. 
These parameters appear both in the lineshapes used to generate the histograms of the resonances as well as in the corresponding analytical expressions of the phases. 
Therefore, for each parameter variation in a given lineshape, we generate a new distribution of the corresponding resonance, and use the same parameter value in the analytical phase expression. 
For each variation we perform a new fit to the nominal \mKpi \splot distribution. 
The largest effect is due to the line-shape parameters of the $K^*_0(1430)$ part of the LASS parametrization, while effects coming from the $\Rhoz$ and $K^{*}(892)^{0}$ line-shape parameters are negligible.   

To account for systematic effects due to the weights of kaonic resonances used to construct the PDF, we generate $10^4$ sets of weights from the full \mKpipi correlation matrix of fit fractions (taking into account the corresponding statistical and systematic uncertainties). 
Then, using each of these sets of weights as a new parametrization of the PDF, we perform a fit to the \mKpi spectrum. From the results of these fits we obtain a distribution for each free parameter and for each of the fit fractions. 
We take the values at plus and minus 34.1\% of the integral of the corresponding distribution on either side of the value obtained using the nominal fit model as the signed uncertainties, respectively. 

The distortions of the $\Rhoz$ and $K^{*}(892)^{0}$ resonances, taken into account in the fit model by histograms generated using simulated events from exclusive kaonic resonance decays, are correlated with the parameters of the kaonic-resonance lineshapes in the Monte Carlo generator. 
To study systematic effects from the fixed values of these parameters, we generate new simulated event distributions of the $\Rhoz$ and $K^{*}(892)^{0}$ for each kaonic resonance. 
The only significant effect for the $\Rhoz$ distribution is found in the $\Kl \to K\Rhoz$ decay channel.
To estimate the systematic uncertainty coming from the \Kl\ resonance parameters, we vary its mean and width, taken from Ref.~\cite{Agashe:2014kda}, within the uncertainties obtained from the fit to the \mKpipi spectrum. 
For each variation we generate a new PDF to perform a fit to the nominal \mKpi \splot distribution. 

To account for systematic effects coming from the \splot extraction procedure on the parameters of the fit to the \mKpi spectrum, we use the procedure described in Appendix~B.

\subsubsection{Branching fractions}
\label{sec:CC_BRSyst}

\begin{table}[b!] 
\begin{center}
\caption{\label{tab:CC_BR_InputBR} Input branching fractions with their corresponding uncertainties taken from Ref.~\cite{Agashe:2014kda} and used in the branching fractions computation.}
\begin{tabular*}{\columnwidth}{@{\extracolsep{\fill}} l c cc}
\hline	
\hline
Mode	&&&			$\BR({\rm Mode})  $ \\
\hline
 $\Y4S \to \BpBm$ &&& $0.513 \pm 0.006 $ \\
\hline
 $K_1(1270)^+ \to \Kp \pip \pim $ &&& $0.329 \pm 0.034 $ \\
 $K_1(1400)^+ \to \Kp \pip \pim $ &&& $0.422 \pm 0.027 $ \\
 $K^*(1410)^+ \to \Kp \pip \pim $ &&& $\,\,\,0.407 \pm 0.041 $\footnote{Since only upper and lower limits are reported in Ref.~\cite{Agashe:2014kda} for $\BR(K^*(1410) \to K \rho)$ and $\BR(K^*(1410) \to K^*(892) \pi)$, respectively, we take the $\BR(K^*(1410) \to K \rho)$ value as the reported upper limit for the calculation of $\BR(K_1(1400)^+ \to \Kp \pip \pim )$, to which we assign a total uncertainty of 10\%.} \\
 $K^*_2(1270)^+ \to \Kp \pip \pim $ &&& $0.139 \pm 0.007  $ \\
 $K^*(1680)^+ \to \Kp \pip \pim $ &&& $0.238 \pm 0.019  $ \\
 \hline
$\rho(770)^0 \to \pip \pim$ &&& $0.990 \pm 0.001 $ \\
 $K^*_0(1430)^0 \to \Kp \pim $ &&& $0.620 \pm 0.067 $ \\
\hline
\hline
\end{tabular*}
\end{center}
\end{table}

To assign systematic uncertainties on the yield for the CR signal category due to the fixed parameters in the fit to \mes, \DeltaE and $\fisher$, we use the same procedure as the one described in Sec.~\ref{sec:CC_mKpipiSyst}.
For each new fit, we obtain a new value of the CR signal event category yield. 
Using the method described in Ref.~\cite{Barlow:2003sg} and assuming no correlations among the fixed parameters, we combine each of the negative (positive) difference between the new fit value and nominal fit value of each free parameter, and take the resulting values as negatively (positively) signed uncertainties.

We use 0.6\% as the systematic uncertainty on $N_{\BB}$, corresponding to the uncertainty on the official \BB count for the full \babar\ dataset~\cite{McGregor:2008ek}.
Similarly to Ref.~\cite{Aubert:2005xk}, to account for possible differences between data and simulation in the tracking and particle identification efficiencies, we assign for each charged particle in the final state a systematic uncertainty of 0.24\% and 1\%, respectively.

The high energy photon selections applied in the present analysis are identical to those used in Ref.~\cite{Aubert:2005xk}, except for the additional likelihood ratio vetoes applied against $\piz$ and $\eta$ decays. 
We adopt a 2\% uncertainty for the requirement on the distance of the reconstructed photon energy cluster and the other energy clusters in the calorimeter, and a 1\% uncertainty due to the $\piz$ and $\eta$ vetoes, similarly to Ref.~\cite{Aubert:2005xk}.

The input branching fractions, as well as the corresponding uncertainties, used in the computation of the branching fractions, are taken from Ref.~\cite{Agashe:2014kda} and are summarized in Table~\ref{tab:CC_BR_InputBR}.

\section{Time-dependent analysis of \boldmath{\MyChanelPiPi} DECAYS}
\label{sec:TD_analysis}

In Sec.~\ref{sec:dt}, we describe the proper-time PDF used to extract the time-dependent \CP\ asymmetries.
In Sec.~\ref{sec:td_selection}, we describe the selection requirements used to obtain the signal candidates and to suppress backgrounds. 
In Sec.~\ref{sec:td_likelihood}, we describe the fit method and the approach used to account for experimental effects. 
In Sec.~\ref{sec:TD_fitresults}, we present the results of the fit and finally, in Sec.~\ref{sec:td_systematics}, we discuss systematic uncertainties.

\subsection{Proper-time PDF}
\label{sec:dt}

The time-dependent \CP\ asymmetries are functions of the proper-time difference $\deltat =t_{\rm rec} - t_{\rm tag}$ between a fully reconstructed \MyChanel decay ($\Brec^0$) and the other $B$ meson decay in the event ($\Btag^0$), which is partially reconstructed. 
The time difference \deltat is obtained from the measured distance between the decay-vertex positions of $\Brec^0$ and $\Btag^0$. The distance is transformed to \deltat using the boost $\beta\gamma = 0.56$ of the $e^+e^-$ system. 

The PDF for the decay rate is 
\begin{eqnarray}
\label{eq:dtmeas}
\lefteqn{{\cal P}^i_{\rm sig}(\Delta t,\sigma_{\deltat};q_{\rm tag},c) =}\\
\nonumber & & \frac{e^{-\left|\deltat\right|/\tau_{\Bz}}}{4\tau_{\Bz}} \Bigg[ 1 + q_{\rm tag}\frac{\Delta D_{c}}{2} \\
\nonumber & & \qquad\qquad \;\;\; + q_{\rm tag}\langle D\rangle_{c} \Bigl({\cal S}\sin(\Delta m_d\deltat)-\;{\cal C}\cos(\Delta m_{d}\deltat)\Bigr)  \\
\nonumber & & \qquad\qquad \;\;\; \Bigg] \;\otimes\; {\cal R}^c_{\rm sig}(\deltat,\sigma_{\deltat}), \nonumber
\end{eqnarray}
where $\tau_{\Bz}$ is the mean \Bz lifetime, $\Delta m_{d}$ is the mixing frequency~\cite{Aubert:2007hm}, ${\cal S}$ $({\cal C})$ is the magnitude for mixing-induced (direct) \CP violation, $q_{\rm tag}= 1(-1)$ for $B_{\rm tag}=\Bz$ ($B_{\rm tag}=\Bzb$), ${\langle D \rangle}_c$ is the average tagging imperfection for determining the correct \B flavor using tagging category $c$ and $\Delta D_c$ is the difference between $D_c$ for \Bz and \Bzb tags. 
We use a \B-flavor tagging algorithm~\cite{Aubert:2004zt} that combines several signatures, such as particle type, charges, momenta, and decay angles of charged particle in the event to achieve optimal separation between the two \B flavors, producing six mutually exclusive tagging categories. 
We assign the untagged events into a seventh category. 
Although these events do not contribute to the measurement of the time-dependent \CP asymmetry, they do provide additional sensitivity for the measurement of direct \CP violation~\cite{Gardner:2003su}. 
The exponential decay distribution modulated by oscillations due to mixing is convolved with the per-event $\Delta t$ resolution function ${\cal R}^c_{\rm sig}(\deltat,\sigma_{\deltat})$, which is parametrized by three Gaussian functions that depend on $\Delta t$ and its error $\sigma_{\deltat}$. 
The parameters of the resolution function can vary for each tagging category.

\subsection{Event selection and backgrounds}
\label{sec:td_selection}

The reconstruction of \MyChanelPiPi candidates is identical to that of \MyControlChanel candidates except for replacing the $\Kp$ with a $\KS$. The $\KS \to \pip\pim$ candidate is required to have a mass within $11\mevcc$ of the nominal $\KS$ mass, and a signed lifetime significance of at least five standard deviations.
The latter requirement ensures that the decay vertices of the \Bz and the $\KS$ are well separated.
In addition, combinatorial background is suppressed by requiring the cosine of the angle between the $\KS$ flight direction and the vector connecting the \Bz and the $\KS$ vertices to be greater than $0.995$.
Moreover, the \Bz candidates with $ \left|\Delta t\right| > 20 \ps $ are rejected, and so are candidates for which the uncertainty on \deltat is larger than $2.5 \ps$. 
The additional selection criteria $0.6<m_{\pi\pi}<0.9\gevcc$, $m_{K\pi}<0.845\gevcc$ or $m_{K\pi}>0.945\gevcc$ are applied for consistency with the corresponding requirements in the dilution factor calculation.

The set of variables used to build the Fisher discriminant in the analysis of \MyControlChanel decays (see Sec.~\ref{sec:CCEventSelection}) is also found to be optimal here. 
Therefore, we only update the coefficients in the linear combination to optimize the separation between signal and continuum background events. 
The requirement on the Fisher discriminant output value is optimized to minimize the statistical uncertainty on the \CP asymmetry parameters, \Ceff and \Seff. 
Furthermore, we again use the likelihood ratio, $\mathcal{L}_{\mathcal{R}}$, defined in Eq.~\eqref{eq:LR}, in order to reduce backgrounds from mis-reconstructed \piz and $\eta$ mesons.

We use simulated events to study the \B backgrounds. Only the channels with at least one event expected after selection are considered. 
We observe that the main \B backgrounds originate from \btosgam processes. 
\B background decays are grouped into classes of modes with similar kinematic and topological properties.
However, we distinguish \B backgrounds with different proper time distributions (see Sec.~\ref{sec:TD_deltatPDF}).

Table \ref{tab:TD_bbackground} summarizes the seven $B$-background classes that are used in the fit.
If the yield of a class is allowed to vary in the fit the quoted number of events corresponds to the fit results.
For the other classes, the yields are estimated from efficiencies derived from the simulation together with the world average branching fractions~\cite{Agashe:2014kda,Amhis:2014hma}.
When a $B$-background class contains a collection of many individual decay modes, as for the two generic $B$ backgrounds originating from either $B^+$ or $B^0$ mesons, respectively, the expected numbers of selected events are estimated solely from Monte Carlo.
The yield of the $B^+ \rightarrow \KS \pip \gamma$ class, which has a clear signature in \mes, is free to vary in the fit. The remaining background yields are fixed.

\begin{table*}[htbp]
\begin{center}
\caption{ \label{tab:TD_bbackground}
Summary of \B-background modes included in the fit model to \MyChanelPiPi decays.
If the yield is left free in the fit, the listed number of events corresponds to the fit results. 
Otherwise the expected number is given, which take into account the branching fractions (if applicable) and efficiencies.
The functions used to parametrize the \B-background PDFs are also given. The term ``Exp.'' corresponds to the exponential function. 
The PDFs for the $\Delta t$ distributions are discussed in Sec.~\ref{sec:TD_deltatPDF}. The terms ``$X_{su(sd)}(\nrightarrow K \pi)$'' denote all decays to $X_{su(sd)}$ final states except for the $K \pi$ final state.
}
\setlength{\tabcolsep}{0.0pc}
\begin{tabular*}{\textwidth}{@{\extracolsep{\fill}}lccccc}
\hline\hline
\multirow{2}{*}{Mode}                                 &   \multicolumn{3}{c}{PDFs}  & \multirow{2}{*}{Varied}                    & \multirow{2}{*}{Number of events} \\
& \mes &  \DeltaE & \fisher\ & \\
\hline\\[-9pt]
\multirow{2}{*}{\TDBbkgCb}      &   \multirow{2}{*}{ARGUS}  &  Chebychev  & \multirow{2}{*}{Gaussian} & \multirow{2}{*}{no}                              & \multirow{2}{*}{$ 94 \pm 17  $} \\
& &  ($2^{\mathrm{nd}}$ order) & & & \\
\hline\\[-9pt]
\multirow{2}{*}{\TDBbkgCa}      &  \multirow{2}{*}{ARGUS}  &  Chebychev & \multirow{2}{*}{Gaussian} & \multirow{2}{*}{no}                              & \multirow{2}{*}{$ 51 \pm 12  $} \\
& &($2^{\mathrm{nd}}$ order) & & & \\
\hline\\[-9pt]
\TDBbkgBa         & \multicolumn{2}{c}{Two-dimensional} & \multirow{2}{*}{Gaussian} & \multirow{2}{*}{yes}                       & \multirow{2}{*}{$ 42	 \pm	22  $} \\
\TDBbkgBb  & \multicolumn{2}{c}{nonparametric} & & & \\
\hline\\[-9pt]
\multirow{2}{*}{$B^0 \to \{\text{neutral generic decays}\}$} 	& \multirow{2}{*}{ARGUS}  &  Chebychev & \multirow{2}{*}{Gaussian} & \multirow{2}{*}{no}  & \multirow{2}{*}{$35 \pm 	13$}  \\
& & ($2^{\mathrm{nd}}$ order)& & & \\
\hline\\[-9pt]
\multirow{2}{*}{$B^+ \to \{\text{charged generic decays}\}$} 	& \multirow{2}{*}{ARGUS}  &  Chebychev & \multirow{2}{*}{Gaussian} & \multirow{2}{*}{no}  & \multirow{2}{*}{$34 \pm 	13$}  \\
& & ($2^{\mathrm{nd}}$ order)& & & \\
\hline\\[-9pt]
 \TDBbkgAc          &   \multirow{2}{*}{ARGUS} & Chebychev  & \multirow{2}{*}{Gaussian} & \multirow{2}{*}{no}                  & \multirow{2}{*}{$ 30	 \pm	11  $} \\
\TDBbkgAd &   & ($2^{\mathrm{nd}}$ order)& & & \\
\hline\\[-9pt]
\TDBbkgAa 		&   \multirow{2}{*}{ARGUS} &  Chebychev & \multirow{2}{*}{Exp.}                   & \multirow{2}{*}{no}                   & \multirow{2}{*}{$ 4  \pm 3  $} \\
\TDBbkgAb & &($1^{\mathrm{st}}$ order) & & & \\
\hline\hline
\end{tabular*}
\end{center}
\end{table*}
%

\subsection{The maximum-likelihood fit}
\label{sec:td_likelihood}
We perform an unbinned extended maximum-likelihood fit to extract the \MyChanelPiPi event yields along with the time-dependent \CP asymmetry parameters \S and \C. 

The PDFs in the fit depend on the variables: $\mes$, $\DeltaE$, $\fisher$, $\deltat$, and $\sigma_{\deltat}$. 
The selected on-resonance data sample is assumed to consist of signal, continuum background, and backgrounds from $B$ decays. 
The likelihood function ${\cal L}_i$ for event $i$ is the sum 
\begin{equation}
{\cal L}_i = \sum_j{N_j {\cal P}^i_j(\mes,\Delta E,{\fisher},\Delta t,\sigma_{\deltat};q_{\rm tag},c)},
\end{equation}
where $j$ stands for the event species (signal, continuum background, one for each $\B$ background category) and $N_j$ is the corresponding yield.

The PDF for the event species $j$ evaluated for event $i$ is given by the product of individual PDFs:
\begin{align}
&{\cal P}^i_j(\mes,\Delta E,{\fisher}, \Delta t,\sigma_{\deltat};q_{\rm tag},c) = \\
\nonumber & {\cal P}^i_j(\mes)\:{\cal P}^i_j(\Delta E)\:{\cal P}^i_j({\fisher})\:{\cal P}^i_j(\Delta t,\sigma_{\deltat};q_{\rm tag},c).
\end{align}

The total likelihood is given by
\begin{equation}
\label{eq:td_Likelihood}
 {\cal L}=\exp(-\sum_j N_j)\prod_i {\cal L}_i.
\end{equation}

Using isospin symmetry, we assume that the fraction and phase of each $K\pip\pim$ resonance channel in the $\Bz$ decay is the same as that in the $\Bp$ decay. 
Therefore, we model the PDFs for signal events with a mixture of exclusive samples from simulated events weighted according to the branching fractions extracted from the analysis of \MyControlChanel.

\subsubsection{$\Delta t$ PDFs}
\label{sec:TD_deltatPDF}

The signal PDF for $\Delta t$ is given in Eq.~\eqref{eq:dtmeas}.
The parameters of the resolution function, as well as ${\langle D \rangle}_c$, $\Delta D_c$ and $q_c$ are taken from the analysis of $\B \to c \bar c K^{(*)}$ decays~\cite{Aubert:2007hm}. 
The same resolution function parameters, ${\langle D \rangle}_c$ and $\Delta D_c$ are used for both correctly and mis-reconstructed signal events. 
The total yield of signal events (i.e. the sum of correctly and mis-reconstructed events) is a free parameter in the fit. 
Using simulated events, we assign a fraction of mis-reconstructed events to each tagging category and fix these fractions in the fit to the data.

For backgrounds from charged \B meson decays, the \deltat PDF is modeled as an exponential decay with an effective lifetime, $\tau_j$,
\begin{eqnarray}
\label{eq:dtBBp}
\lefteqn{{\cal P}^i_{\Bpm}(\Delta t,\sigma_{\deltat};q_{\rm tag},c) =}\\
\nonumber && \frac{e^{-|\deltat|/\tau_j}}{4 \tau_j} \times \Bigg[ \left( \frac{1 - q_{\rm tag} A_{j}}{2} \right) \omega^c \\
\nonumber && \qquad\qquad \quad \; \;\,  + \left( \frac{1 + q_{\rm tag} A_{j}}{2} \right) \left( 1-\omega^c \right) \\ 
\nonumber &&  \qquad\qquad \quad \;\, \Bigg] \otimes {\cal R}^c_{\Bpm}(\deltat, \sigma_{\deltat}), \nonumber
\end{eqnarray} 
where the index $j$ refers to the background event category, $A_{j}$ is the asymmetry accounting for possible differences between \Bz and \Bzb tags and $\omega^c$ is the mis-tag rate for tagging category $c$. 
For the background from neutral \B meson decays to flavor eigenstates (i.e. $\Bz \to K^{\pm} \pi^{\mp} \g$), a \deltat PDF similar to that for charged \B backgrounds is used, where mixing terms are added
\begin{eqnarray}
\label{equ:bzFlvLike}
 \lefteqn{{\cal P}_{\Bz_{\rm Flv}}(\Delta t,\sigma_{\deltat};q_{\rm tag},c) = }\\
\nonumber &&\frac{e^{-|\deltat|/\tau_j}}{4 \tau_j}\Bigg[ \left( \frac{1 - q_{\rm tag} A_{j}}{2} \right) \omega^c  \left(1 - \cos(\deltamd\deltat)\right)\\
\nonumber && \qquad\qquad \;\; + \left( \frac{1 + q_{\rm tag} A_{j}}{2} \right) \left( 1-\omega^c \right) \left(1 + \cos(\deltamd\deltat)\right) \\ 
\nonumber && \qquad\qquad\; \Bigg] \otimes {\cal R}^c_{\Bz_{\rm Flv}}(\deltat, \sigma_{\deltat}).
\end{eqnarray} 
For backgrounds from neutral \B meson decays to \CP eigenstates, we account for possible \CP violation effects using a similar \deltat PDF as for signal with an effective lifetime
\begin{eqnarray}
\label{equ:bzCPLike}
\lefteqn{{\cal P}^i_{\Bz_{\CP}}(\Delta t,\sigma_{\deltat};q_{\rm tag},c) =}\\
\nonumber & & \frac{e^{-\left|\deltat\right|/\tau_{j}}}{4\tau_{j}} \Bigg[ 1 + q_{\rm tag}\frac{\Delta D_{c}}{2} \\
\nonumber & & \qquad\qquad \;\;\; + q_{\rm tag}\langle D\rangle_{c} \Bigl({\cal S}\sin(\Delta m_d\deltat)-\;{\cal C}\cos(\Delta m_{d}\deltat)\Bigr)  \\
\nonumber & & \qquad\qquad \;\;\; \Bigg] \;\otimes\; {\cal R}^c_{\Bz_{\CP}}(\deltat,\sigma_{\deltat}).
\end{eqnarray}
Each \B background \deltat PDF is convolved with a similar resolution function as the signal one.

We describe the \deltat\ PDF for the continuum background as a combination of ``prompt'' decays and ``lifetime'' decays coming from charmed mesons
\begin{eqnarray}
\label{equ:bg_Deltat_PDF}
\lefteqn{\mathcal{P}_{{\rm bg}}(\deltat, \sigma_{\deltat}) = }\\
\nonumber & & \Bigg[f_{\rm p}\delta(\deltat' - \deltat) + (1 - f_{\rm p})\exp\left(-\frac{|\deltat|}{\tau_{{\rm bg}}}\right) \Bigg] \otimes \mathcal{R}_{{\rm bg}},
\end{eqnarray}
where $f_{\rm p}$ corresponds to the fraction of prompt events and $\tau_{{\rm bg}}$ corresponds to an effective lifetime. 
The resolution function, $\mathcal{R}_{{\rm bg}}$, is defined as the sum of a ``core'' and an ``outlier'' Gaussian function. 
The outlier Gaussian function has the bias fixed to $b_{\rm out} = 0$, while the width and the bias of the core Gaussian function are scaled by the event-by-event uncertainty on \deltat.
The small contribution from $\epem \to \ccbar$ events is well described by the tails of the resolution function. 

All the continuum background $\Delta t$ PDF parameters, except for $b_{\rm out}$, are extracted from a fit to the off-resonance data sample.
All ${\langle D \rangle}_c$ and $\Delta D_c$ values, tagging category fractions and asymmetries, and all the $\sigma_{\Delta t}$ parameters are fixed in the fit to the data. 
All resolution function parameters are fixed in the fit except for that of the continuum background for which the mean and width of the core Gaussian function as well as the width and the fraction of the outlier Gaussian function are free parameters in the fit.
Furthermore, the \S and \C parameters for signal are left free in the fit, while those for the \CP-eigenstate neutral \B backgrounds are fixed to zero. 

\subsubsection{Description of the other variables}
\label{sec:td_likeDiscrim}

The \mes\ distribution of CR signal events is parametrized by the CB function defined in Eq.~\eqref{eq:CB_function}. The \DeltaE\ distribution of CR signal events is parametrized by a modified Gaussian defined in Eq.~\eqref{eq:cruijff}.
The $\sigma_l$ and $\sigma_r$ parameters are free in the fit to the data, while the other parameters are fixed to values determined from simulated events.
Correlations between \mes and \DeltaE in CR signal events are taken into account through a two-dimensional conditional PDF identical to the one used in the analysis of \MyControlChanel. 
The dependences of the CB parameters $\mu$ and $\sigma$ on \DeltaE are parametrized by two second-order polynomials for which all the parameters are left free in the fit to the data, while the dependences of $\alpha$ and $n$ are parametrized by first- and second-order polynomials, respectively, for which all the parameters are fixed to values determined from fits performed to simulated events.

The $\fisher$ distribution of CR signal events is parametrized by a Gaussian function for which the mean is left free in the fit to the data.
No significant correlations were found between $\fisher$ and either \mes or \DeltaE.

All mis-reconstructed signal PDF shape parameters are fixed to values determined from simulated events.
The \mes PDF of mis-reconstructed signal events is parametrized by the sum of a first-order Chebychev polynomial and an ARGUS shape function. The \DeltaE PDF is parametrized by a fourth-order polynomial and $\fisher$ PDF is parametrized by the sum of a Gaussian function and an exponential. 

The \mes, \DeltaE and $\fisher$ PDFs for continuum events are parametrized by an ARGUS shape function, a second-order Chebychev polynomial and an exponential function, respectively. 
The parameters of the second-order Chebychev polynomial are left free in the fit to the data. All the other shape parameters are fixed to the values determined from a fit to the off-resonance data.

The \mes, \DeltaE and $\fisher$ PDFs for all the categories of $B$-background events, given in Table~\ref{tab:TD_bbackground}, are described by parametric functions, except for the $\Bp \to \KS \pip \g$ $B$ background \mes and \DeltaE PDFs, for which significant correlations are present. 
These correlations are taken into account through a nonparametric two-dimensional PDF, defined as a histogram constructed from a mixture of $\TDBbkgBa$ and $\TDBbkgBb$ simulated events. 
All shape parameters of the $B$-background PDFs are fixed to values determined from simulation.

No significant correlations were found among the fit variables for the other event species in the fit.

\subsubsection{Branching fraction determination}
\label{sec:td_BFdet}

The branching fraction to the $\KS \pip \pim \g$ final state is determined from the fitted yield of the correctly-reconstructed signal event category, $N_{\rm sig}^{\rm CR} = N_{\rm sig}\times {\rm f}^{\rm CR}$, the weighted signal efficiency, $\langle{\epsilon}^0\rangle$, and the number of neutral \B events, $N_{\Bz}$ 
\begin{equation}
\BR(\Bz \to \KS \pip \pim \g) = \frac{N_{\rm sig}^{\rm CR}}{\langle{\epsilon}^0\rangle \times N_{\Bz}}.
\label{equ:CC_BR_tot_formula}
\end{equation}
where $\langle{\epsilon}^0\rangle = 0.0553 ^{+ 0.0010}_{-0.0009}$ is obtained from Eq.~\eqref{equ:weighted_effi_formula} replacing the efficiencies $\epsilon^+_{k}$ by those of the neutral kaonic resonances listed in Table~\ref{tab:TD_Cut_Efficiency_By_Mode} and, assuming isospin symmetry, using the FFs listed in Table~\ref{tab:AMPmKpipi}. 
The small value of $\langle{\epsilon}^0\rangle$ compared to that of $\langle{\epsilon}^+\rangle$ is due to the additional requirements on \mpipi and \mKpi (see Sec.~\ref{sec:td_selection}).
The term ${\rm f}^{\rm CR} = 0.728 \pm 0.004$ is the fraction of correctly-reconstructed signal events.
The term $N_{\Bz}$ is obtained from the total number of $\BB$ pairs in the full \babar\ dataset, $N_{\BB}$, and the corresponding $\Y4S$ branching fraction taken from Ref.~\cite{Agashe:2014kda}
\begin{eqnarray}
\label{eq:n_BBbar}
N_{\Bz} & = & 2\times N_{\BB} \times \BR(\Y4S \to \BzBzb) \\
& =& (458.7 \pm 6.3) \times 10^6. \nonumber
\end{eqnarray}

\begin{table}[htbp]
\caption{\label{tab:TD_Cut_Efficiency_By_Mode} 
Efficiencies $\epsilon^0_k$ for correctly-reconstructed signal candidates for each kaonic resonance from simulations without the applied requirement $\mKpipi< 1.8\gevcc$. 
The efficiencies in the neutral mode are significantly smaller to the ones in the charged mode (see Table~\ref{tab:CC_Cut_Efficiency_By_Mode}) due to the additional requirements on \mpipi and \mKpi. The difference between the $\epsilon^0$ values is due to the difference in branching fractions of each kaonic resonance to the $K^*(892)^+\pim$ and $\KS\rho(770)^0$ decay modes. 
}
\vspace{10pt}
\begin{tabular*}{\columnwidth}{@{\extracolsep{\fill}}l c}
\hline
\hline
\multirow{2}{*}{\Kres}  &	\multirow{2}{*}{$\epsilon^0_k$} \\
&\\
\hline
$\Kl^0$			& $0.0631 \pm 0.0003$ \\
$\Kll^0$		& $0.0335 \pm 0.0003$ \\
$\KstarX^0$ 	& $0.0318 \pm 0.0005$ \\
$\Kstarlllmy^0$	& $0.0471 \pm 0.0002$ \\
$\KstarlVmy^0$ 	& $0.0742 \pm 0.0004$ \\
\hline
\hline
\end{tabular*}
\end{table}
%

\subsection{Results}
\label{sec:TD_fitresults}

\begin{figure*}[t!]
\begin{tabular}{cc}
	\includegraphics[height= 5.6cm] {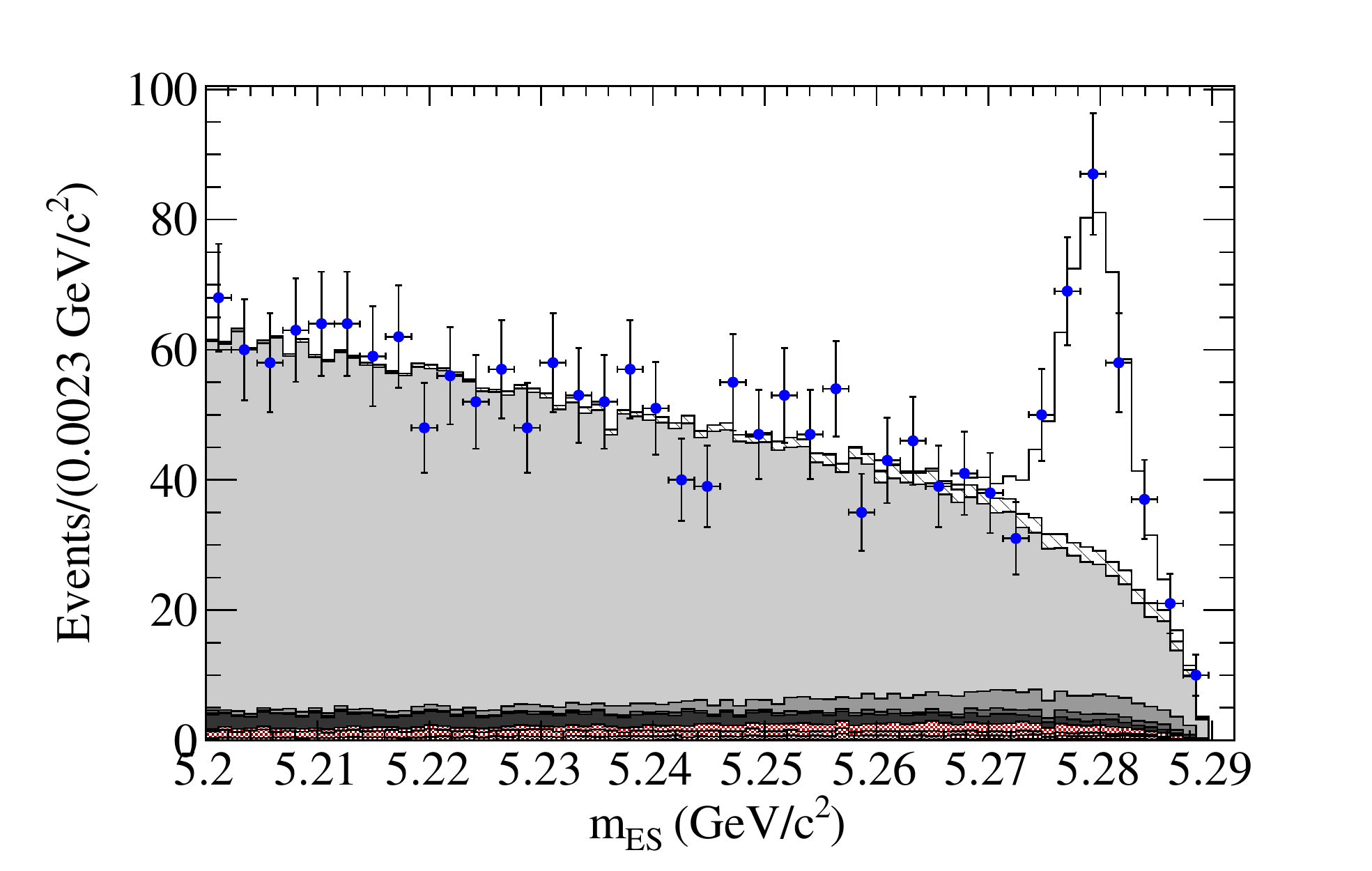}
	&
	\includegraphics[height= 5.6cm] {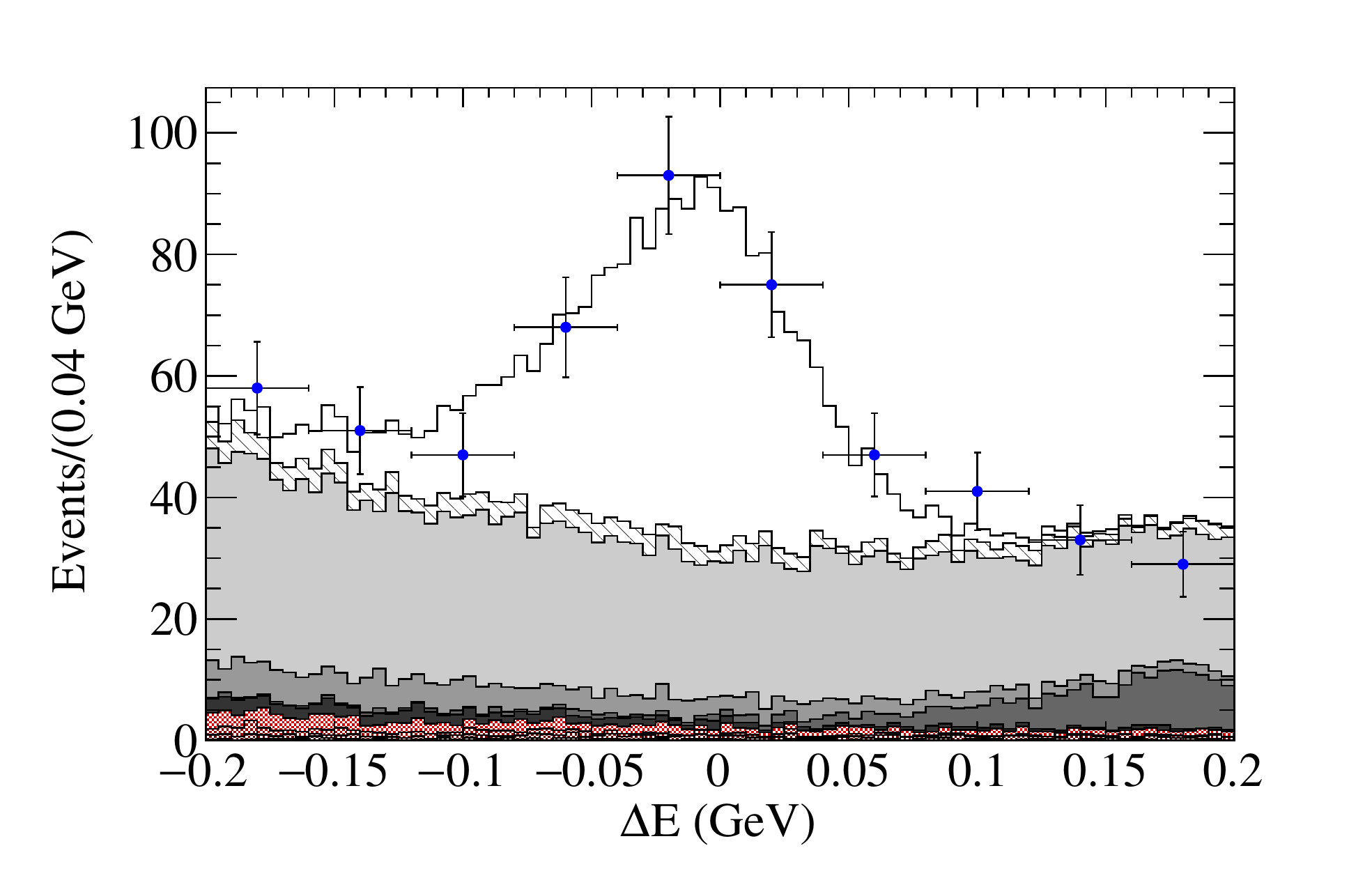}\\
\vspace{-10pt}
\includegraphics[height= 5.6cm] {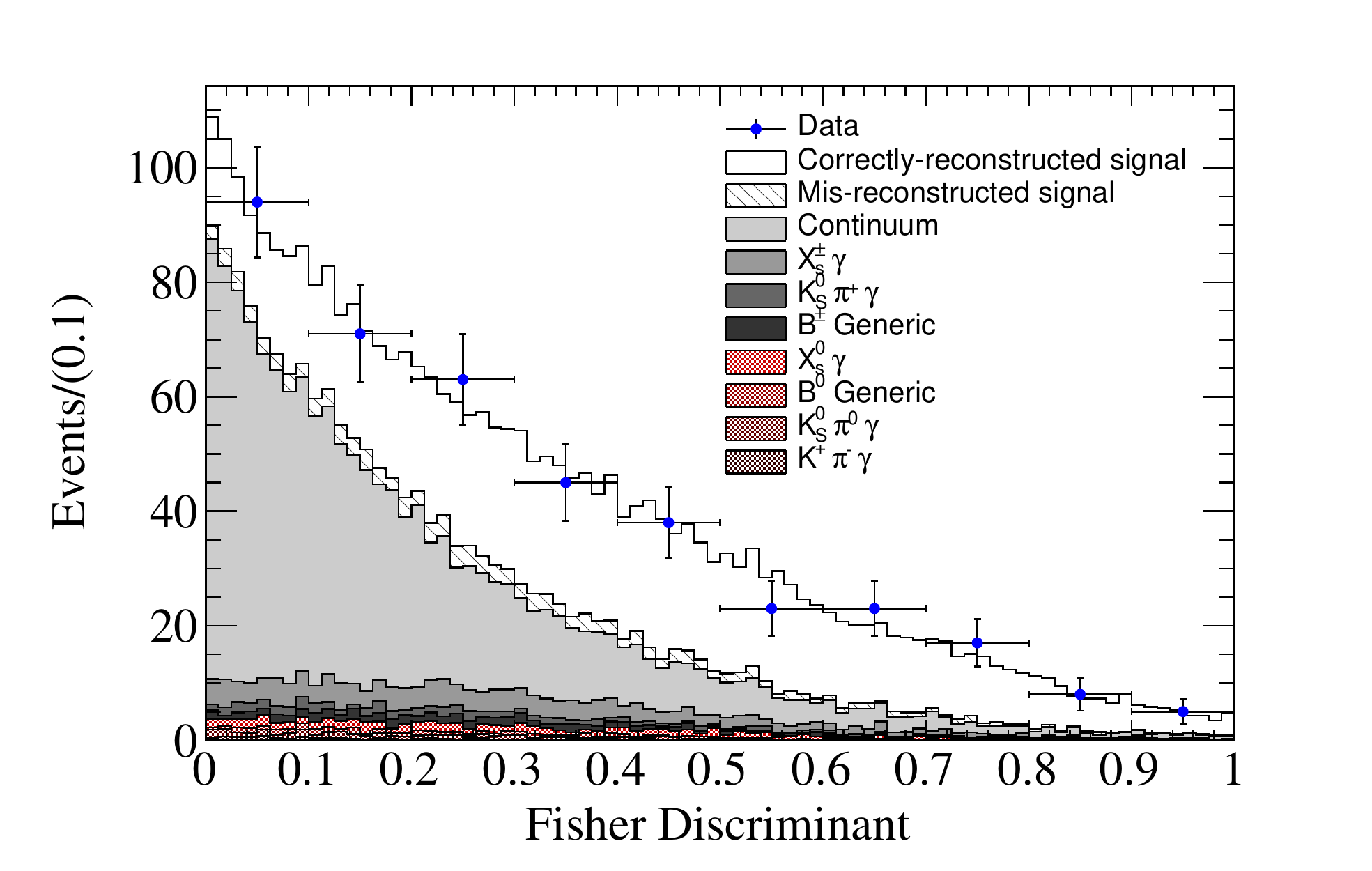}	
&
	\includegraphics[height= 5.6cm] {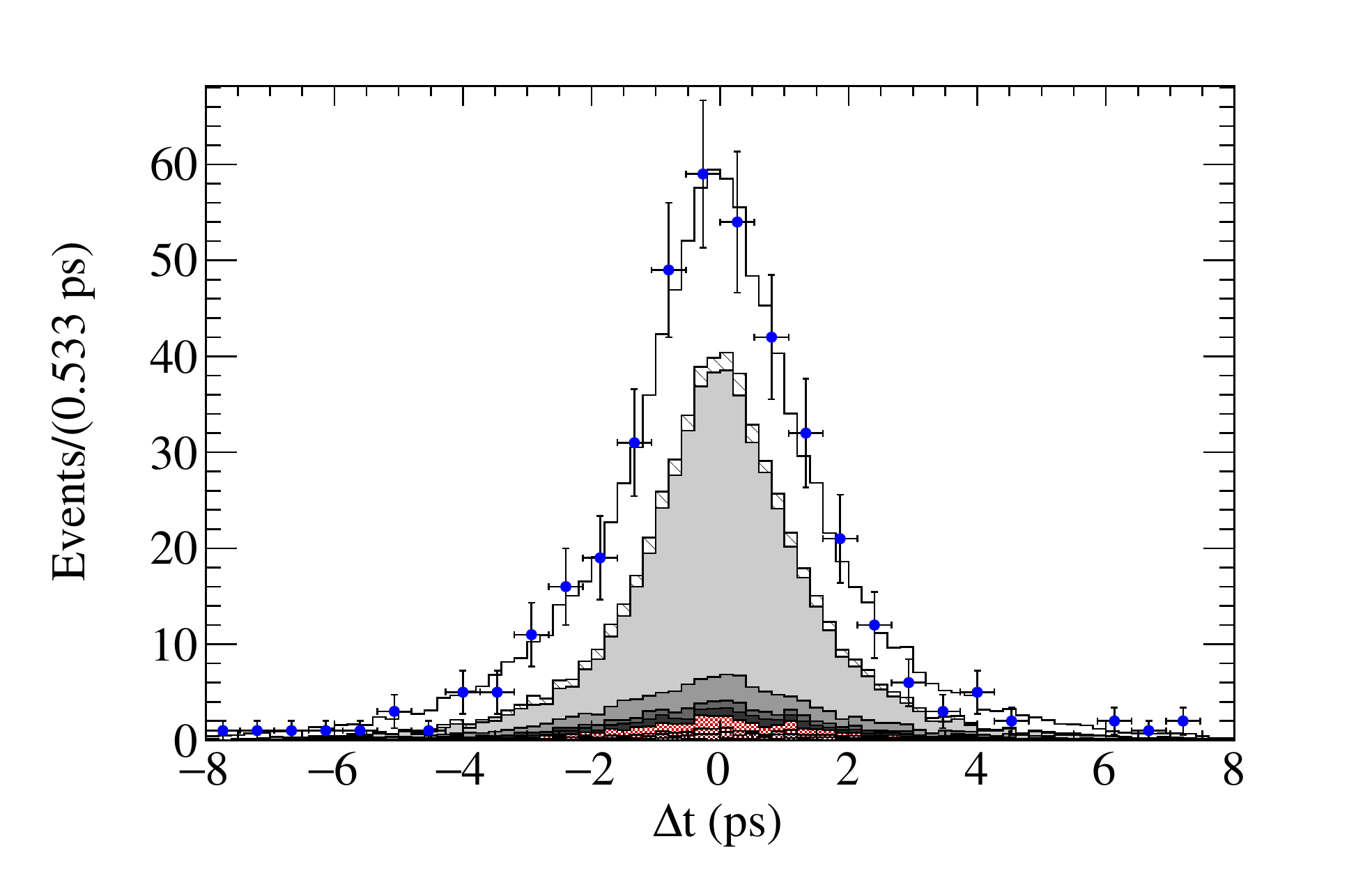}\\
\end{tabular}
\caption{
Distributions of \mes (top left), \DeltaE (top right), the Fisher Discriminant (bottom left), and \deltat (bottom right), showing the results of the fit to the \MyChanelPiPi data sample. 
The distributions have their signal/background ratio enhanced
by means of the following requirements:
$ -0.15 \leq \DeltaE \leq 0.10 \gev$ (\mes); $\mes > 5.27 \gevcc$ (\DeltaE); $\mes > 5.27 \gevcc \, , \, -0.15 \leq \DeltaE \leq 0.10 \gev $ (Fisher and \deltat).
Points with error bars show the data. 
The projection of the fit result is represented by stacked histograms, where the shaded areas represent the background contributions, as described in the legend. 
Some of the contributions are hardly visible due to their small fractions. 
Note that the same order is used for the various contributions in both the stacked histograms and the corresponding legend.}
\label{fig:FitProjNeutral}
\end{figure*}

Requiring $m_{\KS\pi\pi} \leq 1.8\gevcc$, the unbinned maximum-likelihood fit to the data for the \break \MyChanelPiPi decay mode yields $N_{\rm sig} = 243 \pm 24^{+21}_{-17}$ events and in turn a branching fraction of 
\begin{equation}
\BR(\Bz \to \Kz \pip \pim \g) = (20.5	\pm	2.0	^{+	2.6	}_{-	2.2	})\times 10^{-6}, 
\end{equation}
where, the first uncertainty is statistical and the second is systematic.
This result is in good agreement with the previous world average~\cite{Agashe:2014kda}.
The same convention holds for results in Eqs.~\eqref{eq:SeffRes} to~\eqref{eq:SrhoRes}.
The systematic uncertainties are discussed in detail in Sec.~\ref{sec:TD_BRSyst}. 
To check the presence of biases on the parameters of interest, 351 pseudo experiments were generated with embedded signal events drawn from fully simulated MC samples and analyzed.
No significant biases were found.
Figure~\ref{fig:FitProjNeutral} shows signal-enhanced distributions of the four discriminating variables in the fit: \DeltaE, \mes, $\fisher$, and \deltat.
The result of the fit to the data for the time-dependent \CP violation parameters in signal events is
\begin{eqnarray}
\label{eq:SeffRes}
\Seff & =& \phantom{-} 0.14 \pm 0.25 \pm 0.03, \\
\label{eq:CeffRes}
\Ceff & =&  -0.39 \pm 0.20^{+0.03}_{-0.02}. 
\end{eqnarray}
To obtain the value of \Srho, we divide \Seff by the dilution factor given in Eq.~\eqref{eq:dilutionFactor} and obtain
\begin{equation}
\label{eq:SrhoRes}
 \Srho = -0.18 \pm 0.32^{+ 0.06}_{-0.05}.
\end{equation}

Table~\ref{tab:TD_correlationMatrix} shows the correlation matrix for the statistical
uncertainty obtained from the fit to the data.

%

\subsection{Systematic uncertainties}
\label{sec:td_systematics}

\subsubsection{\CP asymmetry parameters}
\label{sec:TD_CP_systematics}

\begin{table}[b!]
\begin{center}
\caption[Systematic uncertainties on the time-dependent \CP-asymmetry parameters]{\label{tab:TD_syst_S_and_C_mESDeltaEFisher} Systematic uncertainties on the time-dependent \CP-asymmetry parameters resulting from the fixed parameters in the fit to \mes, \DeltaE, $\fisher$ and \deltat.}
\vspace{5pt}
\begin{tabular*}{\columnwidth}{@{\extracolsep{\fill}} l c c}
\hline
\hline
Parameter	&	$+$ signed deviation	&	$-$ signed deviation	\\
\hline					
\Seff	&	0.025	&	0.027	\\
\Ceff	&	0.027	&	0.022	\\
\hline					
\hline					
\end{tabular*}
\end{center}
\end{table}

In order to assign systematic uncertainties due to the fixed parameters in the fit to \mes, \DeltaE, $\fisher$ and \deltat, we vary each of the fixed parameters within its uncertainty, which are taken from different sources that are detailed below, and re-perform the fit.
The fixed shape parameters of \mes, \DeltaE and $\fisher$ PDFs are varied according to the uncertainties obtained in the fit to the simulated event samples from which they are extracted. 
Since the \mes-\DeltaE distribution of \TDBbkgB background events is described by a two-dimensional histogram, we fluctuate the bin contents using the same procedure as described in Sec.~\ref{sec:CC_systematics}. 
The fixed yields are varied according to the corresponding branching fraction uncertainties taken from Ref.~\cite{Agashe:2014kda}. 
For the categories describing a sum of modes, the fraction of each mode is varied according to the relative branching fraction uncertainties taken from Ref.~\cite{Agashe:2014kda}. 
The mis-reconstructed signal fractions are varied according to the uncertainties due to the sample size of the simulated events and the signal branching fraction uncertainties in Ref.~\cite{Agashe:2014kda}. 
The fixed yields of \BzBzb and \BpBm generic \B backgrounds, describing a sum of several small contributions from various \B-background modes, are varied according to the uncertainties due to the sample size of the simulated events. 
The fixed parameters of the \deltat PDFs are varied according to the uncertainties that are either taken from other \babar\ measurements or are extracted from simulated event distributions. 
Using the method described in Ref.~\cite{Barlow:2003sg} and assuming no correlations among the fixed parameters, we combine each of the negative (positive) difference between the new fit value and nominal fit value of each of the time-dependent \CP-asymmetry parameters, and take the resulting values as negatively (positively) signed uncertainties.
The corresponding values are given in Table~\ref{tab:TD_syst_S_and_C_mESDeltaEFisher}. 
Note that these uncertainties are small compared to the statistical uncertainties.

\subsubsection{Branching fraction}
\label{sec:TD_BRSyst}

We take the same sources of systematic uncertainties as described in Sec.~\ref{sec:CC_BRSyst} when applicable. 
A few sources, which are described below, differ from the analysis of \MyControlChanel decays.

From the procedure described in Sec.~\ref{sec:TD_CP_systematics}, and assuming no correlations among the fixed parameters, we combine each of the negative (positive) difference between the new fit value and nominal fit value of each of the total signal yield, and take the resulting values as negatively (positively) signed uncertainties.

Using Eq.~\eqref{eq:n_BBbar}, we compute the number of \BzBzb pairs using as input the branching fraction: $ \BR(\Y4S \to \BzBzb) = 0.487 \pm 0.006 $ taken from Ref.~\cite{Agashe:2014kda}.
The branching fraction $\BR(\Kz \to \KS \to \pip \pim)$ is well measured~\cite{Agashe:2014kda} and we assign no systematic uncertainty due to this input.
We apply a systematic uncertainty of 0.7\% due to the $\KS$ reconstruction efficiency, as estimated using simulated events.

\section{SUMMARY}
\label{sec:Summary}

We have presented a measurement of the time-dependent \CP asymmetry in the radiative-penguin decay $\Bz \to \KS \pip \pim \gamma$, using a sample of $470.9$ million $\FourS \to \BB$ events recorded with the \babar\ detector at the \pep2 \epem storage ring at SLAC. 
Using events with $m_{K\pi\pi}<1.8\gevcc$, $0.6<m_{\pi\pi}<0.9\gevcc$ and with $m_{K\pi}<0.845\gevcc$ or $m_{K\pi}>0.945\gevcc$, we obtain the \CP-violating parameters $\Seff = 0.14 \pm 0.25 \pm 0.03$ and $\Ceff= -0.39 \pm 0.20^{+0.03}_{-0.02}$, where the first uncertainties are statistical and the second are systematic.
From this measurement, assuming isospin symmetry, we extract the time-dependent \CP asymmetry related to the \MyChanel decay and obtain $ \Srho = -0.18 \pm 0.32^{+ 0.06}_{-0.05}$.
This measurement of time-dependent asymmetries in radiative \B decays is in agreement with previously published results~\cite{Li:2008qma,Aubert:2008gy,Ushiroda:2006fi} and is of equivalent precision.
In this statistics-limited measurement, no deviation from the SM prediction is observed.

We have studied the decay \MyControlChanel to measure the intermediate resonant amplitudes of resonances decaying to $K\pi\pi$ through the intermediate states $\rho^0\Kp$, $\Kstarz\pip$ and $\swave\pip$. 
Assuming isospin symmetry, these results are used to extract $\Srho$ from $\Seff$ in the neutral decay \MyChanel. 
In addition to the time-dependent \CP asymmetry, we gain information on the $K \pi \pi$ system which may be useful for other studies of the photon polarization. 
We have measured the branching fractions of the different $\Kres \to K\pi\pi$ states and the overall branching fractions of the $\rho^0\Kp$, $\Kstarz\pip$ and $\swave\pip$ components, listed in Tables~\ref{tab:BFmKpipi} and~\ref{tab:BFmKpi}, respectively.
\section*{ACKNOWLEDGMENTS}
\label{sec:acknowledgments}

We thank E. Kou for help throughout the analysis, especially with the computation of the dilution factor, and A. Le Yaouanc for valuable discussions.
We are grateful for the 
extraordinary contributions of our \pep2\ colleagues in
achieving the excellent luminosity and machine conditions
that have made this work possible.
The success of this project also relies critically on the 
expertise and dedication of the computing organizations that 
support \babar.
The collaborating institutions wish to thank 
SLAC for its support and the kind hospitality extended to them. 
This work is supported by the
US Department of Energy
and National Science Foundation, the
Natural Sciences and Engineering Research Council (Canada),
the Commissariat \`a l'Energie Atomique and
Institut National de Physique Nucl\'eaire et de Physique des Particules
(France), the
Bundesministerium f\"ur Bildung und Forschung and
Deutsche Forschungsgemeinschaft
(Germany), the
Istituto Nazionale di Fisica Nucleare (Italy),
the Foundation for Fundamental Research on Matter (The Netherlands),
the Research Council of Norway, the
Ministry of Education and Science of the Russian Federation, 
Ministerio de Ciencia e Innovaci\'on (Spain), and the
Science and Technology Facilities Council (United Kingdom).
Individuals have received support from 
the Marie-Curie IEF program (European Union), the A. P. Sloan Foundation (USA) 
and the Binational Science Foundation (USA-Israel).

\section*{APPENDIX}
\subsection{Extraction of the dilution factor}\label{app:dilutionFactor_calculation}

Using the hypothesis of isospin conservation, we assume that \Bz decays have the same amplitudes as \Bp decays. 
This allows to use the results extracted from the fit to the \mKpi spectrum in $B^+ \to K^+ \pi^+\pi^-\gamma$ decays from the measured amplitudes to obtain the dilution factor for the time dependent analysis.

In the analysis of the \Bp decay, the amplitude of a resonance is modeled in $m_{12}$ as
 \begin{equation}
 F_{\rm res} = c_{\rm res} \sqrt{{\rm H_{res}}(m_{12},m_{23})}e^{i\Phi(m_{12})},
 \end{equation}
where $c_{\rm res}$ is a complex constant, and ${\rm H_{res}}$ is a real distribution, $\sqrt{{\rm H_{res}}(m_{12},m_{23})}e^{i\Phi(m_{12})}$ being the lineshape.
Note that here $m_{12}=\mKpi$ and $m_{23}=\mpipi$.
The total event rate (given here without the $(K\pi)$ S-wave for simplicity) is written as
\begin{equation}
|F|^2 = |F_{\rho} + F_{\Kstar}|^2. 
\end{equation}
In the analysis, we consider the total event rate from \Bp and \Bm in the \mKpi-\mpipi plane. If the charge specific amplitudes are denoted as $F^+_{\rm res}$ and $F^-_{\rm res}$, this implies the underlying assumption
\begin{equation}
|F_{\rho} + F_{\Kstar}|^2 = |F^+_{\rho} + F^+_{\Kstar}|^2 + |F^-_{\rho} + F^-_{\Kstar}|^2,
\end{equation}
or 
\begin{eqnarray}
\label{eq:Bp_Bm_sym}
       \lefteqn{ |F_{\rho}|^2 + |F_{\Kstar}|^2 + 2\Re(F_{\rho}F^{*}_{\Kstar})  =} \\
&& |F^+_{\rho}|^2 + |F^-_{\rho}|^2 + |F^+_{\Kstarp}|^2 + |F^-_{\Kstarm}|^2 \nonumber \\
& & + 2\Re(F^+_{\rho}F^{+*}_{\Kstarp}) + 2\Re(F^-_{\rho}F^{-*}_{\Kstarm}). \nonumber
\end{eqnarray}
Assuming no direct \CP violation in the considered transition:
 \begin{eqnarray}
  F_{\rho} & = &\sqrt{2}F_{\rho}^+=\sqrt{2}F_{\rho}^-,\\
\label{eq:deltaRescat}
  F_{\Kstar} & = & e^{i\delta_{\rm rescat.}}\sqrt{2} F_{\Kstarp}^+ = \sqrt{2} F_{\Kstarm}^-,
 \end{eqnarray}
with $\delta =\delta_{\rm rescat.} = 0$ or $\pi$. Given that we measure a sizable interference between the $\rho$ and the $\Kstar$ (see Table~\ref{tab:mKpi_fit_results}), we keep $\delta_{\rm rescat.} = 0$. Indeed, $\delta_{\rm rescat.} = \pi$ would result in zero interference, as can be deduced from Eqs.~\eqref{eq:Bp_Bm_sym} and~\eqref{eq:deltaRescat}. Identical expressions are obtained for the $(K\pi)$ S-wave terms.\\
Using these conventions, the term $|A_{\rho\KS}|^2$ in Eq.~\eqref{eq:dilutionExpression} can be expressed as
\begin{equation}
|A_{\rho\KS}|^2 = \frac{|F^+_{\rho}|^2 + |F^-_{\rho}|^2}{2} = \frac{|F_{\rho}|^2}{2},
\end{equation}
whose contribution to the dilution factor is
\begin{eqnarray}
&& 	 \frac{1}{2} \int |F_{\rho}|^2  \\
&& = \frac{1}{2} |c_{\rho}|^2 \int_{m_{12}} \int_{m_{23}} |{\rm H_{\rho}}(m_{12},m_{23})|^2 dm_{12} dm_{23} \nonumber \\
&& = \frac{1}{2} {\rm FF_{\rho}},\nonumber
\end{eqnarray}
where ${\rm FF}_{\rho}$ is the measured fit fraction of the $\rho$ resonance in the considered \mKpi-\mpipi domain.\\ 
The term $|A_{\Kstarp\pim}|^2$ is expressed as
\begin{equation}
 \frac{|F^+_{\Kstarp}|^2 + |F^-_{\Kstarm}|^2}{2} = \frac{|F_{\Kstar}|^2}{2},
\end{equation}
and its contribution to the dilution factor is
\begin{eqnarray}
&&   \frac{1}{2} \int |F_{\Kstar}|^2 \\
&& = \frac{1}{2} |c_{\Kstar}|^2 \int_{m_{12}} \int_{m_{23}} |{\rm H_{\Kstar}}(m_{12},m_{23})|^2 dm_{12} dm_{23}\nonumber \\
&& = \frac{1}{2} {\rm FF_{\Kstar}},\nonumber
\end{eqnarray}
where ${\rm FF}_{\Kstar}$ is the measured fit fraction of the $\Kstar$ resonance in the considered \mKpi-\mpipi domain.\\
Analogously, the term $|A_{\swavep\pim}|^2$ is expressed as
\begin{equation}
 \frac{|F^+_{\swavep}|^2 + |F^-_{\swavem}|^2}{2} = \frac{|F_{\swave}|^2}{2},
\end{equation}
and its contribution to the dilution factor is
\begin{eqnarray}
&& \frac{1}{2} \int |F_{\swave}|^2  \\
&& = \frac{1}{2} |c_{\swave}|^2 \int_{m_{12}} \int_{m_{23}} |{\rm H}_{\swave}(m_{12},m_{23})|^2 dm_{12} dm_{23} \nonumber \\
&& = \frac{1}{2} {\rm FF}_{\swave},\nonumber
\end{eqnarray}
where ${\rm FF}_{\swave}$ is the measured fit fraction of the $(K\pi)$ S-wave component in the considered \mKpi-\mpipi domain.\\
The term 2$\Re(A^*_{\rho\KS} A_{\Kstarp\pim})$ is expressed as
\begin{eqnarray}
\label{eq:contrib_interf_rho_Kstar}
 \lefteqn{ \Re(F^{+*}_{\rho} F^+_{\Kstarp}) + \Re(F^{-*}_{\rho} F^-_{\Kstarm}) }  \\
&& = 2\Re\left(\frac{1}{\sqrt{2}} F^{*}_{\rho} \frac{1}{\sqrt{2}} 	F_{\Kstar}\right) \nonumber \\
&& =  \Re\left( F^{*}_{\rho} F_{\Kstar}\right)\nonumber \\
&& = \Re\bigg( c^*_{\rho} c_{\Kstar} \sqrt{{\rm H_{\rho}}(m_{12},m_{23}){\rm H_{\Kstar}}(m_{12},m_{23})} \nonumber \\ 
&& \quad\quad \times e^{i(\Phi_{K^{*}}(m_{12}) - \Phi_{\rho}(m_{23}))} \bigg). \nonumber
\end{eqnarray}
With the notation $c_{\rm res} = \alpha_{\rm res}e^{i\phi_{\rm res}}$, the contribution of the terms in Eq.~\eqref{eq:contrib_interf_rho_Kstar} to the dilution factor is given by Eq.~\eqref{eq:dilFactor_InterfKstarRho}, where $\mathrm{FF}^{\mathrm{interf.}}_{K^*-\rho}$ is the measured fit fraction of the interference between the $\Kstar$ and the $\rho$ resonances in the considered \mKpi-\mpipi domain, with the convention $\alpha_{\Kstar} = 1$ and $\phi_{\Kstar} = 0$.
Analogously, the term 2$\Re(A^*_{\rho\KS} A_{\swavep\pim})$ is expressed as
\begin{eqnarray}
\label{eq:contrib_interf_rho_swave}
 \lefteqn{ \Re(F^{+*}_{\rho} F^+_{\swavep}) + \Re(F^{-*}_{\rho} F^-_{\swavem}) }  \\
&& = 2\Re\left(\frac{1}{\sqrt{2}} F^{*}_{\rho} \frac{1}{\sqrt{2}} 	F_{\swave}\right) \nonumber \\
&& =  \Re\left( F^{*}_{\rho} F_{\swave}\right)\nonumber \\
&& = \Re\bigg( c^*_{\rho} c_{\swave} \sqrt{{\rm H_{\rho}}(m_{12},m_{23}){\rm H_{\swave}}(m_{12},m_{23})} \nonumber \\ 
&& \quad\quad \times e^{i(\Phi_{\swave}(m_{12}) - \Phi_{\rho}(m_{23}))} \bigg), \nonumber
\end{eqnarray}
whose contribution to the dilution factor is given by Eq.~\eqref{eq:dilFactor_InterfSwaveRho}, where $\mathrm{FF}^{\mathrm{interf.}}_{\swave-\rho}$ is the measured fit fraction of the interference between the $\swave$ and the $\rho$ resonances in the considered \mKpi-\mpipi domain.
\begin{widetext}
\begin{eqnarray}
&&\alpha_{\rho}\alpha_{\Kstar}\int_{m_{12}} dm_{12} \int_{m_{23}} dm_{23} \sqrt{{\rm H_{\rho}}(m_{12},m_{23}){\rm H_{\Kstar}}(m_{12},m_{23})} \cos\left(\phi_{\rho} - \phi_{\Kstar} + \Phi_{\rho}(m_{23}) - \Phi_{K^{*}}(m_{12}) \right) \nonumber \\
&& = \alpha_{\rho} \Bigg[ \int_{m_{12}} dm_{12} \cos\left(\phi_{\rho} - \label{eq:dilFactor_InterfKstarRho}
\Phi_{K^{*}}(m_{12}) \right) \int_{m_{23}} dm_{23} \sqrt{{\rm H_{\rho}}(m_{12},m_{23}){\rm H_{\Kstar}}(m_{12},m_{23})} \cos\left(\Phi_{\rho}(m_{23}) \right)  \\
&& \phantom{=} \phantom{\alpha_{\rho} \Bigg[} -  \int_{m_{12}} dm_{12} \sin\left(\phi_{\rho} - \Phi_{K^{*}}(m_{12}) \right) \int_{m_{23}} dm_{23} \sqrt{{\rm H_{\rho}}(m_{12},m_{23}){\rm H_{\Kstar}}(m_{12},m_{23})} \sin\left(\Phi_{\rho}(m_{23}) \right) \bigg] \nonumber \\
&& = \frac{1}{2} \mathrm{FF}^{\mathrm{interf.}}_{K^*-\rho}.\nonumber
\end{eqnarray}
\begin{eqnarray}
&&\alpha_{\rho}\alpha_{\swave}\int_{m_{12}} dm_{12} \int_{m_{23}} dm_{23} \sqrt{{\rm H_{\rho}}(m_{12},m_{23}){\rm H_{\swave}}(m_{12},m_{23})} \cos\left(\phi_{\rho} - \phi_{\swave} + \Phi_{\rho}(m_{23}) - \Phi_{K^{*}}(m_{12}) \right) \nonumber \\
&& = \alpha_{\rho} \Bigg[ \int_{m_{12}} dm_{12} \cos\left(\phi_{\rho} - \label{eq:dilFactor_InterfSwaveRho}
\Phi_{\swave}(m_{12}) \right) \int_{m_{23}} dm_{23} \sqrt{{\rm H_{\rho}}(m_{12},m_{23}){\rm H_{\swave}}(m_{12},m_{23})} \cos\left(\Phi_{\rho}(m_{23}) \right)  \\
&& \phantom{=} \phantom{\alpha_{\rho} \Bigg[} -  \int_{m_{12}} dm_{12} \sin\left(\phi_{\rho} - \Phi_{\swave}(m_{12}) \right) \int_{m_{23}} dm_{23} \sqrt{{\rm H_{\rho}}(m_{12},m_{23}){\rm H_{\swave}}(m_{12},m_{23})} \sin\left(\Phi_{\rho}(m_{23}) \right) \bigg] \nonumber \\
&& = \frac{1}{2} \mathrm{FF}^{\mathrm{interf.}}_{\swave-\rho}.\nonumber
\end{eqnarray}
\end{widetext}
%

\subsection{\splot technique}\label{app:splot}

The \splot technique corresponds to a background subtracting method. It takes place in the context of a unbinned extended maximum-likelihood fit, making use of the discriminating variables denoted $y$. 
The aim of \splot technique is to unfold the true distribution, $\mathrm{M}_\mathrm{n}(x)$, of a variable $x$, whose distributions are unknown for signal events.
An estimate of the $x$-distribution, denoted $_s\tilde{\mathrm{M}}_\mathrm{n}$, can be defined as the sum of the \sweights in each bin, as described in Ref.~\cite{Pivk:2004ty}.
If one or more event categories have their yields fixed in the maximum-likelihood fit, we need to apply a correction to reproduce a good estimate of the $x$-distribution.
This correction consists of adding to the $_s\tilde{\mathrm{M}}_\mathrm{n}$ histogram the normalized distributions of each fixed category scaled by the factor $c_{\rm n} = N_{\rm n} - \sum_j {\rm V}_{{\rm n}j}$, where V is the covariance matrix resulting from the fit and $N$ the expected yield of category n. 
This procedure, which is used in the present analysis to extract the CR signal \splot, implies that the $x$-distributions of the fixed categories are well known. 
The \mKpipi distributions of the event categories with fixed yields cannot be considered to completely fulfill this criterion since they are taken from simulation. 
Therefore, we perform a new fit to \mes, \DeltaE and $\fisher$, with all the previously fixed-yield categories merged to a single one to check for possible effects on the parameters of the fit to the \mKpipi and \mKpi spectra.
Since the shape of PDFs for the generic \B background and that of the merged category are very similar, we add the former to the latter and consider them as a single ``large background'' category.
This way, we can perform a fit with four event categories (i.e. CR signal, continuum, \BbkgA and this new large background) where all the yields are left free in the fit. 
We observe good agreement between the fitted yields in the present and the nominal fit configurations.
Thus, we extract the CR signal \splot distributions, where no corrections need to be applied since no event category yield is fixed in this configuration. 
We perform a fit to the new \mKpipi (\mKpi) \splot distributions, using the nominal \mKpipi (\mKpi) fit model, and take the deviation from the nominal value of each free parameter as the corresponding signed uncertainty.

\begin{widetext}

\begin{sidewaystable}[p!]
\begin{center}
\caption[Systematic uncertainties on the time-dependent \CP-asymmetry parameters]{\label{tab:TD_correlationMatrix} Statistical correlation matrix for the parameters from the fit to the \MyChanelPiPi data. The entries are given in percent. When $0.0$ is quoted, the corresponding value is less than $ 0.05 \%$. Since the matrix is symmetric, all elements above the diagonal are omitted.
The notations ``$\tilde{\mathrm{G}}$'', ``CB'' and ``Ch.'' correspond to the modified Gaussian function given in Eq.~\eqref{eq:cruijff}, the Crystal Ball function given in Eq.~\eqref{eq:CB_function} and the Chebychev polynomial, respectively; $\KS \pip \g$ stands for the $B$ background category composed of \TDBbkgBa and \TDBbkgBb decays, while \conti stands for the continuum background.}
\vspace{5pt}
\begin{tabular}{c|c||l|r r r r|r r|r r r r r r|r r|r|r r|r r r}
\hline
\hline
	&	 Fit  	&	 \multirow{2}{*}{Fit Parameter} 	&	 \multirow{2}{*}{$b_{\rm core}$}	&	 \multirow{2}{*}{$s_{\rm core}$}	&	 \multirow{2}{*}{$s_{\rm outlier}$}	&	 \multirow{2}{*}{$f_{\rm outlier}$ }	&	 \multirow{2}{*}{\S  }  	&	 \multirow{2}{*}{\C }  	&	 \multirow{2}{*}{$\textrm{CB}_{\mu}^{p_0}$ }	&	 \multirow{2}{*}{$\textrm{CB}_{\mu}^{p_1}$ }	&	 \multirow{2}{*}{$\textrm{CB}_{\mu}^{p_2}$} 	&	 \multirow{2}{*}{$\textrm{CB}_{\sigma}^{p_0}$ }	&	 \multirow{2}{*}{$\textrm{CB}_{\sigma}^{p_1}$ }	&	 \multirow{2}{*}{$\textrm{CB}_{\sigma}^{p_2}$ }	&	 \multirow{2}{*}{$\tilde{\mathrm{G}}_{\sigma_R}$ }	&	 \multirow{2}{*}{$\tilde{\mathrm{G}}_{\sigma_L}$ }	&	 \multirow{2}{*}{$\textrm{G}_{\mu}$ }	&	 \multirow{2}{*}{$\textrm{Ch.}^{p_0}$ }	&	 \multirow{2}{*}{$\textrm{Ch.}^{p_1}$ }	&	 \multirow{2}{*}{Signal} 	&	 \multirow{2}{*}{\conti}	&	\multirow{2}{*}{$\KS \pip \g$}   	\\
	&	 variable	&		&		&		&		&		&		&		&		&		&		&		&		&		&		&		&		&		&		&		&		&	  	\\
\hline
\multirow{4}{*}{\rotatebox{90}{\conti}}		&	 \multirow{6}{*}{$\Delta t$}	&	$b_{\rm core}$ 	&	100.0 \\
	&		&	$s_{\rm core}$ 	&	4.7	&	100.0 \\
		&	 						&	$s_{\rm outlier}$ 	&	1.4	&	7.3	&	100.0 \\
		&	 						&	$f_{\rm outlier}$ 	&	6.6	&	60.9	&	35.3	&	100.0 \\
\cline{1-1}	\cline{3-8}

\multirow{2}{*}{\rotatebox{90}{Sig.}}		&		&	\S 				&	0.1	&	0.3	&	$-$0.1	&	0.2	&	100.0 \\
	&		&	\C 			&	1.1	&	$-$0.1	&	$-$0.3	&	0.1	&	$-$8.5	&	100.0 \\
\cline{1-10}
\multirow{9}{*}{\rotatebox{90}{Signal}}		&	 \multirow{6}{*}{\mes}	&	$\textrm{CB}_{\mu}^{p_0}$ 	&	$-$0.2	&	0.3	&	0.2	&	0.2	&	$-$0.9	&	$-$2.2	&	100.0 \\
		&	 					&	$\textrm{CB}_{\mu}^{p_1}$ 	&	$-$0.7	&	$-$0.8	&	0.2	&	$-$0.1	&	4.9	&	$-$5.5	&	$-$1.3	&	100.0 \\
		&	 					&	$\textrm{CB}_{\mu}^{p_2}$ 	&	$-$0.9	&	0.0	&	0.1	&	$-$0.3	&	1.9	&	$-$3.8	&	$-$39.5	&	56.9	&	100.0 \\
		&	 					&	$\textrm{CB}_{\sigma}^{p_0}$ 	&	$-$0.3	&	0.3	&	$-$0.4	&	0.2	&	2.3	&	5.3	&	$-$11.7	&	$-$2.8	&	19.9	&	100.0 \\
		&	 					&	$\textrm{CB}_{\sigma}^{p_1}$ 	&	$-$0.2	&	0.1	&	$-$0.3	&	0.0	&	4.9	&	0.9	&	$-$4.4	&	5.8	&	8.3	&	15.1	&	100.0 \\
		&	 					&	$\textrm{CB}_{\sigma}^{p_2}$ 	&	0.6	&	$-$1.4	&	$-$0.1	&	$-$0.2	&	1.0	&	$-$1.4	&	17.3	&	7.0	&	$-$38.7	&	$-$39.5	&	46.7	&	100.0 \\
\cline{2-16}
		&	 \multirow{2}{*}{\DeltaE}	&	$\tilde{\mathrm{G}}_{\sigma_R}$ 	&	1.1	&	$-$2.2	&	0.0	&	0.1	&	$-$1.0	&	$-$0.8	&	6.0	&	0.1	&	$-$17.1	&	$-$15.3	&	$-$10.0	&	18.1	&	100.0 \\
		&	 						&	$\tilde{\mathrm{G}}_{\sigma_L}$ 	&	1.1	&	0.3	&	$-$0.2	&	1.1	&	$-$1.8	&	2.3	&	4.3	&	$-$22.0	&	$-$27.8	&	$-$2.3	&	16.3	&	23.2	&	33.1	&	100.0 \\
\cline{2-18}
		&	 Fisher 	&	$\textrm{G}_{\mu}$ 	&	$-$0.6	&	2.3	&	$-$0.1	&	0.7	&	$-$0.3	&	$-$6.8	&	0.3	&	4.0	&	3.2	&	$-$2.2	&	$-$0.7	&	$-$3.0	&	$-$4.7	&	$-$10.1	&	100.0 \\
\cline{1-19}
\multirow{2}{*}{\rotatebox{90}{\conti}}		&	  \multirow{2}{*}{\DeltaE} 	&	$\textrm{Ch.}^{p_0}$ 	&	0.2	&	0.3	&	0.2	&	1.4	&	0.2	&	0.4	&	0.9	&	0.1	&	$-$2.7	&	$-$0.1	&	$-$1.5	&	1.9	&	7.5	&	2.5	&	$-$2.6	&	100.0 \\
		&	 						&	$\textrm{Ch.}^{p_1}$ 	&	0.2	&	1.4	&	0.0	&	1.0	&	0.2	&	1.9	&	$-$1.3	&	$-$1.9	&	$-$0.3	&	4.4	&	2.2	&	$-$0.4	&	$-$4.0	&	3.0	&	$-$4.1	&	$-$3.6	&	100.0 \\
\cline{1-21}
\multirow{3}{*}{\rotatebox{90}{Yields}}		&	  	&	Signal 				&	1.3	&	$-$4.4	&	$-$0.2	&	$-$0.5	&	$-$0.7	&	7.4	&	0.2	&	$-$7.3	&	$-$15.1	&	9.2	&	3.3	&	11.9	&	24.1	&	23.3	&	$-$20.6	&	8.3	&	6.0	&	100.0 \\
		&	 	&	\conti 			&	$-$0.1	&	7.9	&	0.6	&	3.1	&	0.3	&	$-$1.5	&	0.4	&	$-$0.1	&	2.2	&	$-$1.3	&	0.3	&	$-$2.3	&	$-$5.7	&	$-$0.2	&	5.3	&	0.6	&	2.0	&	$-$12.0	&	100.0 \\
		&	 	&	$\KS \pip \g$   	&	$-$1.2	&	$-$8.5	&	$-$0.8	&	$-$5.6	&	$-$1.1	&	$-$3.2	&	$-$1.4	&	8.4	&	10.1	&	$-$5.5	&	$-$5.1	&	$-$5.9	&	$-$5.7	&	$-$24.4	&	7.6	&	$-$19.0	&	$-$9.7	&	$-$23.1	&	$-$21.7	&	100.0 \\
\hline
\hline
\end{tabular}
\end{center}
\end{sidewaystable}

\end{widetext}

\clearpage


\end{document}